\documentclass[aps,a4paper, preprint, superscriptaddress,preprintnumbers,floatfix,
nofootinbib,amsmath,amssymb]{revtex4}
\usepackage{graphicx}
\usepackage{epsfig}
\usepackage{amsmath}
\usepackage{amsfonts}
\usepackage{amssymb}
\usepackage{url}
\usepackage{hyperref}
\usepackage{subfigure}
\usepackage{hhline}
\usepackage{color}
\usepackage{bm}
\usepackage{cancel}
\newcommand{\beqa}{\begin{eqnarray}}
\newcommand{\eeqa}{\end{eqnarray}}
\newcommand{\beq}{\begin{equation}}
\newcommand{\eeq}{\end{equation}}

\newcommand{\nn}{\nonumber}
\newcommand{\bmt}{\begin{pmatrix}}
\newcommand{\emt}{\end{pmatrix}}
\usepackage[toc,page]{appendix}
\usepackage{comment}
\newcommand{\be}{\begin{equation}}
\newcommand{\ee}{\end{equation}}
\newcommand{\bea}{\begin{eqnarray}}
\newcommand{\eea}{\end{eqnarray}}

\def\Bbar{\overline{B}}

\def\nubar{{\overline{\nu}}}

\def\A{\mathcal{A}}
\def\O{\mathcal{O}}

\def\th{{\theta}}
\begin{document}
\title{$B_c \to D^{(*)}\tau \bar \nu_\tau$ processes in an effective field theory approach}

\author{Anupama Bhol}
\email{anupama.phy@gmail.com}
 \affiliation{Govt. Women's College Baripada-757001, India}                                      
  \author{Suchismita Sahoo}
\email{suchismita8792@gmail.com}
 \affiliation{Department of Physics, Central University of Karnataka, Kalaburagi-585367, India}
  \author{Soram Robertson Singh}
\email{robsoram@gmail.com}
 \affiliation{Department of Physics, Manipur University, Canchipur - 795003, India }

\begin{abstract}
We present a compressive study on rare semileptonic $B_c \to D^{(*)}\tau \bar \nu_\tau$ decays involving $b \to u \tau \bar \nu_\tau$  quark level transitions in an effective field theory approach. We consider the presence of an additional (pseudo)vector and (pseudo)scalar type interactions which can be either complex or real and constrain the new couplings using the existing data on $R_{D^{(*)}}$, $R_{J/\psi}$, $R_\pi^l$, Br($B_{u,c} \to \tau \bar \nu_\tau$) and Br($B \to\pi \tau \bar \nu_\tau$)  parameters. In order to segregate the sensitivity of  new coefficients, we check the effects of these  couplings on the branching ratios,  lepton non-universality and various angular observables of $B_c \to D^{(*)}\tau \bar \nu_\tau$ processes.

\end{abstract}
\maketitle

\section{Introduction}
The standard model (SM) of particle physics is successful in understanding the fundamental particles, their interactions and in explaining various experimental observations with precise prediction of a wide variety 
of phenomena. But still it leaves some phenomena such as dark matter, dark energy, neutrino mass, matter antimatter asymmetry in the universe unexplained. In fact it is not a complete theory. There remains a huge possibility of physics beyond the SM to explain the deficiencies contained within in it. Again in last few years some tensions have been observed by dedicated experiments such as BABAR, Belle, and more recently by LHCb in various B meson decays with the flavor changing charged current (FCCC) and the flavor changing 
neutral current (FCNC) quark level transitions which account for the lepton flavor universality violation (LFUV) to look for new physics effects.

The lepton flavor universality (LFU) of the standard model (SM) demands the strength of couplings between 
all electroweak gauge bosons ($\gamma,$ $Z\,\,$and $W^{\pm} $) and the three families of leptons 
($e^-,$ $\mu^-\,$\, and $\tau^{-}$) to be equal and the only differences arise due to the mass hierarchy, since $m_e < m_\mu << m_\tau$. Recently, various anomalies have been reported in the sector of lepton flavor universality, that enforces researchers to study beyond the SM. The observed ratio of branching fractions of $R_{D^{(*)}}, R_{J/\psi}$ in the semileptonic decays of $B \to D^{(*)} \,l\, \bar{\nu}_l$ and 
$B_c \to {J/\Psi}\, l \bar{\nu}_l$ respectively provide the hints of LFU violation, are defined as 
\bea
R_{D^{({\ast})}} = \frac{\mathcal {\rm Br}( B \to D^{({\ast})}\tau \bar{\nu}_\tau)}{\mathcal {\rm Br}( B \to D^{({\ast})}\, l \bar{\nu}_l)}, \qquad\qquad R_{J/\Psi} = \frac{\mathcal {\rm Br}( B_c \to {J/\Psi}\tau \bar{\nu}_\tau)}{\mathcal {\rm Br}( B_c \to {J/\Psi}\, l \bar{\nu}_l)}.
\eea
The ratios of branching fractions, $R_{D^{(*)}}$ have been measured by BaBaR~\cite{Lees:2012xj,Lees:2013uzd}, Belle~\cite{Huschle:2015rga,Hirose:2016wfn,Hirose:2017dxl,Abdesselam:2019dgh} collaborations as well as LHCb~\cite{Aaij:2015yra,Aaij:2017uff,Aaij:2017deq} where $l= \,e $ or $\mu $. The average value of all the measurements of $R_{D}$ and $R_{D^{\ast}}$ as found by Heavy Flavor Averaging Group (HFLAV)~\cite{HFLAV},
\bea
R_{D}^{~\rm{avg}}=0.340 \pm 0.027 \pm 0.013,      \qquad   \qquad R_{D^{\ast}}^{~\rm{avg}}=0.295 \pm 0.011 \pm 0.008,
\eea
exceed the arithmetic average of the latest Standard Model (SM) predictions~\cite{Lattice:2015rga,Na:2015kha,Aoki:2016frl,Bigi:2016mdz,Fajfer:2012vx,Bernlochner:2017jka,Bigi:2017jbd,Jaiswal:2017rve} by 1.4 $\sigma$ and 2.5 $\sigma$ respectively.
 \bea
 R_{D}^{~\rm{SM}}=0.299 \pm 0.003,      \qquad   \qquad R_{D^{\ast}}^{~\rm{SM}}=0.258 \pm 0.005 .  
 \eea
Similarly in $B_c \to {J/\Psi}\, l \bar{\nu}_l$ decays, the measured ratio $R_{J/\Psi}$~\cite{Aaij:2017tyk} by LHCb
\bea
R_{J/\Psi}=0.71 \pm 0.17 \pm 0.18 
\eea
 shows a deviation of around 1.7 $\sigma$ from the SM prediction~\cite{Cohen:2018dgz,Dutta:2017xmj,Ivanov:2005fd,Wen-Fei:2013uea}
 \bea 
 R_{J/\Psi}^{~\rm{SM}}= 0.289 \pm 0.01
 \eea
 at the 95 $\%$  CL.
Moreover, 1.4 standard deviation is observed in $\mathcal{B}(B\to \tau \nu)$ decay~\cite{Ciezarek:2017yzh} mediated via $b\to u l\nu$ charged current transition which again implies intriguing hints of LFU violating new physics (NP) beyond the SM. Similarly, deviation is also reported in the observable $R^l_\pi $ ~\cite{Tanabashi:2018oca} with
 $b\to u l \bar{\nu}$ transition process can be defined as,
 \bea
 R^l_\pi = \frac{Br(B^- \to \tau^- \bar {\nu}_\tau)}{Br(B^0 \to \pi^+ l^- \bar {\nu}_l)}
 \eea 

Since the theoretical uncertainties due to the CKM matrix elements and hadronic form factors cancel out 
to a large extent in the observables like ratios of branching fractions, this leads to predict with higher accuracy. Therefore, the lepton flavor universality violating studies
are the most powerful tools to probe new physics beyond the standard
model. There have been a lot of works in the last few years to understand the nature of
NP that can be responsible for such deviations.

Experimentally, $B_c$ meson, first observed by The CDF Collaboration at Fermilab ~\cite{Abe:1998fb,
Ackerstaff:1998zf}, is unique in the SM as it is the only known heavy meson consisting of two heavy 
quarks, a b (bottom) and a c (charm) quark, of different flavors and charges. At present, a few measurements of its properties from Tevatron data~\cite{Abe:1998fb,Abulencia:2005usa,Abulencia:2006zu,Aaltonen:2007gv} exist before the operation of the LHC. The LHCb experiments assure the first detailed study of $B_c$ meson.  More precise measurements of its mass and lifetime are now feasible, and several decay channels have been witnessed for the first time.The decay modes of $B_c$ meson are different from decay modes of $B_{u,d,s}$ mesons in the heavy quark limit. However heavy flavor and spin symmetries must be reconsidered as both the constituent quarks are heavy in case of $B_c$ meson. 
Since the accessible kinematic range is broader in the decays of $B_c$ meson than for the decays of $B_s$ 
and $B_d$ meson, many weak decays are kinematically allowed in case of $B_c$ meson while restricted in the other meson system. 
Furthermore, it can only decay weakly because its mass is below the $B\,D$ threshold.
Besides that, the $B_c$ meson can decay via $ b \to (u, d, c, s)$ and $ c \to (u, d, s)$ transition decays.
Thus it offers a very rich laboratory for studying various decay channels which
are important both theoretically and experimentally. These $B_c$ meson decays provide complimentary decay channels to similar decays in the other B mesons and the possibility to extract the CKM parameter $V_{ub}$ 
as well. The precise measurements for such semileptonic $B_c$ decays can play a significant role in testing the SM and in searching for the signal of the new physics (NP) beyond the SM. 

Further the lifetime of $B_c$ meson put severe constraint on scalar NP couplings~\cite{Alonso:2016oyd}. According to the SM the rate of $B_c \to \tau \nu $ does not exceed the fraction of the total width which leads to a very strong bound on new-physics scenarios involving scalar interactions. The lifetime of 
$B_c$ meson, $\tau_{B_c} =0.52^{+0.18}_{-0.12}\, {\rm ps} $~\cite{Chang:2000ac} is found to be consistent 
with the experimental value of $\tau_{B_c} =0.507(9) ps$~\cite{Tanabashi:2018oca}. The branching ratio
 of $B_c\to \tau \nu \leq 5 \% $ as predicted by various 
 SM calculations~\cite{Bigi:1995fs,Beneke:1996xe,Chang:2000ac}, 
 but this can be relaxed up to $30 \%$ after using different the input parameters. Again LEP data taken at 
 the Z peak desires $\mathcal{BR}(B_c\to \tau \nu) \leq 10 \%$~\cite{Akeroyd:2017mhr}.

Precise measurements of the branching ratios and the ratio of branching fraction  of hadrons, definitely 
play an important role in constraining the new physics couplings. The measured branching ratios of decays 
such as $B_u \to \tau\nu$, $B_c \to \tau\nu$ and the ratio of branching fraction $R_D^{(\ast)}$, $R_{J/\Psi}$ all provide constraints on the coupling constants and the masses of new physics particles, and often such constraints are very significant and severe than those that are resulted from direct searches at the LHC. A few works on the potential of the $B_c$ meson to probe the presence of new physics interactions have been performed. In particular the presence of a charged Higgs boson $(H^{\pm})$~\cite{Du:1997pm,Mangano:1997md,Akeroyd:2008ac} or supersymmetric particles with specific R-parity violating couplings~\cite{Baek:1999ch,Akeroyd:2002cs} can reinforce branching ratio of $B_{c,(u)} \to \tau\nu$ decay.

 The decay amplitude of $B_c \to D^{(*)} \,\tau\, \bar{\nu}_\tau $ computation includes both leptonic and hadronic matrix elements. The evalution of hadronic matrix element depends on various meson to meson transition form factors. Several approaches exist in literature where semileptonic decays of $B_c$ meson 
 have been investigated extensively \cite{Dhir:2009ub,Du:1988ws,Colangelo:1992cx,Nobes:2000pm,Ivanov:2000aj,Kiselev:2000pp,Kiselev:2002vz,Ebert:2003cn,Ebert:2003wc,Wang:2008xt} in the framework of Bauer-Stech-Wirbel relativistic quark model, the QCD sum rules, the covariant light front quark model, the relativistic constituent quark model, and the non relativistic QCD. We follow the covariant light front quark model~\cite{Wang:2008xt} for the $B_c \to D^{(\ast)}$ transition form factors. Recently the problem has been addresed by authors in Ref \cite{Dutta:2018vgu,Leljak:2019fqa} within the SM. In this paper,we consider the presence of an additional vector and scalar type interactions which can be either complex or real beyond the SM. Since, the recent measurements propose that, there is a possibility to find new physics in the third generation leptons only. However, more experimental studies are required to confirm the existence of NP. The decay processes with a tau lepton in the final state are more sensitive to new physics than the 
processes with first two generation leptons due to the large mass of the tau lepton. Therefore,
 decays with a tau lepton in the final state can act as an excellent probe of new physics as these are 
 easily affected by non-SM contributions arising from the violation of lepton flavor universality
 among all the leptonic and semileptonic decays. In this context we will focus here on anomalies present in $B_c\to D^{(*)} \,\tau\, \bar{\nu}_\tau $ meson decays mediated via $b\to u\,\tau\,\nu_{\tau} $ charged current interactions in the presence of new physics
 and constrain the new physics couplings from the $\chi^2$ fit of $R_{D^{(*)}}$, $R_{J/\psi}$, $R_\pi^l$, Br($B_{u,c} \to \tau \bar \nu_\tau$) and Br($B \to\pi \tau \bar \nu_\tau$). Then we study effect of these couplings on the branching ratios,  lepton non-universality and various angular observables of 
 $B_c \to D^{(*)}\tau \bar \nu_\tau$ processes.

The  paper is organised as follows. We present the effective Hamiltonian associated with semileptonic decays involving $b \to u \tau \nu_\tau$  quark level transitions and numerical fit to new coefficients  in section II.  In section III, we provide the differential decay distribution for the 
$B_c \to D^{(*)}\tau \bar \nu_\tau$ semileptonic decays and we write down expressions of various observables pertaining to $B_c \to D^{(*)}\tau \bar \nu_\tau$ decays. The numerical analysis of all the physical observables of $B_c \to D^{(*)}\tau \bar \nu_\tau$ decay modes in the presence of scalar and vector type couplings using $\chi^2$ fit and discussions are given in section IV and section V summarize our estimated results.

\section{Effective Hamiltonian and Numerical fit to new coefficients}

The most general effective Hamiltonian responsible for rare semileptonic processes mediated via $b\to u \tau \bar \nu_l$ transition (neutrinos are left handed only) is given by \cite{Sakaki:2013bfa},
\bea \label{ham-bu}
\mathcal{H}_{\rm eff}=\frac{4G_F}{\sqrt{2}} V_{cb} \Big [ \left(\delta_{l\tau} + V_L \right) \mathcal{O}_{V_L}^l + V_R \mathcal{O}_{V_R}^l +  S_L \mathcal{O}_{S_L}^l +  S_R \mathcal{O}_{S_R}^l + T \mathcal{O}_T^l \Big ],
\eea
where $l=e,\mu,\tau$ are the flavor of neutrinos,  $\O_C$'s $(C=V_{L(R)},~S_{L(R)},~T)$ are the six dimensional effective  operators,  $C$'s are their corresponding Wilson coefficients, which though vanish in the SM, but can have nonzero values in the presence of new physics. 

\begin{figure}[htb]
\includegraphics[width=0.4\textwidth]{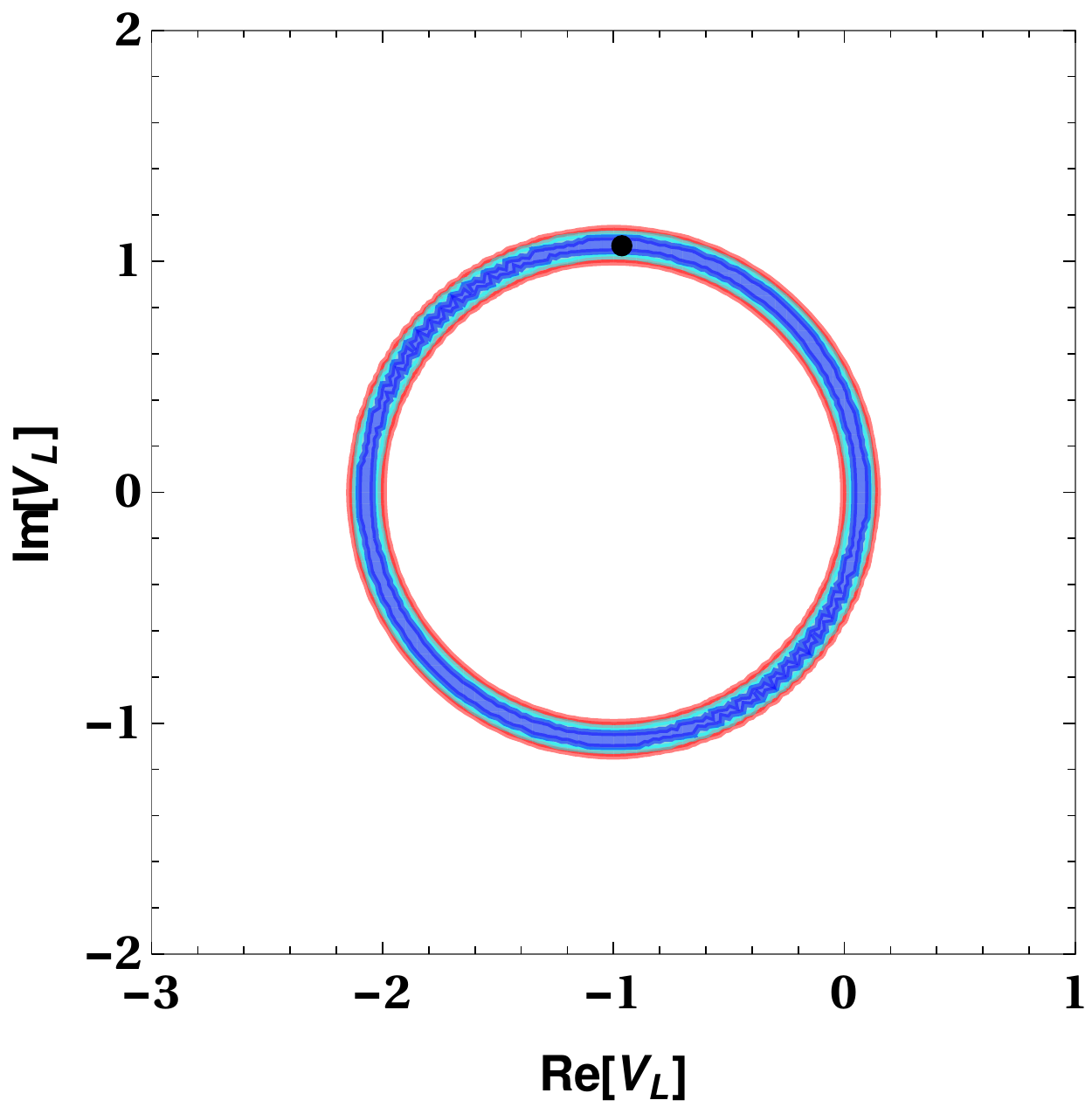}
\quad
\includegraphics[width=0.42\textwidth]{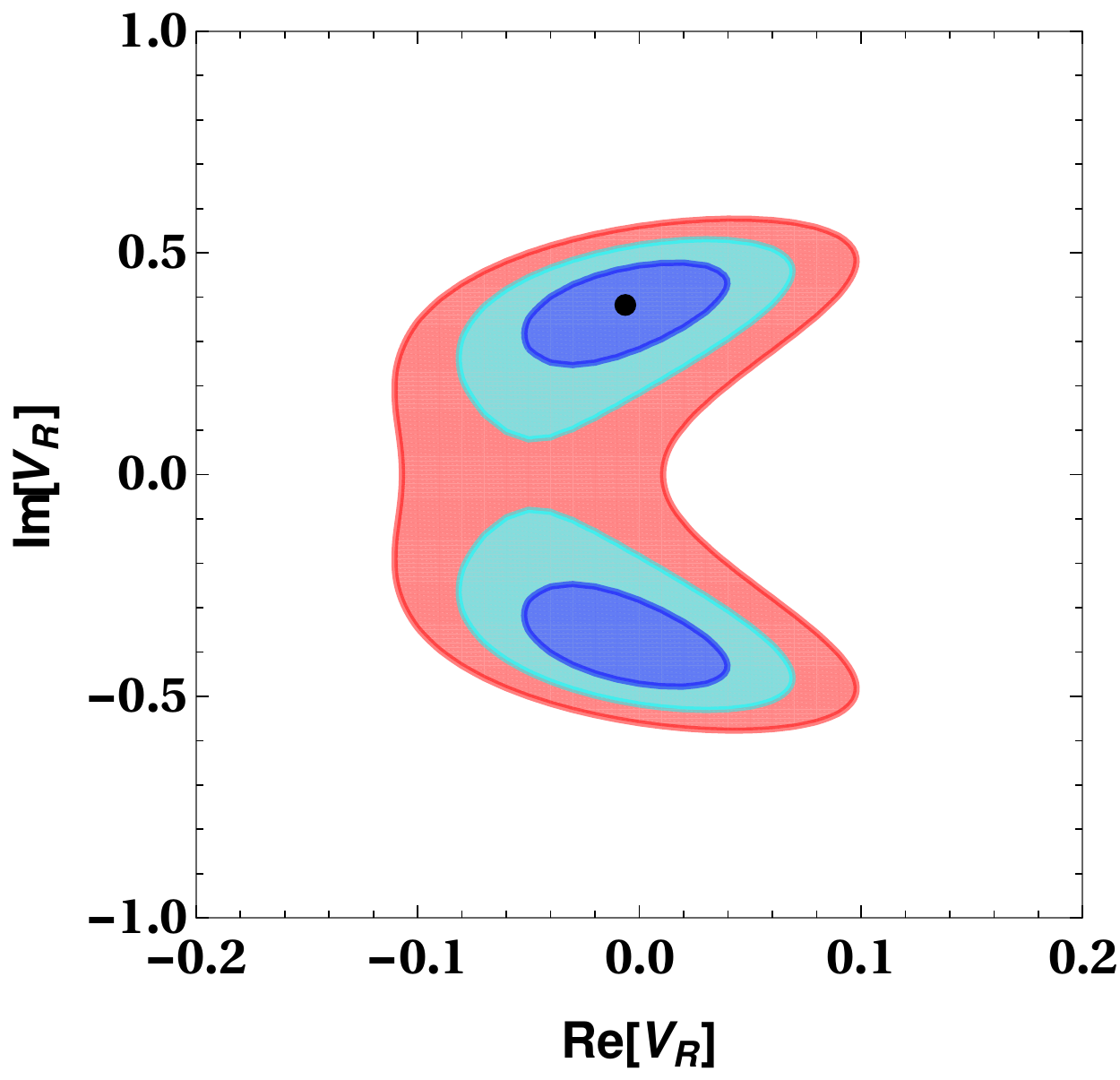}
\quad
\includegraphics[width=0.42\textwidth]{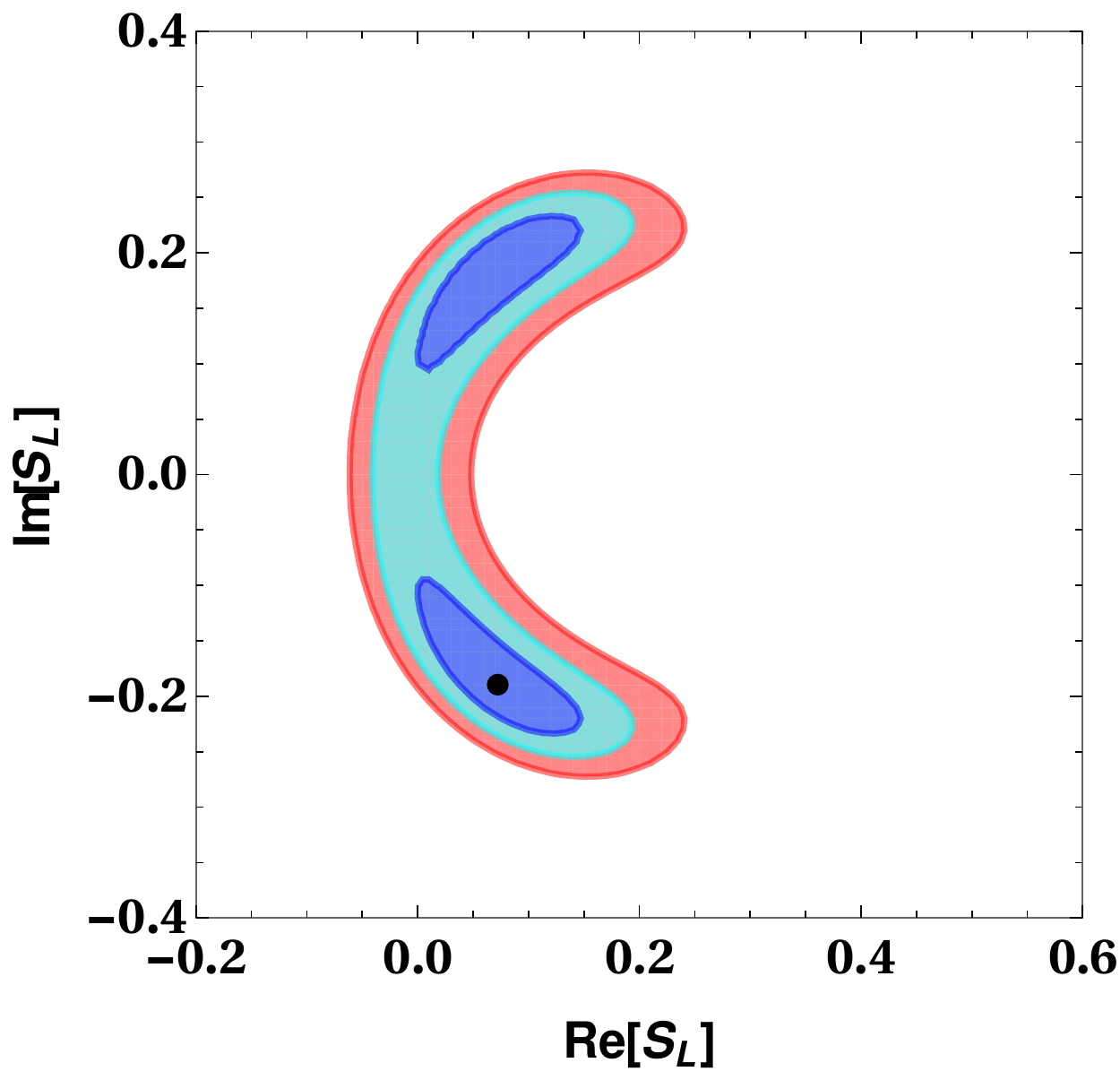}
\quad
\includegraphics[width=0.42\textwidth]{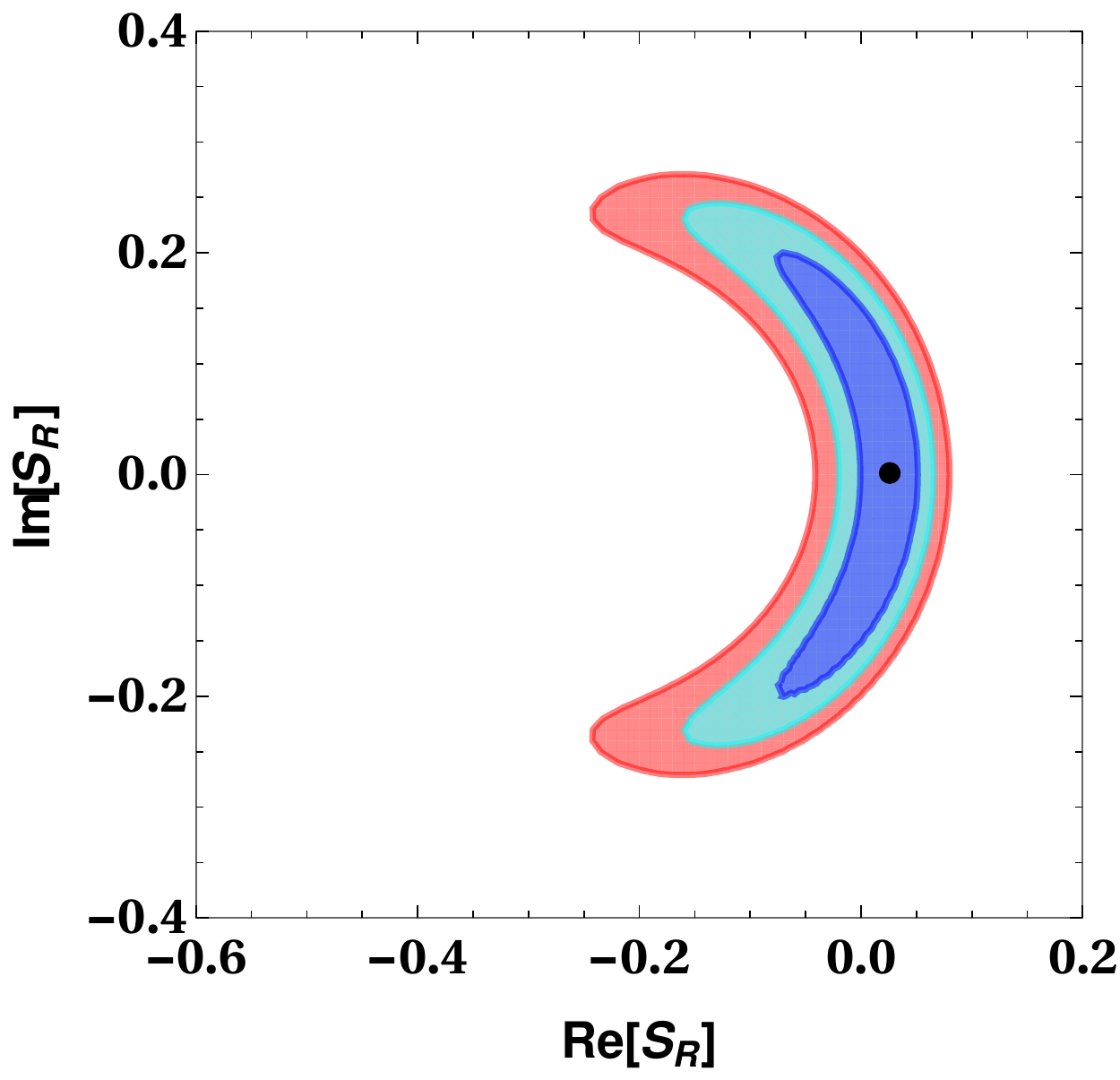}
\caption{Constraints on real and imaginary parts of the new coefficients obtained from $R_{D^{(*)}}$, $R_{J/\psi}$, $R_\pi^l$, Br($B_{u,c} \to \tau \bar \nu_\tau$) and Br($B \to\pi \tau \bar \nu_\tau$) observables. Here the black dots represent the best-fit values. }\label{Fig:Case-A}
\end{figure}

With the assumption  that the coupling of $b \to u$ and $b \to c$ transitions are same,   we $\chi^2$ fit the new coefficients to  $R_{D^{(*)}}$, $R_{J/\psi}$, $R_\pi^l$, Br($B_{u,c} \to \tau \bar \nu_\tau$) and Br($B \to\pi \tau \bar \nu_\tau$) observables,  defined as 
\bea
\chi^2(C)=\sum_i \frac{(\mathcal{O}_i^{\rm th}(C)-\mathcal{O}_i^{\rm Expt})^2}{(\Delta \mathcal{O}_i^{\rm Expt})^2+(\Delta \mathcal{O}_i^{\rm SM})^2}\,.
\eea
where $\mathcal{O}_i^{\rm th}(C)$ are the  theoretical predictions of observables, $\mathcal{O}_i^{\rm Expt}$ are  the respective experimental  central values, and $\Delta \mathcal{O}_i^{\rm Expt}$  ($\Delta \mathcal{O}_i^{\rm SM}$)  represent corresponding  experimental (SM) uncertainties. The constraint on individual real 
and complex new coefficients associated with $b \to c \tau \bar \nu_\tau$ from the $\chi^2$ fit to 
$R_{D^{(*)}}$, $R_{J/\psi}$ and Br($B_{c} \to \tau \bar \nu_\tau$)  is  presented in Ref. \cite{Sahoo:2019hbu}, and the complex new parameters linked to $b \to u \tau \bar \nu_\tau$ from the $\chi^2$ fit to  $R_\pi^l$, Br($B_{u} \to \tau \bar \nu_\tau$) and Br($B \to\pi \tau \bar \nu_\tau$) is computed in Ref. \cite{Sahoo:2019hbu}. Since the fit to tensor coefficient with the inclusion of an additional $R_\pi^l$, Br($B_{u} \to \tau \bar \nu_\tau$) and Br($B \to\pi \tau \bar \nu_\tau$) observables don't change the predictions in \cite{Sahoo:2019hbu}\,, we focus only on vector and scalar type coefficients. We consider new coefficients which are classified  as
\begin{itemize}
\item \textbf{Case A}: Existence of only  individual  new complex  coefficient.
\item \textbf{Case B:} Existence of only two  new real coefficients.
\end{itemize}
In Table \ref{Tab:Input-Fit}\,, we have quoted the  experimental and SM values of all the observables  
used in the fitting \cite{Zyla:2020zbs}.  
\begin{table}[htb]
\centering
\caption{Experimental and theoretical values of the observables used in the fitting} \label{Tab:Input-Fit}
\begin{tabular}{|c|c|c|}
\hline
Observables &~Experimental values~~&~SM Predictions~ \\
\hline
\hline
$R_D$ & $0.340 \pm 0.027 \pm 0.013$  &     $0.299 \pm 0.003$ \\
$R_{D^*}$ &  $ 0.295 \pm 0.011\pm0.008 $  &  $0.258 \pm 0.005 $\\
$R_{J/\psi}$  & $ 0.71 \pm 0.251 $   &$0.289 \pm 0.01$\\
${\rm Br}(B_c \to \tau \nu)$ & $< 30\%$     & $ (3.6 \pm 0.14) \times 10^{-2}$\\
\hline
$R_\pi^l$  & $0.699 \pm 0.156$  & $0.583\pm 0.055$\\
${\rm Br}(B_u \to \tau \nu)$ & $ (1.09\pm 0.24) \times 10^{-4}$ & $(8.48\pm0.5)  \times 10^{-5}$ \\
${\rm Br}(B^0 \to \pi^+ \tau \nu) $  & $< 2.5 \times 10^{-4}$  & $(9.40 \pm 0.75) \times 10^{-5}$\\
\hline
\end{tabular}
\end{table}
Fig. \ref{Fig:Case-A} depicts the constrained on individual complex couplings, $V_L$ (top-left panel), $V_R$ (top-right panel), $S_L$ (bottom-left panel) and $S_R$ (bottom-right panel) obtained from  $R_{D^{(*)}}$, $R_{J/\psi}$, $R_\pi^l$, Br($B_{u,c} \to \tau \bar \nu_\tau$) and Br($B \to\pi \tau \bar \nu_\tau$) observables. Here blue, cyan and magenta colors stand for $1\sigma$, $2\sigma$ and $3\sigma$, respectively 
and black dots represent the best-fit values. The predicted best-fit values, $\chi^2_{\rm min}/{\rm d.o.f}$ and pull$(=\sqrt{\chi^2_{C,~ \rm min}-\chi^2_{\rm SM,~min}})$ values of  complex coefficients  are presented  in Table \ref{Tab:Best-fit}\,. The minimum value of $\chi^2$ in the SM is $\chi^2_{\rm SM, min}=12.803$. The global fit to the complex $V_L$ and $V_R$ coefficients are  too good i.e., the presence of either $V_L/V_R$ complex coefficients can explain all the $R_{D^{(*)}}$, $R_{J/\psi}$, $R_\pi^l$, Br($B_{u,c} \to \tau \bar \nu_\tau$) and Br($B \to\pi \tau \bar \nu_\tau$) data simultaneously. However, the  
$\chi^2_{S_{L(R)},\rm min}/{\rm d.o.f}$  are found to be greater than $1$ which implies that 
the fit is poor.  

\begin{figure}[htb]
\includegraphics[width=0.42\textwidth]{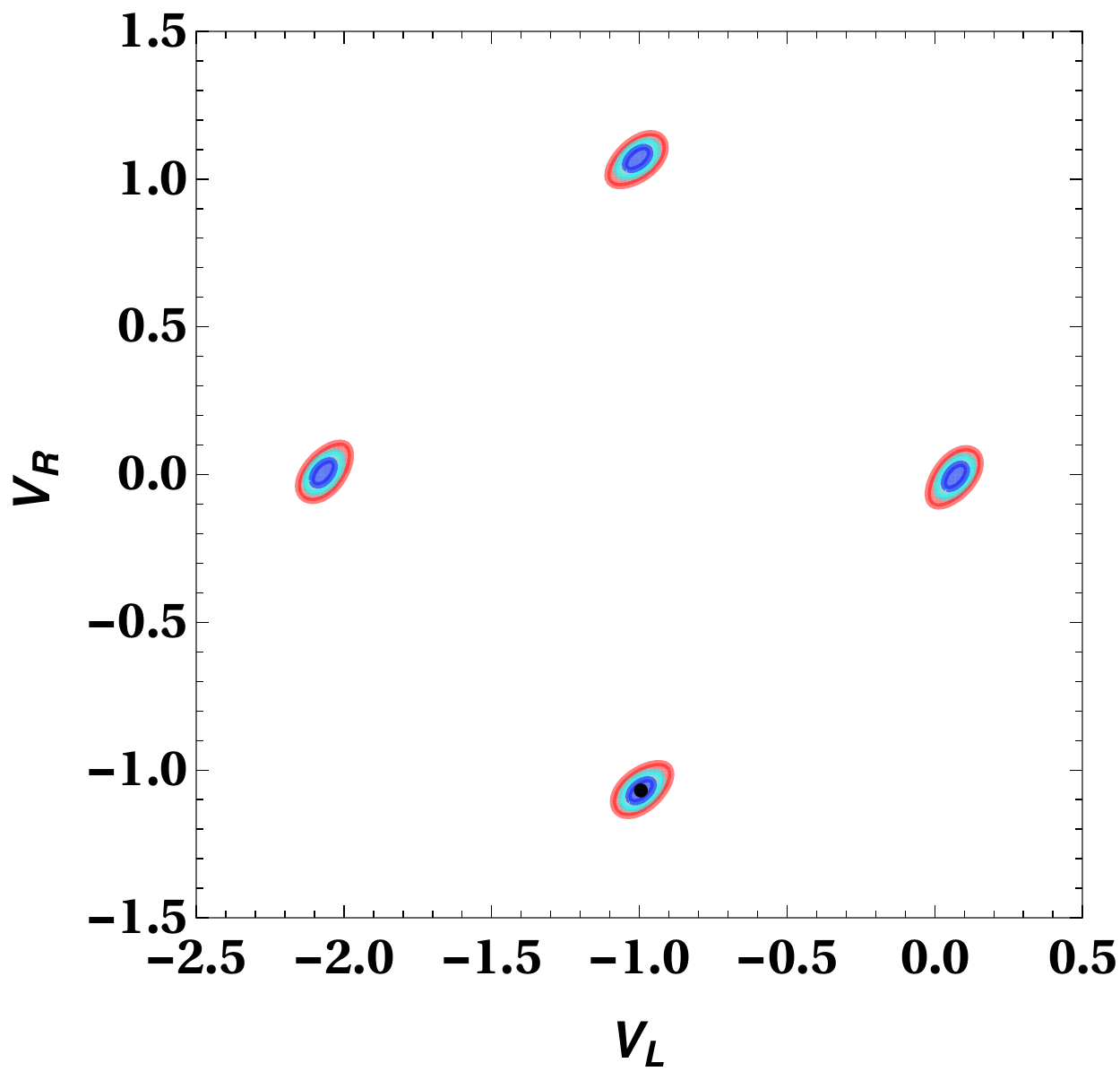}
\quad
\includegraphics[width=0.42\textwidth]{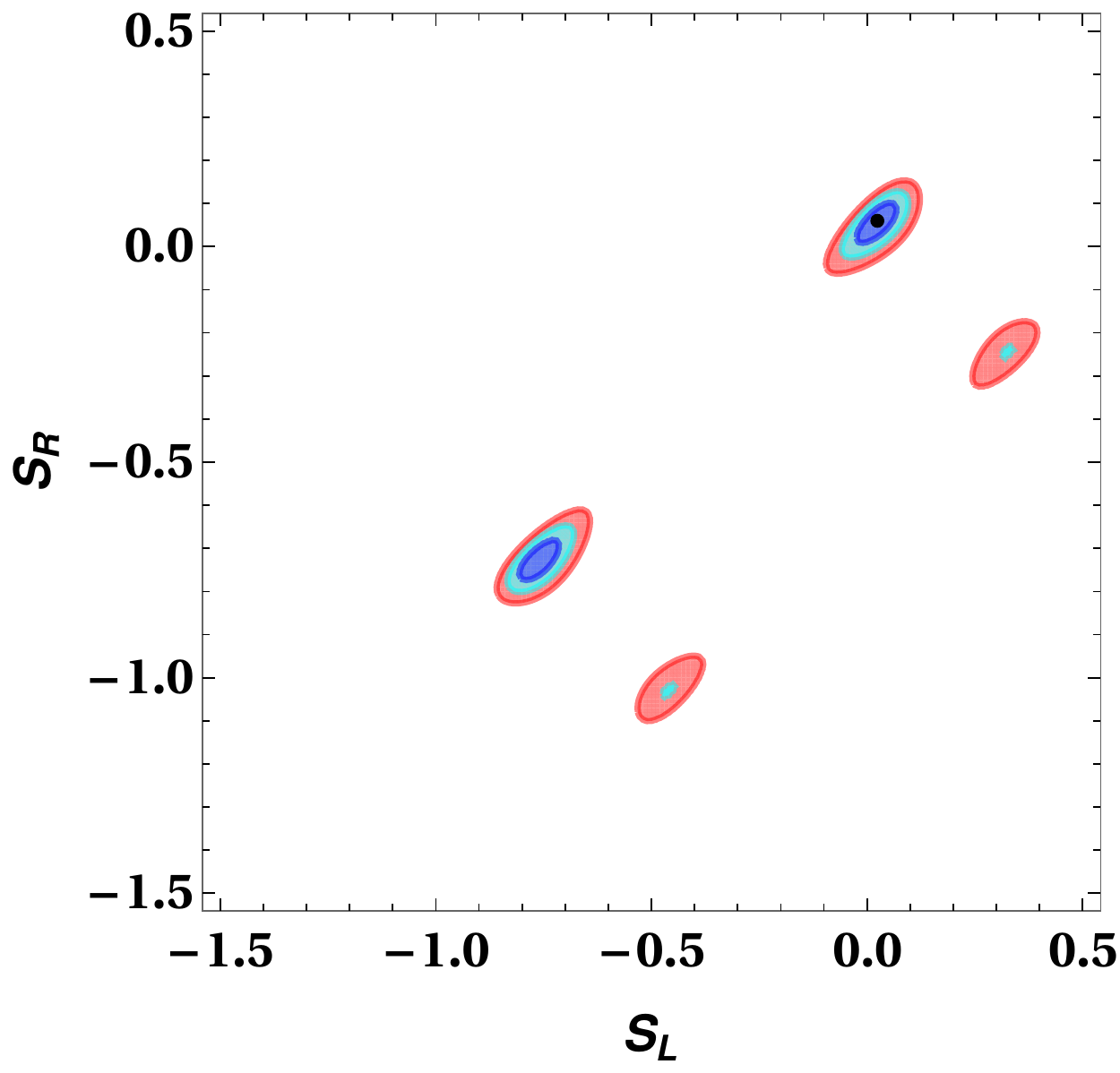}
\quad
\includegraphics[width=0.42\textwidth]{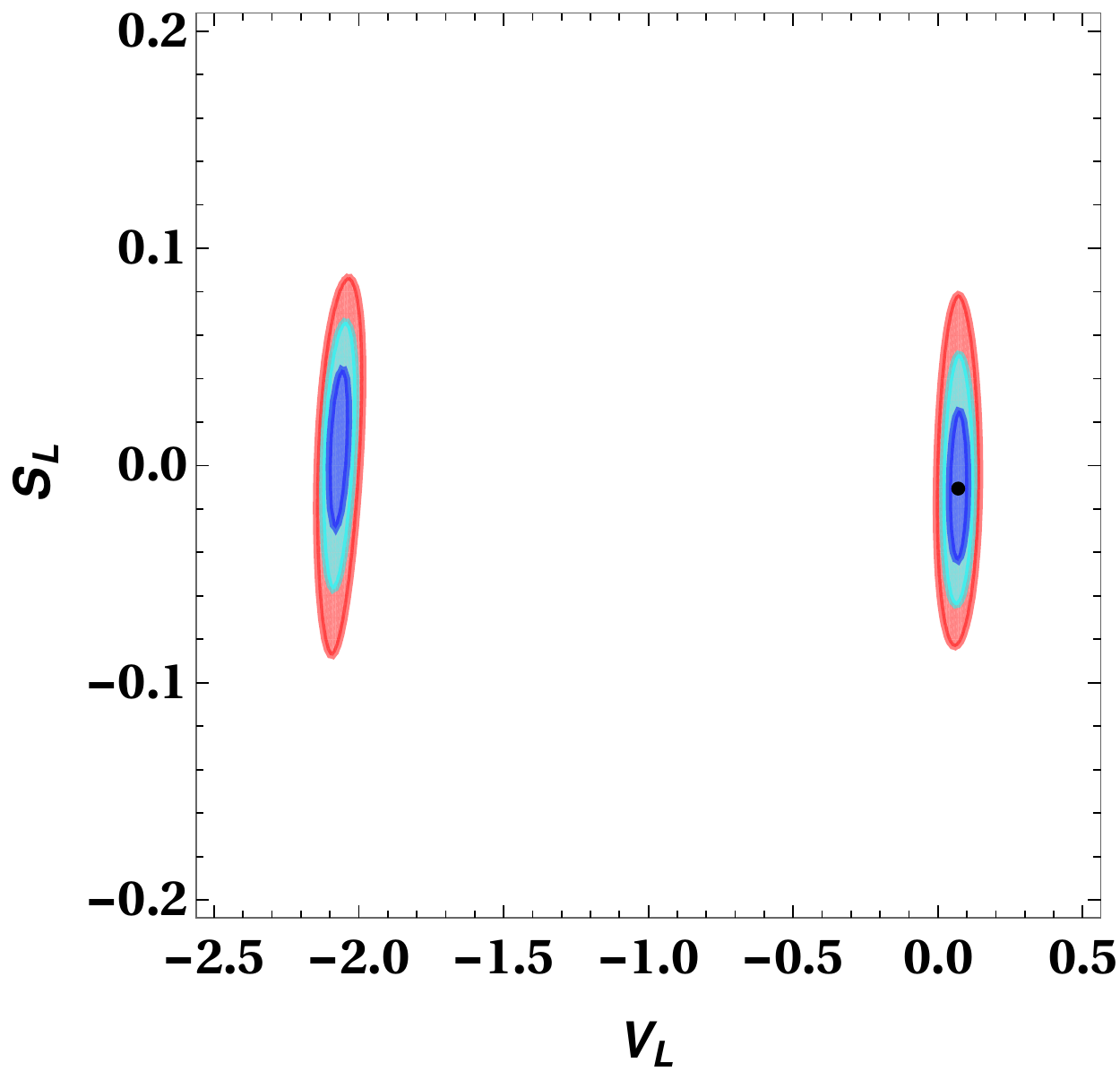}
\quad
\includegraphics[width=0.42\textwidth]{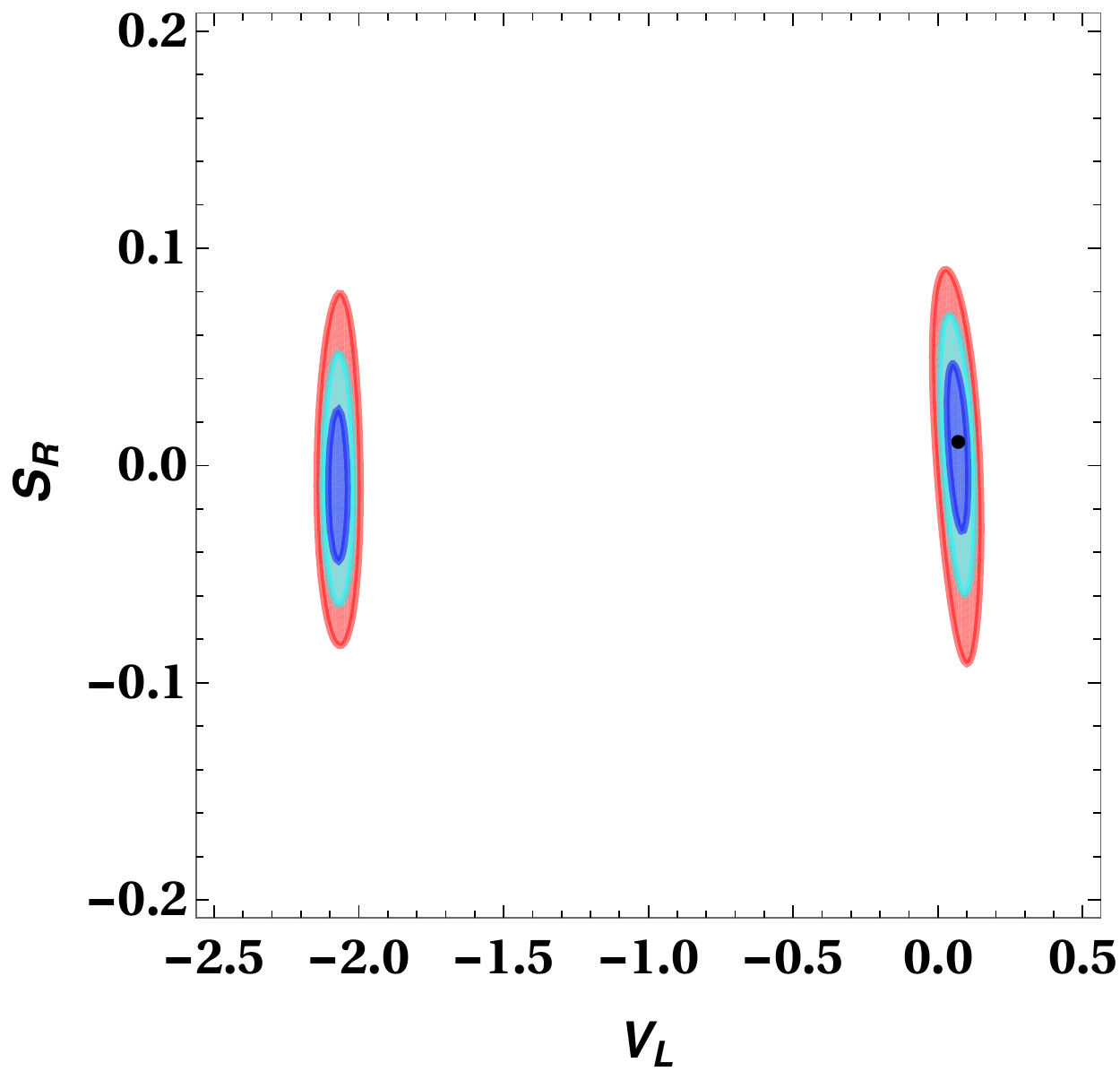}
\quad
\includegraphics[width=0.42\textwidth]{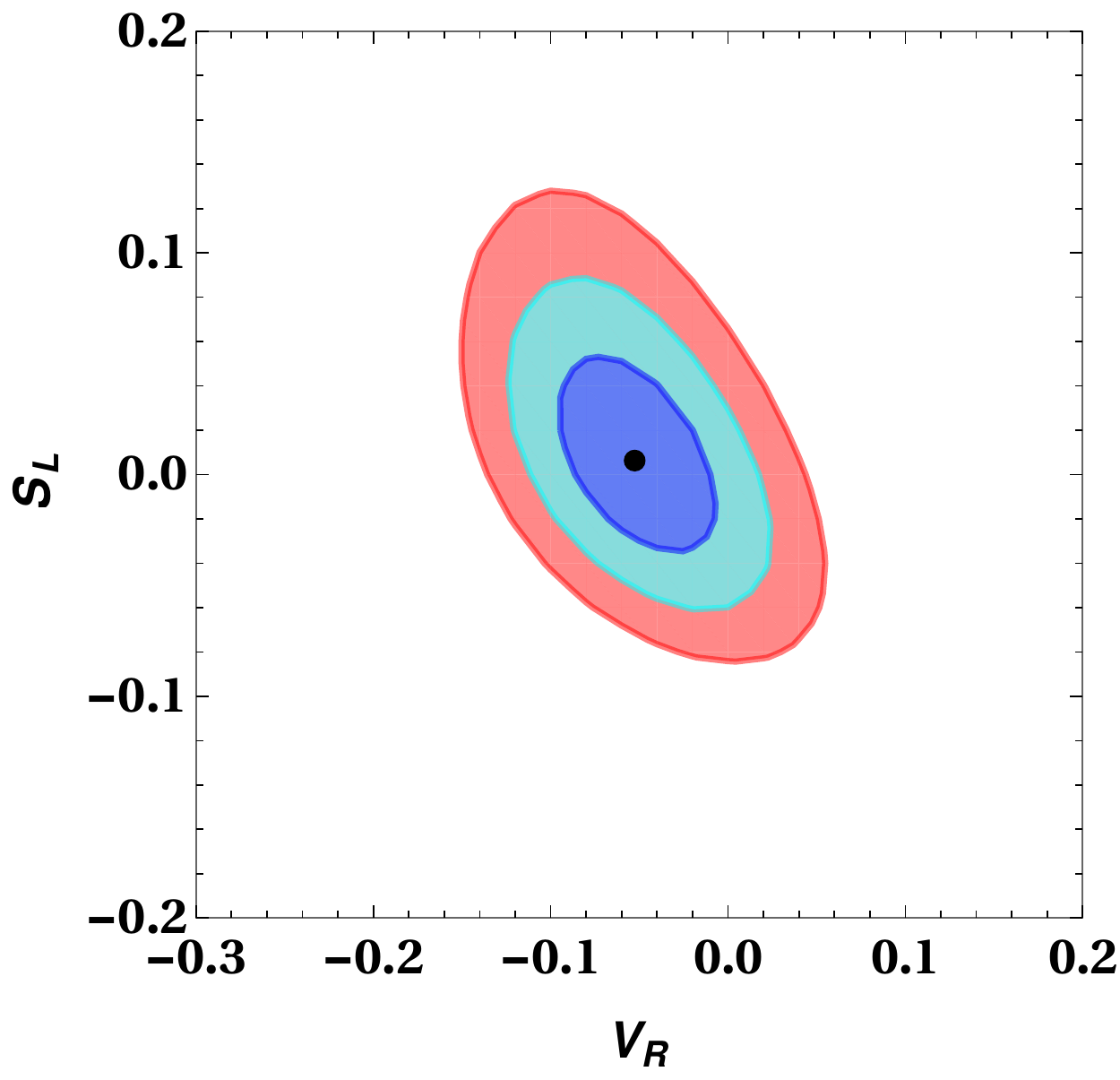}
\quad
\includegraphics[width=0.42\textwidth]{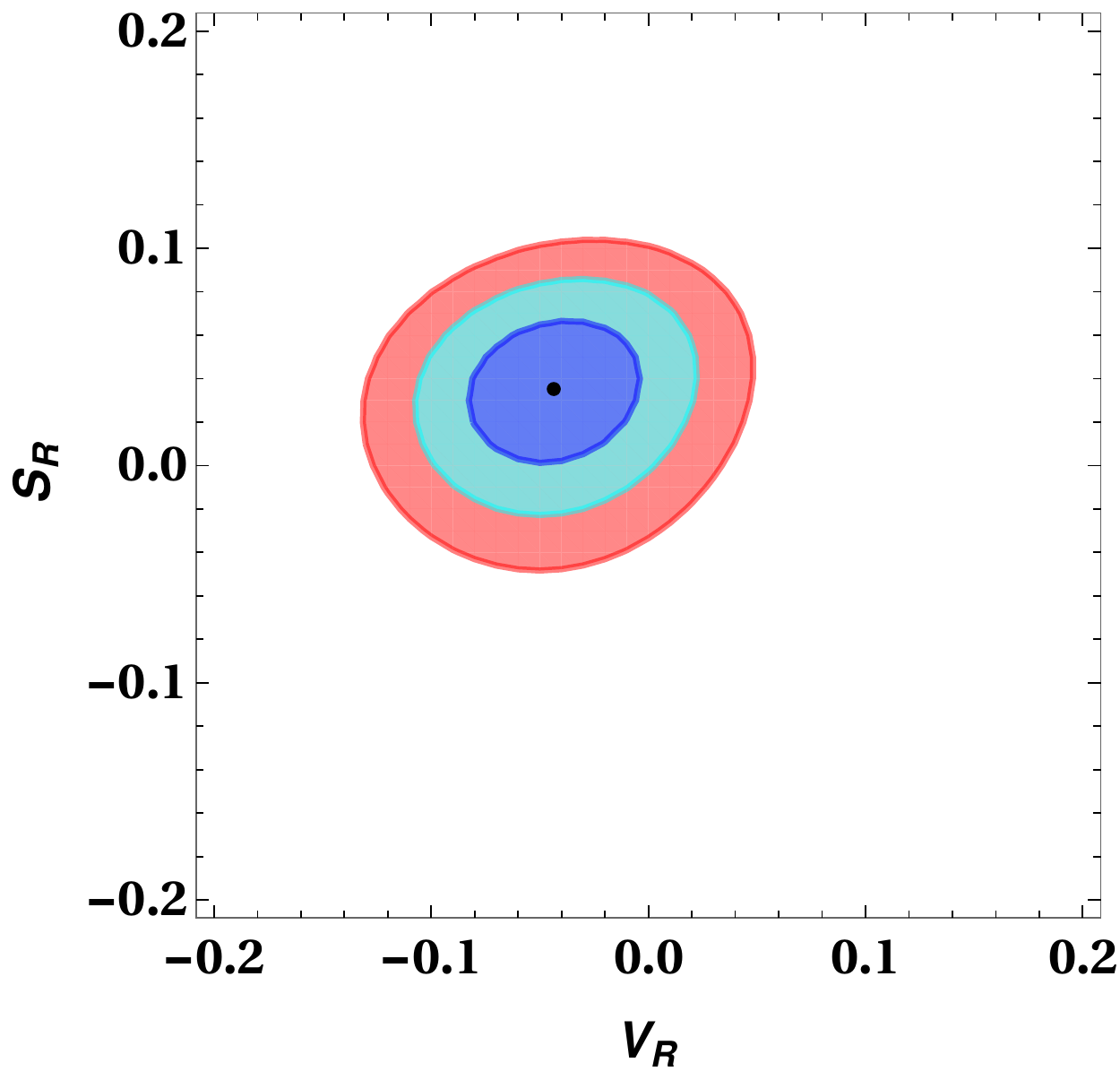}
\caption{Constraints on  new real coefficients obtained from $\chi^2$ fit to  $R_{D^{(*)}}$, $R_{J/\psi}$, $R_\pi^l$, Br($B_{u,c} \to \tau \bar \nu_\tau$) and Br($B \to\pi \tau \bar \nu_\tau$) observables. Here the black dots represent the best-fit values. }\label{Fig:Case-B}
\end{figure}

 \begin{table}[htb]
\begin{center}
\caption{Best-fit,  $\chi^2_{\rm min}/{\rm d.o.f}$ and pull values of new Wilson coefficients.}
\label{Tab:Best-fit}
\begin{tabular}{|c| c | c | c |c|}
\hline
Cases~&~New Wilson coefficients~& ~Best-fit values~&~ $\chi^2_{\rm min}/{\rm d.o.f}$~&~Pull~\\
\hline 
\hline
 Case A~&~$({\rm Re}[V_L],{\rm Im}[V_L])$~&~$(-0.9671,1.0723)$~&~$0.549$~&~$3.171$~\\
&~$({\rm Re}[V_R],{\rm Im}[V_R])$~&~$(-0.0062,0.3815)$~&~$0.543$~&~$3.1755$~\\
&~$({\rm Re}[S_L],{\rm Im}[S_L])$~&~$(0.0736,-0.188)$~&~$1.749$~&~$2.0134$~\\
&~$({\rm Re}[S_R],{\rm Im}[S_R])$~&~$(0.027,0)$~&~$2.1413$~&~$1.447$~\\
\hline
 Case B~&~$(V_L,V_R)$~&~$(-0.995,-1.07)$~&~$0.5425$~&~$3.176$~\\
 &~$(V_L,S_L)$~&~$(0.072,-0.011)$~&~$0.506$~&~$3.205$~\\
&~$(V_L,S_R)$~&~$(0.069,0.01)$~&~$0.518$~&~$3.196$~\\

&~$(V_R,S_L)$~&~$(-0.052,0.0065)$~&~$1.886$~&~$1.836$~\\
&~$(V_R,S_R)$~&~$(-0.044,0.036)$~&~$1.409$~&~$3.0$~\\

&~$(S_L,S_R)$~&~$(0.0236,0.0575)$~&~$1.8415$~&~$1.9$~\\

\hline

\end{tabular}
\end{center}
\end{table}
 
The constrained on various combination of real coefficients (case B), $V_L-V_R$ (top-left panel), 
$S_L-S_R$ (top-right panel), $V_L-S_L$ (middle-left panel),  $V_L-S_R$ (middle-right panel), $V_R-S_L$ 
(bottom-left panel) and $V_R-S_R$ (bottom-right panel) obtained from  $R_{D^{(*)}}$, $R_{J/\psi}$, 
$R_\pi^l$, Br($B_{u,c} \to \tau \bar \nu_\tau$) and Br($B \to\pi \tau \bar \nu_\tau$) observables are presented in Fig. \ref{Fig:Case-B}\,. Table \ref{Tab:Best-fit} contains the best-fit values, 
$\chi^2_{\rm min}/{\rm d.o.f}$ and pull values of various sets of new real  coefficients. 
It should be noticed that, the fit of $V_L-V_R$, $V_L-S_L$ and $V_L-S_R$ sets to all the $7$ observables 
are very well. However, the  $V_R-S_L$, $V_R-S_R$ and $S_L-S_R$ sets provide very poor fit. 
\section{$B_c \to D^{(*)} \tau \bar \nu_\tau$ DECAY OBSERVABLES}
We discuss the branching ratio and angular observables of $B_c \to D^{(*)} \tau \bar \nu_\tau$ processes in this section. The branching ratios of $\bar B \to D \tau \bar \nu_l$ processes with respect to $q^2$ in the presence of new (pseudo)vector and (pseudo)scalar coefficients are given by \cite{Sakaki:2013bfa}
\bea
\frac{d\mathcal{BR}(\bar B \to  D l \bar \nu_l)}{dq^2} &=& \tau_B {G_F^2 |V_{cb}|^2 \over 192\pi^3 M_B^3} q^2 \sqrt{\lambda_D(q^2)} \left( 1 - {m_l^2 \over q^2} \right)^2  \nn \\   && \times \Bigg \lbrace \Big | 1 + V_L + V_R \Big |^2 \left[ \left( 1 + {m_l^2 \over 2q^2} \right) H_{0}^{2} + {3 \over 2}{m_l^2 \over q^2}  H_{t}^{2} \right] \nn \\ && + {3 \over 2} \left |S_L + S_R \right |^2 \, H_S^{2}   +3{\rm Re}\left[ ( 1 + V_L + V_R ) (S_L^* + S_R^* ) \right] {m_l \over \sqrt{q^2}} \, H_S H_{t}   \Bigg \rbrace,  \label{br-exp}
\eea
where 
\bea
\lambda_D =\lambda (M_B^2, M_D^2, q^2), ~~ ~~ {\rm with}~~~~~\lambda(a,b,c)=a^2+b^2+c^2-2(ab+bc+ca)\,,
\eea
 and  $H_{i,\lambda}^s$'s ($\lambda=0,\pm,t$) are the heicity amplitudes. The branching ratios of $\Bbar \to D^* \tau\nubar_l$  with respect to $q^2$ in the presence of new (pseudo)vector and (pseudo)scalar 
 coefficients are given by are given by \cite{Sakaki:2013bfa}
\bea
 {d\mathcal{BR}(\bar B \to  D^* l \bar \nu_l) \over dq^2} &=& \tau_B{G_F^2 |V_{cb}|^2 \over 192\pi^3 M_B^3} q^2 \sqrt{\lambda_{D^*} (q^2)} \left( 1 - {m_l^2 \over q^2} \right)^2 \times \nn  \\ && \bigg \{  \left( \left|1 + V_L \right|^2 + \left| V_R\right|^2 \right)  \left[ \left( 1 + {m_l ^2 \over 2q^2} \right) \left( H_{V, +}^2 + H_{V,-}^2 + H_{V,0}^2 \right) + {3 \over 2}{m_l^2 \over q^2} \, H_{V,t}^2 \right]  \nn \\ && - 2{\rm Re}\left[\left(1+ V_L \right) V_R^* \right] \left[ \left( 1 + {m_l^2 \over 2q^2} \right) \left( H_{V,0}^2 + 2 H_{V,+} H_{V,-} \right) + {3 \over 2}{m_l^2 \over q^2} \, H_{V,t}^2 \right] \nn \\ && +  {3 \over 2} |S_L - S_R|^2 \, H_S^2   + 3{\rm Re}\left [ \left ( 1 + V_L - V_R \right) \left (S_L^* - S_R^* \right) \right ] {m_l \over \sqrt{q^2}} \, H_S H_{V,t}  \bigg \}, 
\eea
where  $\lambda_{D^*}= \lambda (M_B^2, M_{D^*}^2, q^2)$  $H_{i,\lambda}$'s are the helicity amplitudes.

Along with the branching ratios, we also explore the following observables in order to inspect the structure of new physics.
\begin{itemize}
\item $\tau$ forward-backward asymmetry 
\bea
\A_{\rm FB}^{D^{(*)}} = { \int_0^1 {d\Gamma \over d\cos\th}d\cos\th-\int^0_{-1}{d\Gamma \over d\cos\th}d\cos\th \over \int_{-1}^1 {d\Gamma \over d\cos\th}d\cos\th }\,.
\eea
\item Lepton non-universality 
\bea
&&R_{D^{(*)}}^{B_c}=\frac{{\mathcal{BR}}( B_c \to D^{(*)} \tau \bar \nu_\tau)}{{\mathcal{BR}}(B_c \to D^{(*)} l \bar \nu_l)}, ~~~~l=e, \mu.
\eea 
\item $\tau$ polarization asymmetry \cite{Sakaki:2013bfa}
\bea
P_\tau^{D^{(*)}} (q^2) = \frac{ d\Gamma (\lambda_\tau = 1/2)/dq^2 - d\Gamma (\lambda_\tau = -1/2)/dq^2}{d\Gamma (\lambda_\tau = 1/2)/dq^2 + d\Gamma (\lambda_\tau = -1/2)/dq^2}\,.
\eea
\item $D^*$ polarization asymmetry \cite{Biancofiore:2013ki}
\bea
F_{L, T}^{D^*}(q^2) = \frac{d\Gamma_{L, T} \left(B_c \to D^* \tau \bar{\nu} \right)/ dq^2}{d\Gamma \left(B_c \to D^* \tau \bar{\nu} \right) / dq^2}\,.
\eea
\end{itemize}
Like $R_{D^{(*)}}$ observables, the following fascinating observables (with the
denominators involving only the light-lepton modes) are defined in Ref. \cite{Hu:2018veh}\,.
\begin{itemize}
\item $\tau$ forward and backward fractions
\bea
\chi_{1,2}^{D^{(*)}}=\frac{1}{2}R_{D^{(*)}}\left(1+A_{FB}^{D^{(*)}}\right)\,.
\eea
\item $\tau$ spin $1/2$ and $-1/2$ fractions 
\bea
\chi_{3,4}^{D^{(*)}}=\frac{1}{2}R_{D^{(*)}}\left(1+P_\tau^{D^{(*)}}\right)\,.
\eea
\item $D^*$ longitudinal and transverse polarization fractions
\bea
\chi_{5,6}^{D^{*}}=R_{D^{*}}F_{L,T}^{D^*}\,.
\eea
\end{itemize}

\section{NUMERICAL ANALYSIS AND EFFECT OF NEW COUPLINGS ON $B_c \to D^{(*)} \tau \bar \nu_\tau$ DECAY MODES}

For numerical evaluation, we consider all the particle masses, life time of $B_c$ meson and CKM matrix elements from PDG \cite{Tanabashi:2018oca}\,.  The $q^2$ dependence of the $B_c\to D^{(*)}$ form factors is parametrized as 
\bea
F(q^2)=F(0)exp(c_1 \hat{s}+c_2\hat{s}^2)\,,~~~~~ F=F_{+,0}^{B_c\to D}, V^{B_c \to D^*}, A_{0,1,2}^{B_c \to D^*}\,.
\eea
where $\hat{s}=s/M_{B_c}^2$ and the  $F(0),c_1,c_2$  coefficients that are obtained in the covariant light-front quark model are taken from \cite{Wang:2008xt}\,.
\begin{figure}[htb]
\includegraphics[scale=0.52]{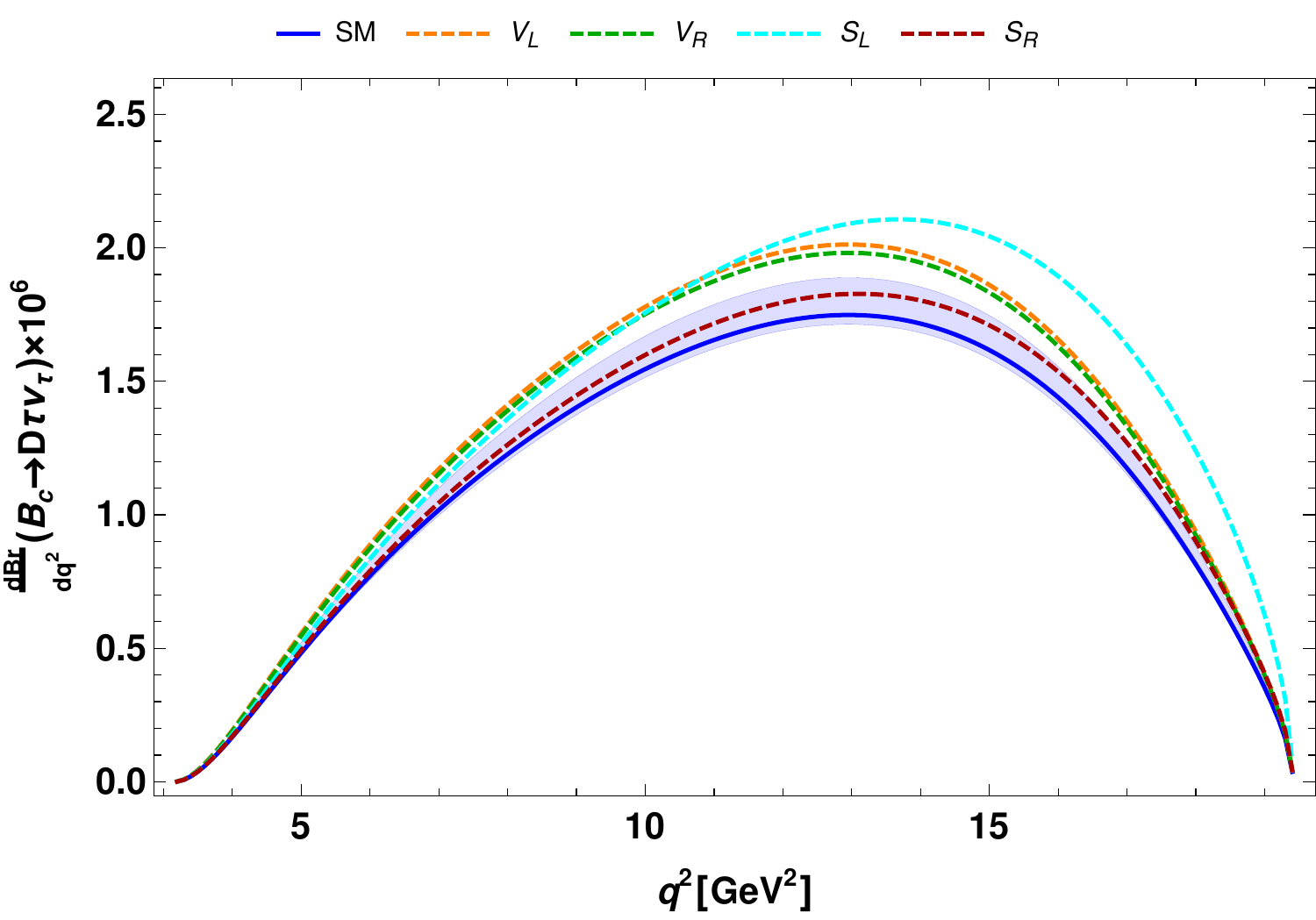}
\quad
\includegraphics[scale=0.45]{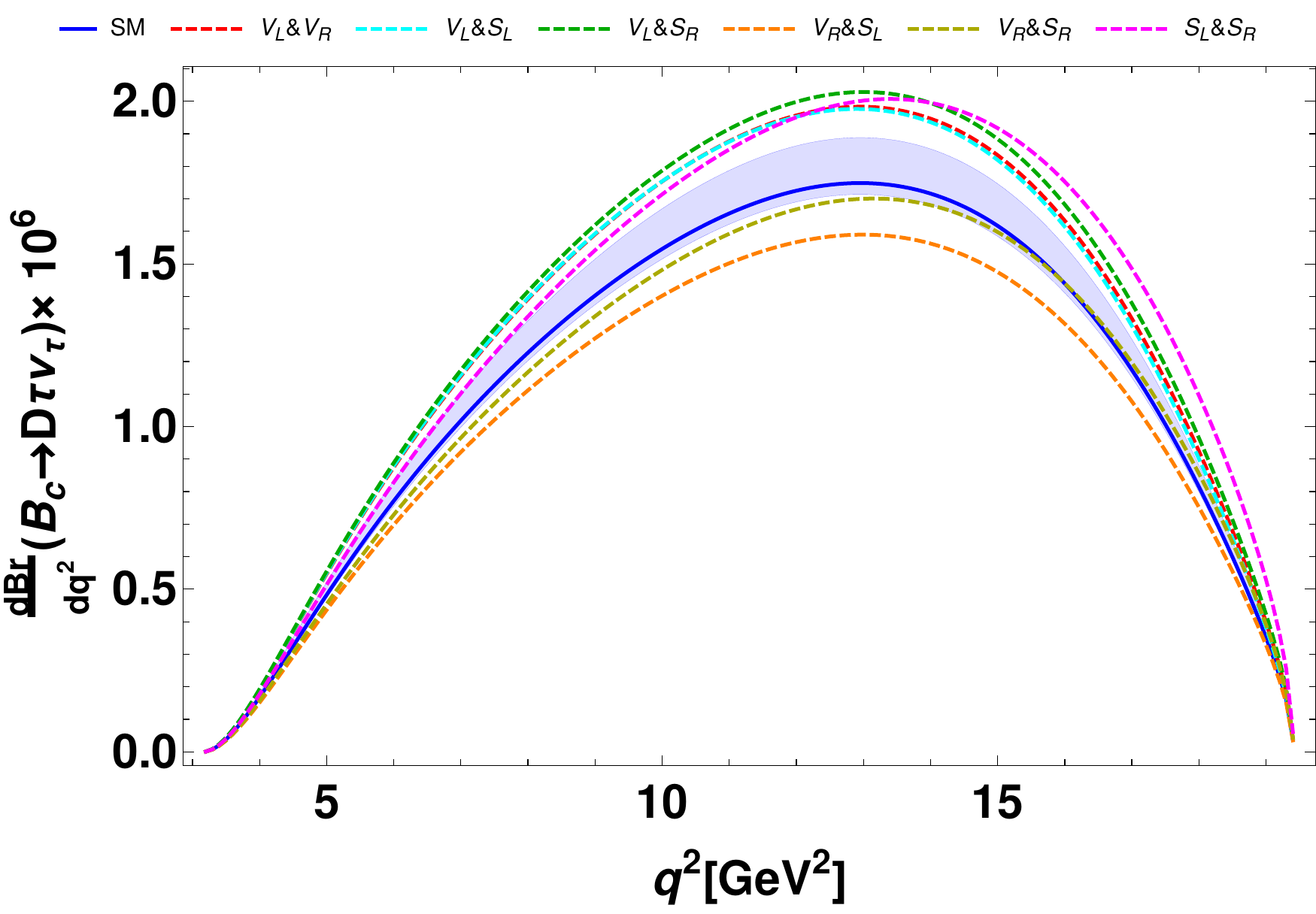}
\quad
\includegraphics[scale=0.52]{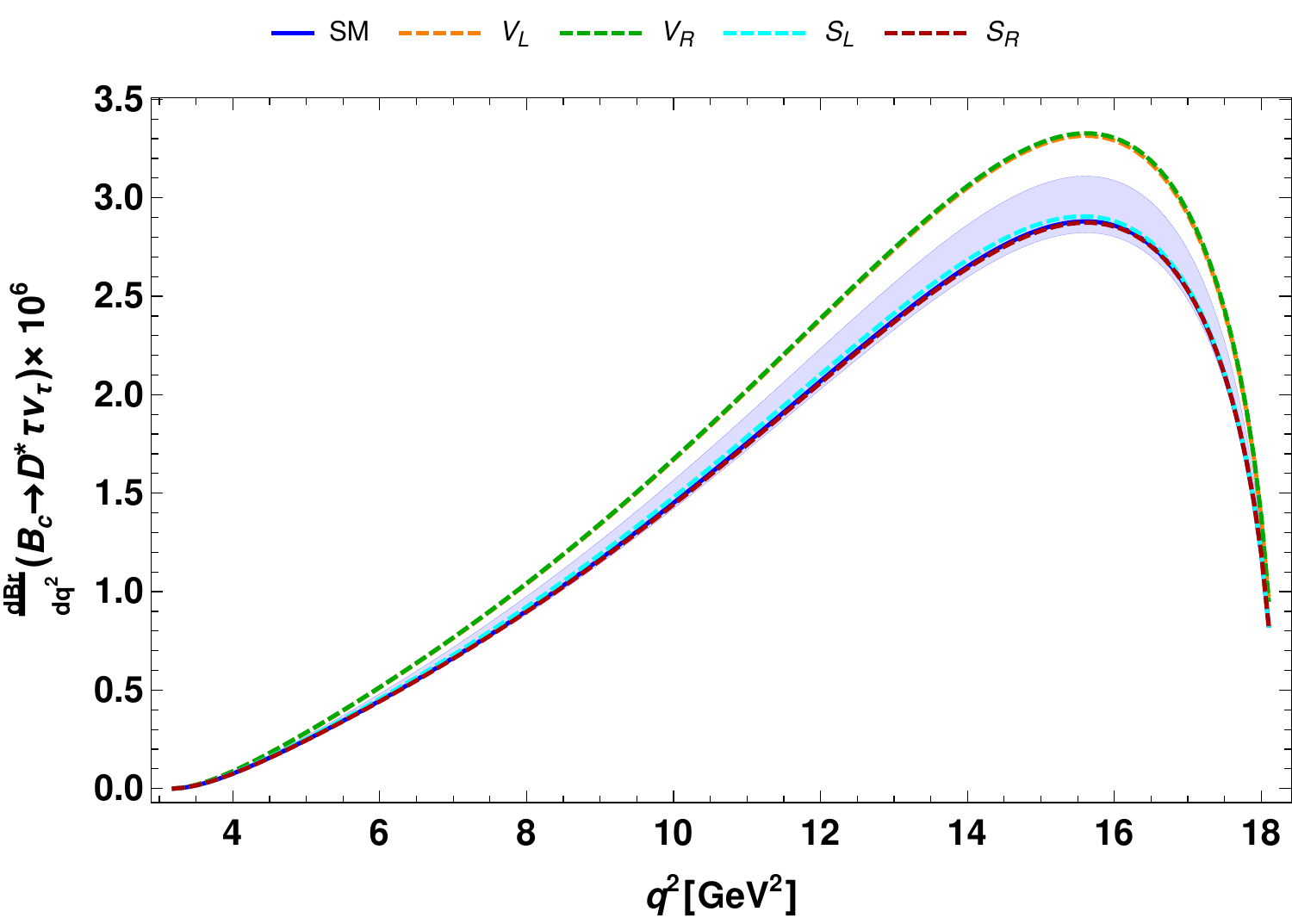}
\quad
\includegraphics[scale=0.45]{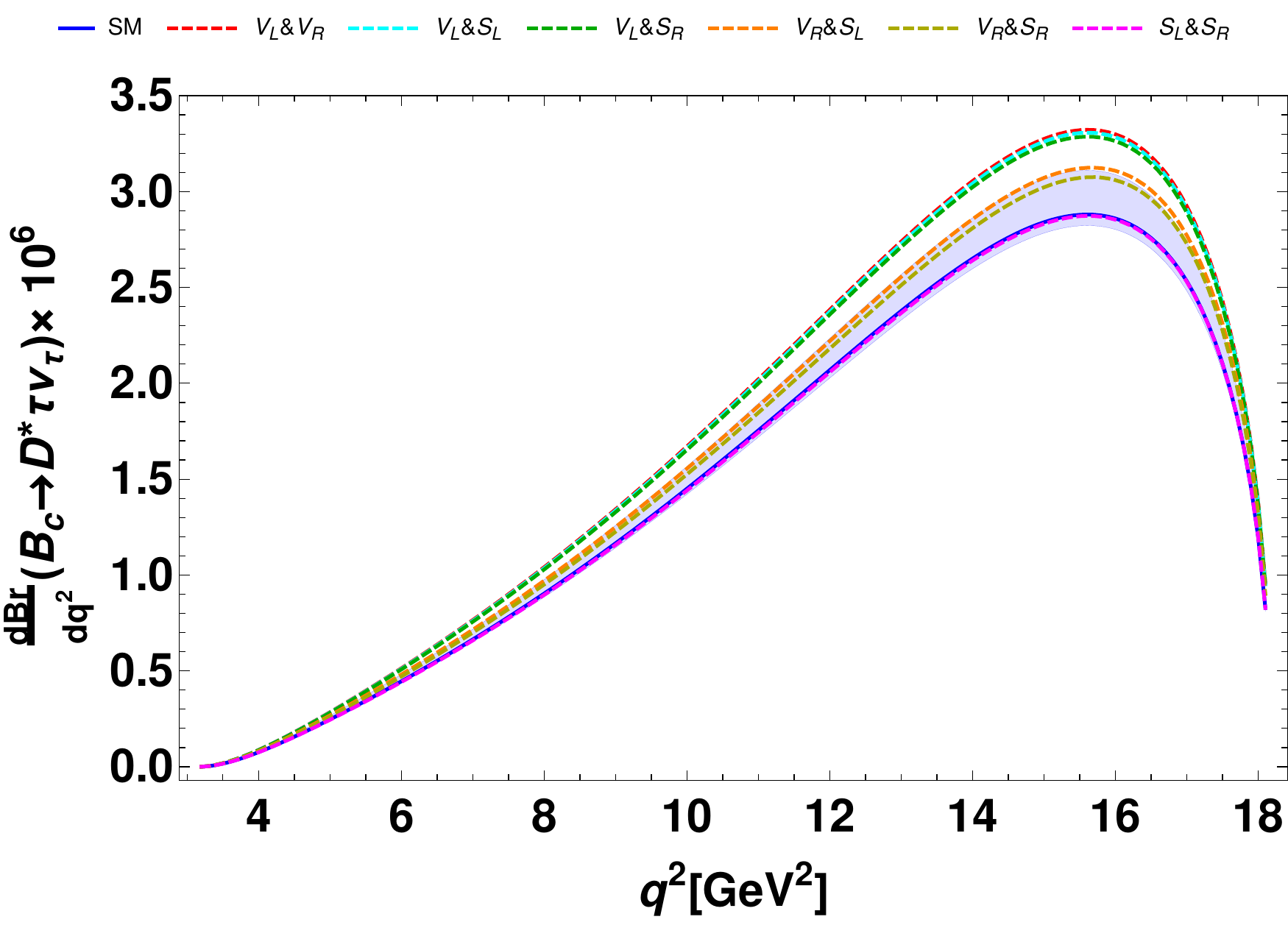}
\caption{The  variation of branching ratio of $B_c\to D\tau \nu_\tau$ (top panel) and $B_c\to D^*  \tau \nu_\tau$ (bottom panel) processes with respect to $q^2$ for case A (left) and case B (right), respectively. Here the solid blue lines (lighter blues bands) represent the SM values (theoretical uncertainties obtained from input parameters).  The orange, dark green, cyan and dark red colors stand for the additional complex $V_L$, $V_R$, $S_L$ and $S_R$ contributions, respectively. The red, cyan, dark green, orange, dark yellow and magenta colors are for the $(V_L, V_R)$, $(V_L, S_L)$, $(V_L, S_R)$, $(V_R, S_L)$, $(V_R, S_R)$ and $(S_L, S_R)$ sets of case B.} \label{Fig:BR}
\end{figure}
\begin{figure}[htb]
\includegraphics[scale=0.52]{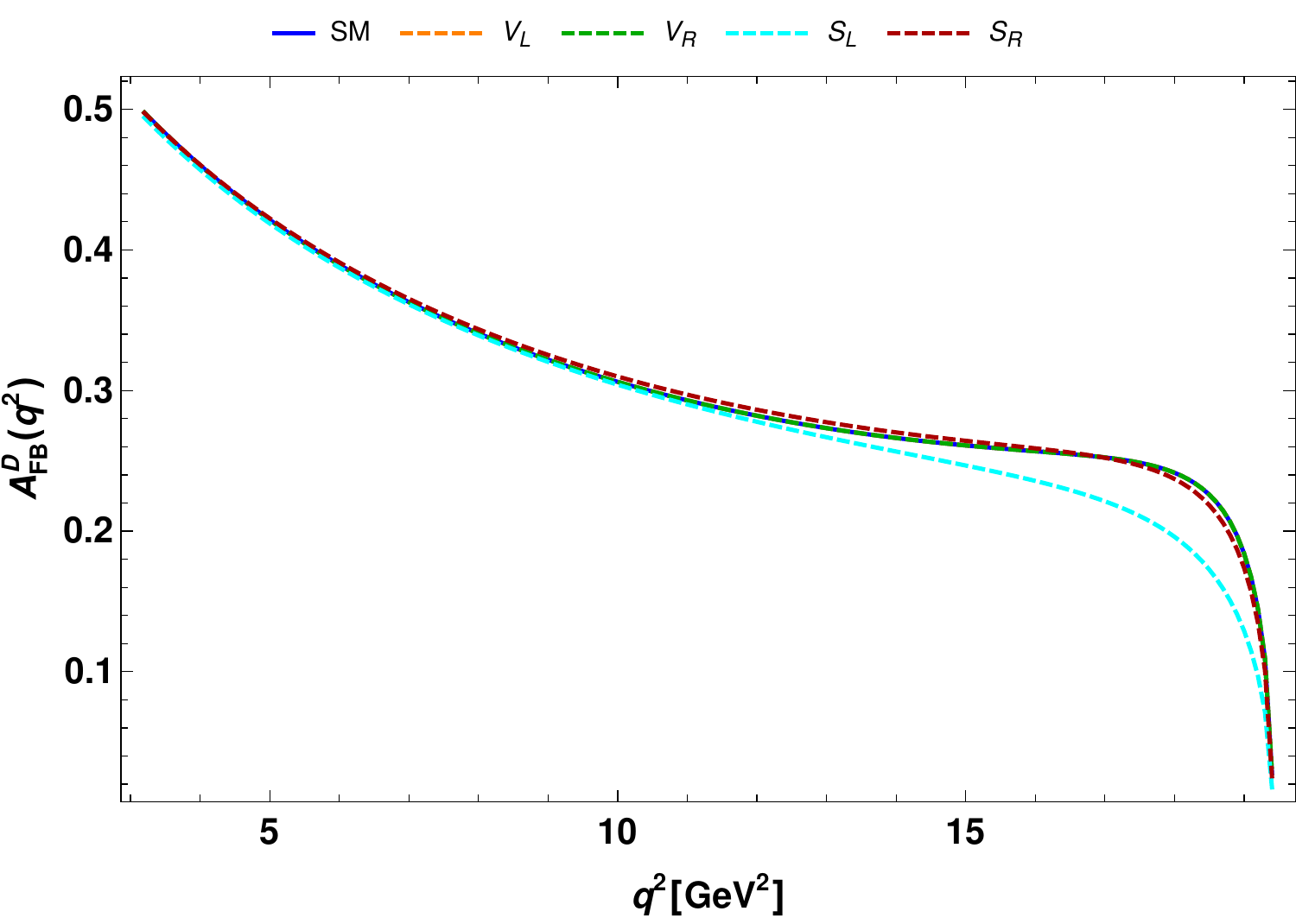}
\quad
\includegraphics[scale=0.45]{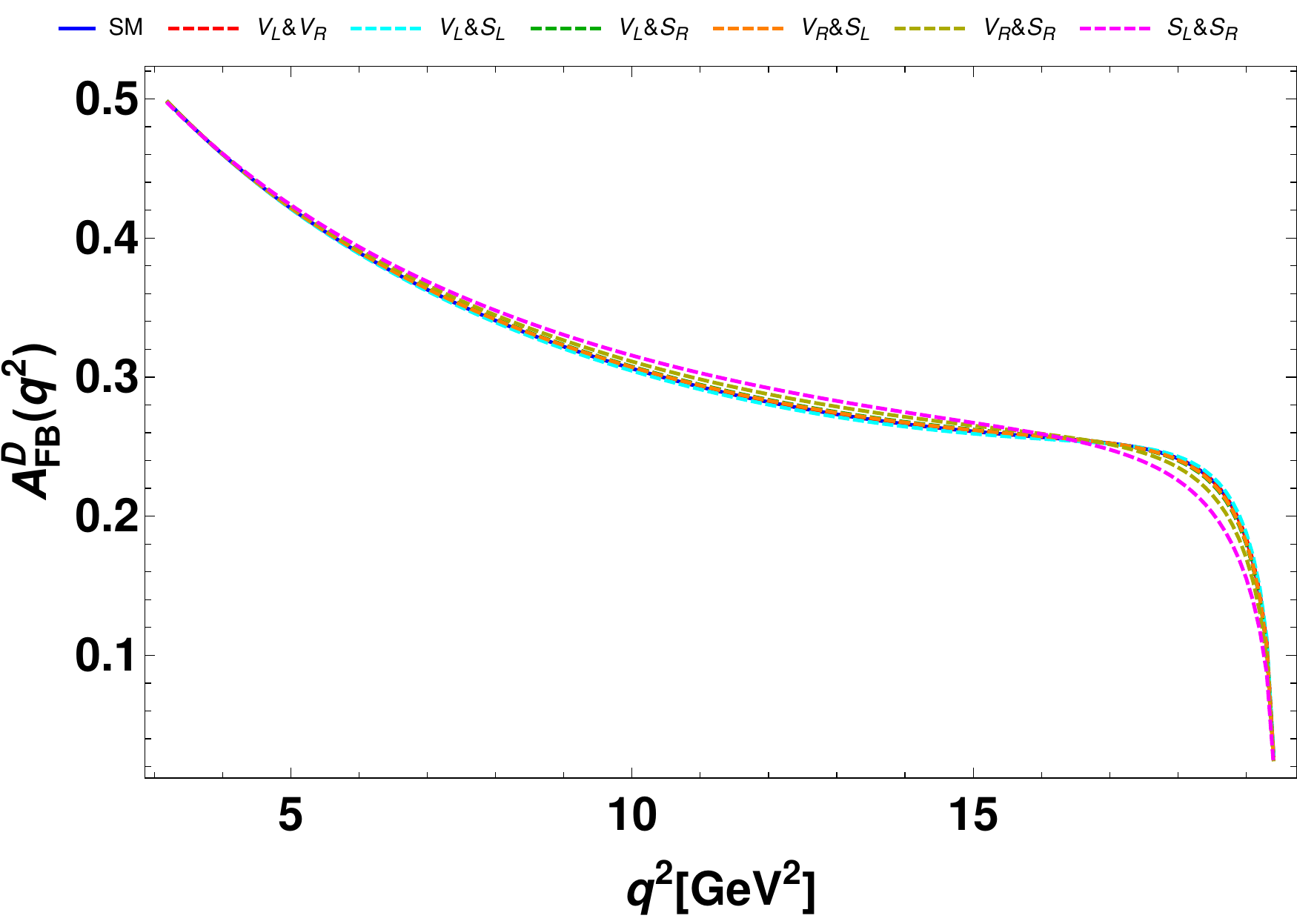}
\quad
\includegraphics[scale=0.52]{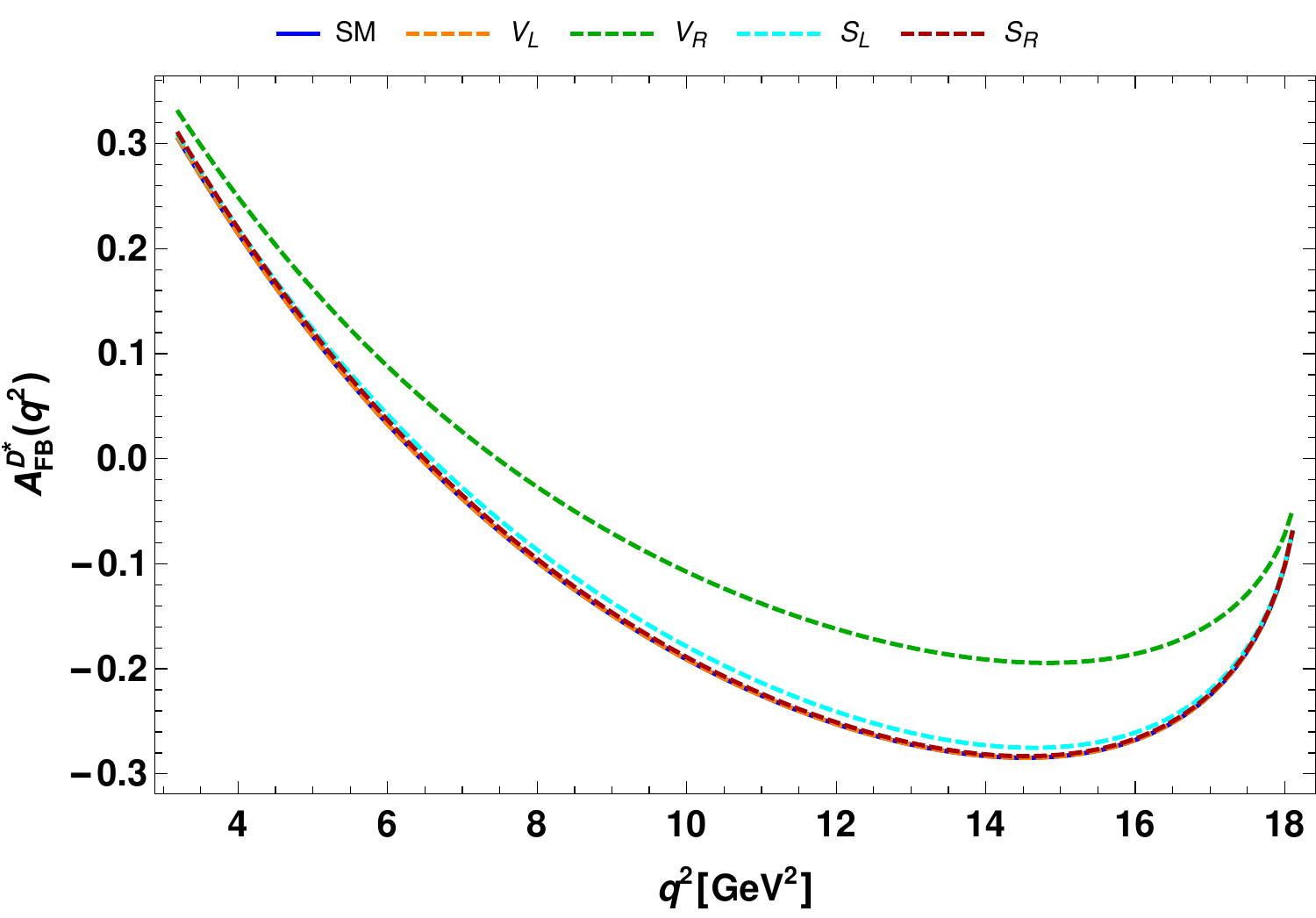}
\quad
\includegraphics[scale=0.45]{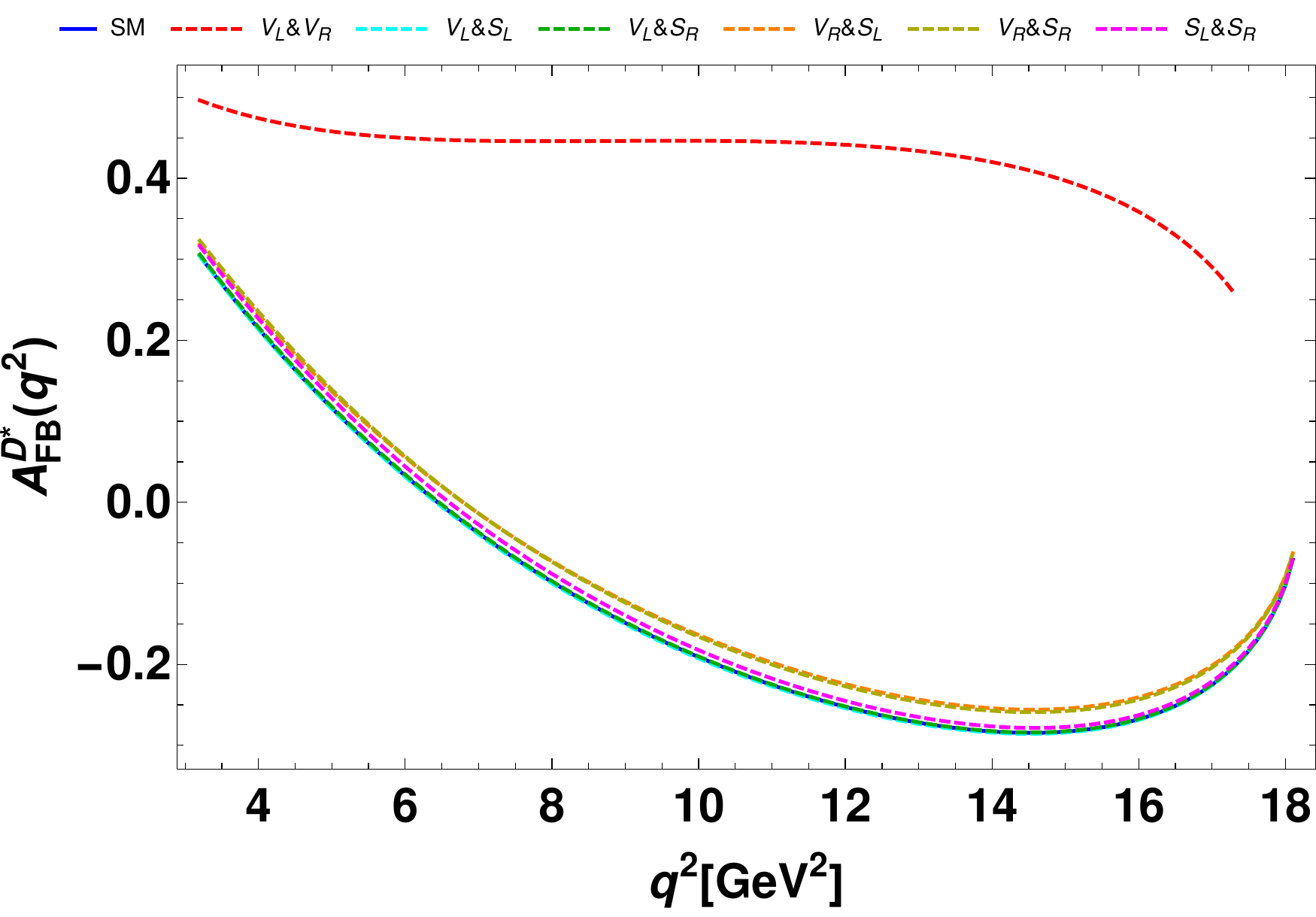}
\caption{The  $q^2$ variation of forward-backward asymmetry of $B_c\to D\tau \nu_\tau$ (top panel) and $B_c\to D^*  \tau \nu_\tau$ (bottom panel) processes  for case A (left) and case B (right), respectively. } \label{Fig:AFB}
\end{figure}
 Using the best-fit values of complex (real) Wilson coefficients from Table \ref{Tab:Best-fit}\,, 
 we show the $q^2$ variation of differential branching ratio ($\frac{d\mathcal{BR}}{dq^2}$), forward-backward 
 asymmetry ($\A_{\rm FB}$), lepton non-universality ($R_{D^{(*)}}^{B_c}$), 
 $\tau$ polarization asymmetry ($P_\tau^{D^{(*)}}$), $D^*$ longitudinal and transverse polarization 
 ($F_{L, T}^{D^*}$), $\tau$ forward and backward fractions ($\chi_{1,2}^{D^{(*)}}$), $\tau$ spin 
 $1/2$ and $-1/2$ fractions ($\chi_{3,4}^{D^{(*)}}$), $D^*$ longitudinal and transverse polarization 
 fractions ($\chi_{5,6}^{D^{*}}$) of $B_c \to D \tau \bar \nu_\tau$ and 
 $B_c \to D^* \tau \bar \nu_\tau$ decay modes in various Figures for both case A and case B.

In Fig. \ref{Fig:BR}\, the solid blue lines represent the SM central values and the lighter blue bands stand for the SM uncertainties which are obtained from the  input parameters used in our analysis. Again, the orange, dark green, cyan and dark red colors are drawn by using the best-fit values of complex $V_L$, $V_R$, $S_L$ 
and $S_R$ coefficients, respectively for case A. Similarly, for case B, the red, cyan, dark green, orange, dark yellow and magenta colors stand respectively for real $(V_L, V_R)$, $(V_L, S_L)$, $(V_L, S_R)$, $(V_R, S_L)$, $(V_R, S_R)$ and $(S_L, S_R)$ combined sets of coefficients. We now wish to see how the differential branching ratios (${d\mathcal{BR}(\bar B \to  D^{(*)} l \bar \nu_l) \over dq^2}$), and other angular observables like 
$\A_{\rm FB}$, $R_{D^{(*)}}^{B_c}$, $P_\tau^{D^{(*)}}$, $F_{L, T}^{D^*}$, $\chi_{1,2}^{D^{(*)}}$, $\chi_{3,4}^{D^{(*)}}$, $\chi_{5,6}^{D^{*}}$ behave
with different NP coefficients in both case A and case B. 

In Fig. \ref{Fig:BR} we observe that the differential branching ratio is not affected by the presence of complex $S_R$ coefficient (top left panel), rather it lies within SM uncertainty blue band while both $V_L$ and $V_R$ coefficients show a small deviation in differential branching ratio, but the presence of complex $S_L$ coefficient provide significant deviation in the branching ratio of $B_c \to D\tau \bar \nu_\tau$ process where the peak of the distribution of differential branching ratio can shift to a higher $q^2$ region once the complex $S_L$ coefficient is introduced in case A.  For case B, the differential branching ratio of
$(B_c \to D \tau \bar \nu_\tau)$ has deviated from the SM prediction for all possible sets of new 
coefficients except $(V_R, S_R)$ set of real coefficients while the peak can be shifted to a higher value $q^2$ 
for ($S_L, S_R)$ set of coefficients (top right panel) as 
shown in magenta color. Similarly the presence of complex $V_L$ and $V_R$ coefficients result in significant and identical deviation in differential branching ratio for $B_c \to D^* \tau \bar \nu_\tau$ decay (bottom-left panel) in case A while in case B, the sets of ($V_L,V_R$), $(V_L, S_L)$ and $(V_L,S_R)$ sets of coefficients give remarkable deviation (bottom-right panel).

Fig. \ref{Fig:AFB} depicts the forward-backward asymmetry  of $B_c \to D \tau \bar \nu_\tau$ and $B_c \to D^* \tau \bar \nu_\tau$ decay modes with respect to $q^2$ in the top panel and bottom panel, respectively. We observe almost no deviation from SM predictions in the forward-backward asymmetry ($\A_{\rm FB}$) of $(B_c \to D \tau \bar \nu_\tau)$ decay for both case A and case B, except a very little deviation for complex $S_L$ coefficient higher $q^2$ region. 
Therefore, one may conclude that
the forward-backward asymmetry does not vary with any of the NP coefficients (except the complex $S_L$ cofficient) for $B_c \to D \tau \bar \nu_\tau$ decay, which is expected, since it is a ratio, the NP 
dependency gets canceled in the ratio. But the forward-backward asymmetry of 
$B_c \to D^* \tau \bar \nu_\tau$ process show profound deviation due to the presence of  
$V_R$ (bottom-left panel) in case A and $(V_L, V_R)$ set of real coefficients (bottom-right panel) in case B. 
Thus, we see the forward-backward asymmetry of $B_c \to D^* \tau \bar \nu_\tau$ decay process is very sensitive to $(V_L, V_R)$ set of coefficients. Again, we find 
$\A_{\rm FB}$ starts with peak value of 0.3 at low $q^2$ and becomes 
minimum with value -0.3 at $q^2 = 15\,$ GeV$^2$ and ends with value -0.05 at high $q^2$ in SM. Thus there is 
a zero crossing 
at $q^2 = 6.5 \, $ GeV$^{2}$. However, in the presence of $(V_L, V_R)$ set of real coefficients (case B) there is no zero crossing and the value of forward-backward asymmetry lies between 0.5 at low $q^2$ and 0.3 at high $q^2$ as observed from bottom-right panel of Fig. \ref{Fig:AFB}.

\begin{figure}[htb]
\includegraphics[scale=0.52]{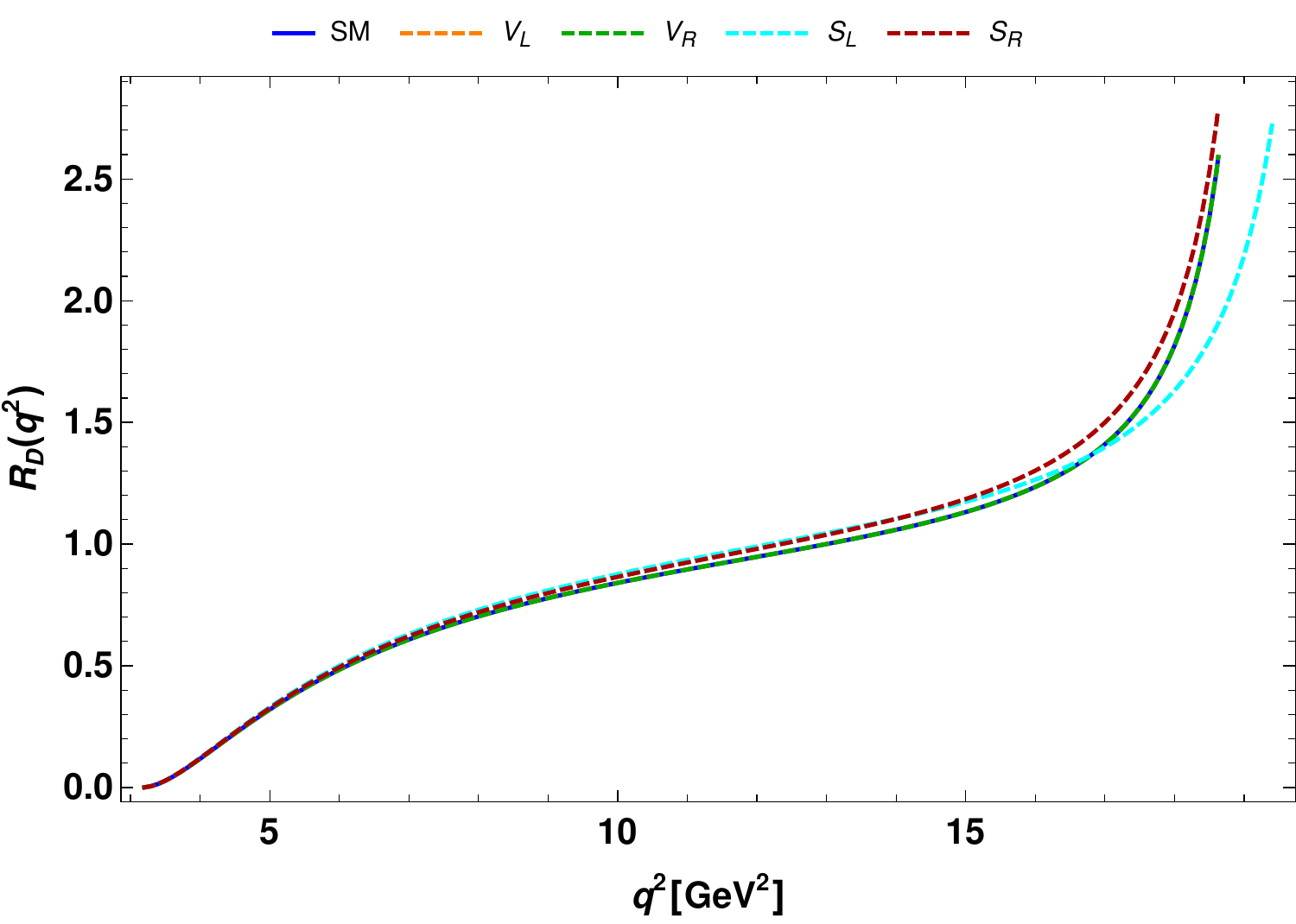}
\quad
\includegraphics[scale=0.45]{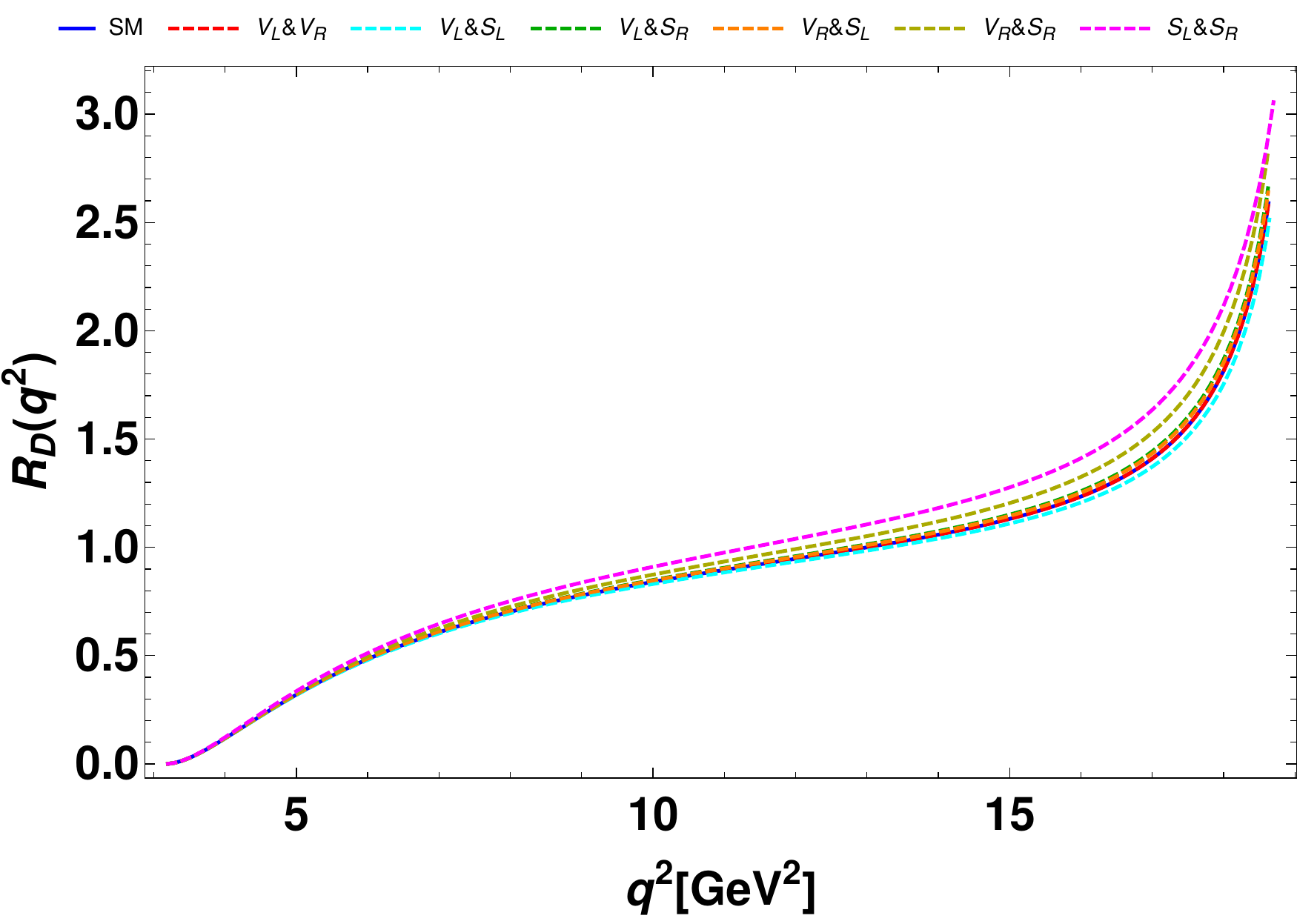}
\quad
\includegraphics[scale=0.52]{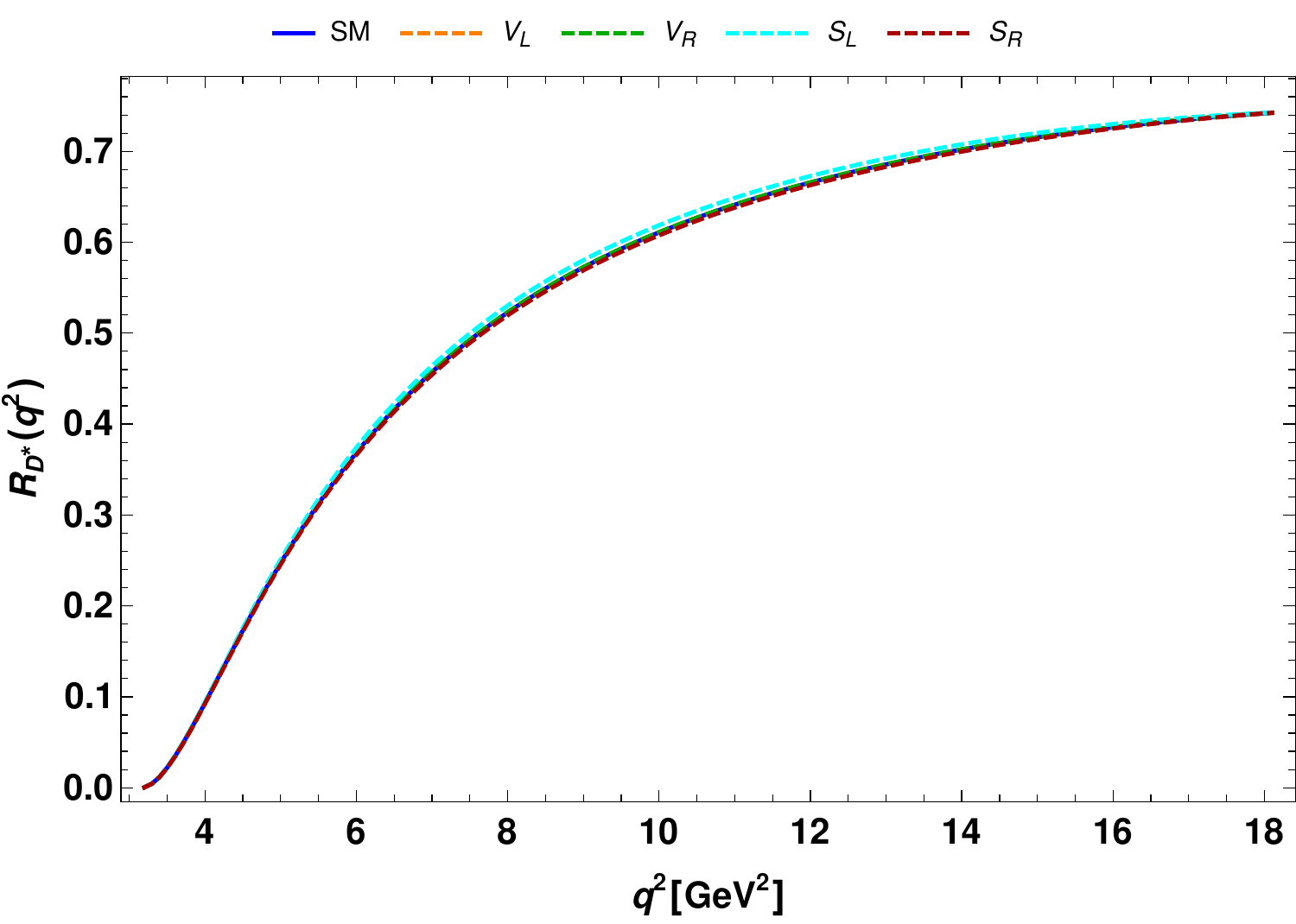}
\quad
\includegraphics[scale=0.45]{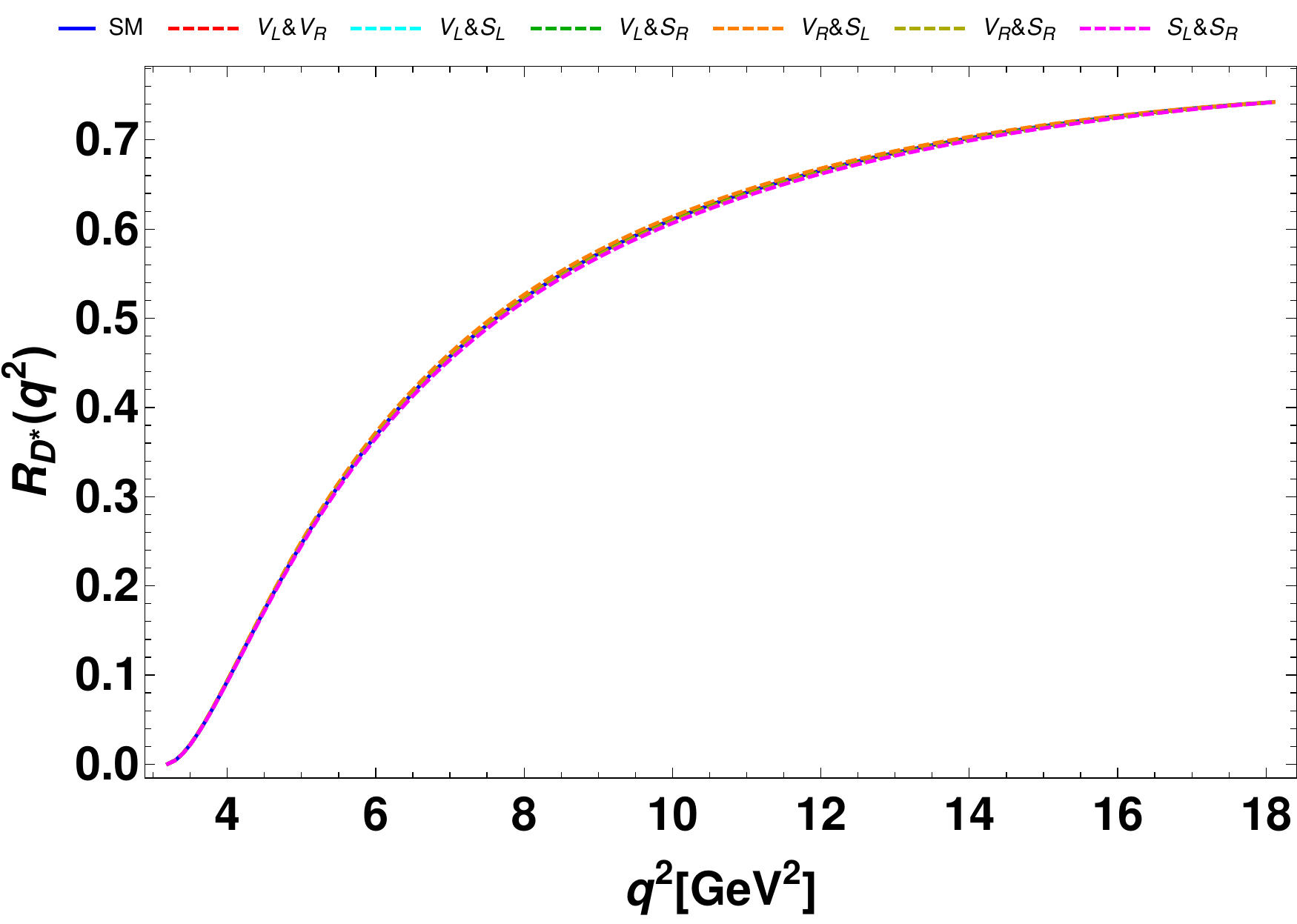}
\caption{The variation of $R_D^{B_c}$ (top panel) and $R_{D^*}^{B_c}$ (bottom panel) parameters with respect to $q^2$ for case A (left) and case B (right), respectively. } \label{Fig:LNU}
\end{figure}

The $q^2$ variation of the lepton non-universality parameters, $R_D^{B_c}$ and $R_{D^*}^{B_c}$ of $B_c \to D \tau \bar \nu_\tau$ and $B_c \to D^* \tau \bar \nu_\tau$ decay modes, for 
complex (left) and real (right) new coefficients are shown in the top and bottom panel respectively in
 Fig. \ref{Fig:LNU}\,. Here, in top left panel, we see very little deviation from SM predictions in lepton non-universality parameter, $R_D^{B_c}$ of $B_c \to D \tau \bar \nu_\tau$ decay mode for complex $S_L$ coefficient in higher $q^2$ region (for $q^2 > 17$ Gev$^2$ ) in case A and for $(V_R, S_R)$, $(S_L, S_R)$ sets of real coefficients in case B (top right panel). Again, we want to emphasize the fact that 
 the lepton non-universality parameter, $R_{D^*}^{B_c}$ of $B_c \to D^* \tau \bar \nu_\tau$ decay modes is not influenced by any of the NP coefficients in both case A and case B as observed in the bottom panel of Fig. \ref{Fig:LNU}\,. It is worth mentioning that, the lepton non-universality parameter for both $B_c \to D \tau \bar \nu_\tau$ and $B_c \to D^* \tau \bar \nu_\tau$ decay modes, does not differ appreciably with the NP coefficients which is anticipated, because, the impact of NP coefficients gets canceled in the ratio of lepton non-universality parameter.
  
\begin{figure}[htb]
\includegraphics[scale=0.52]{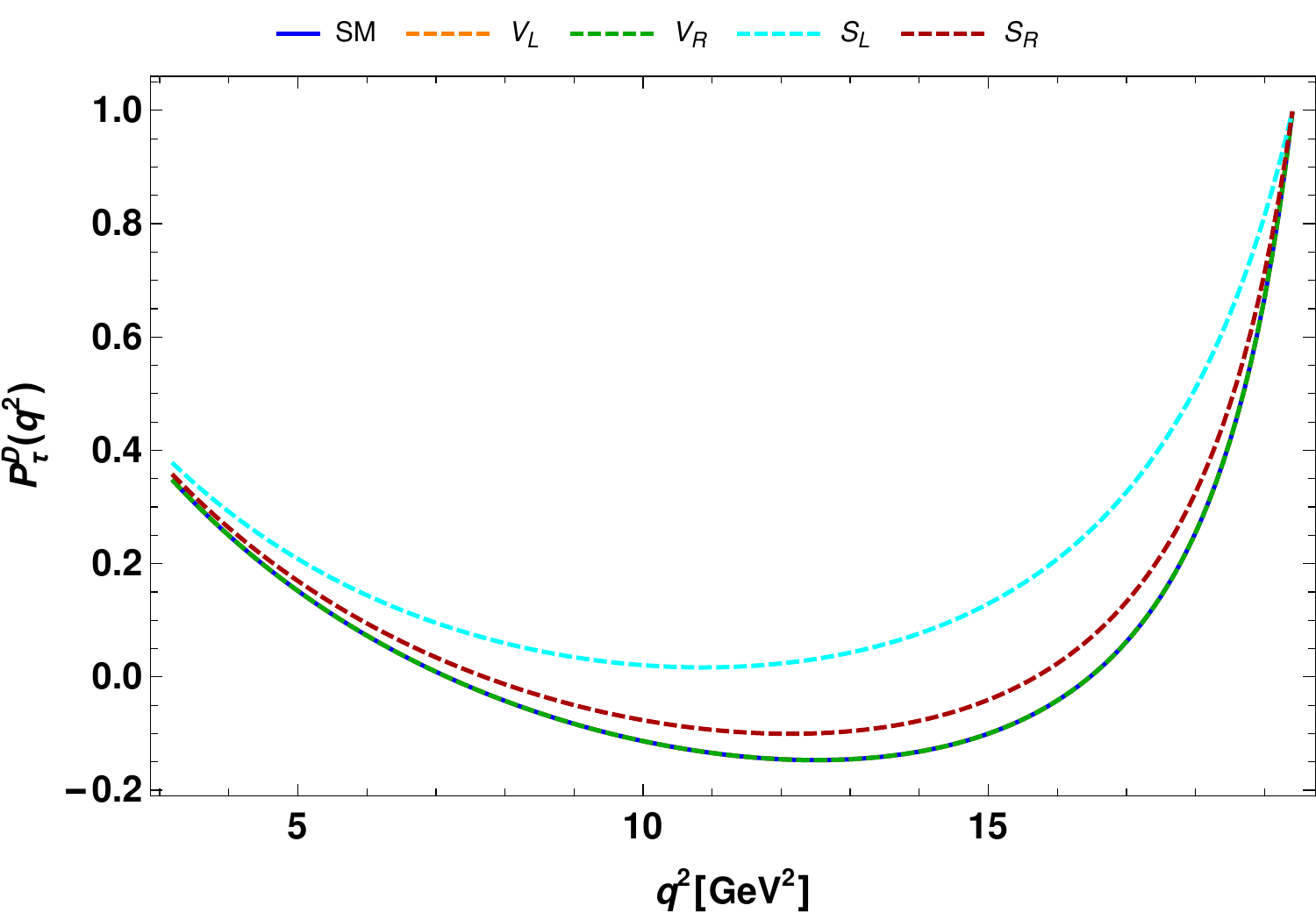}
\quad
\includegraphics[scale=0.45]{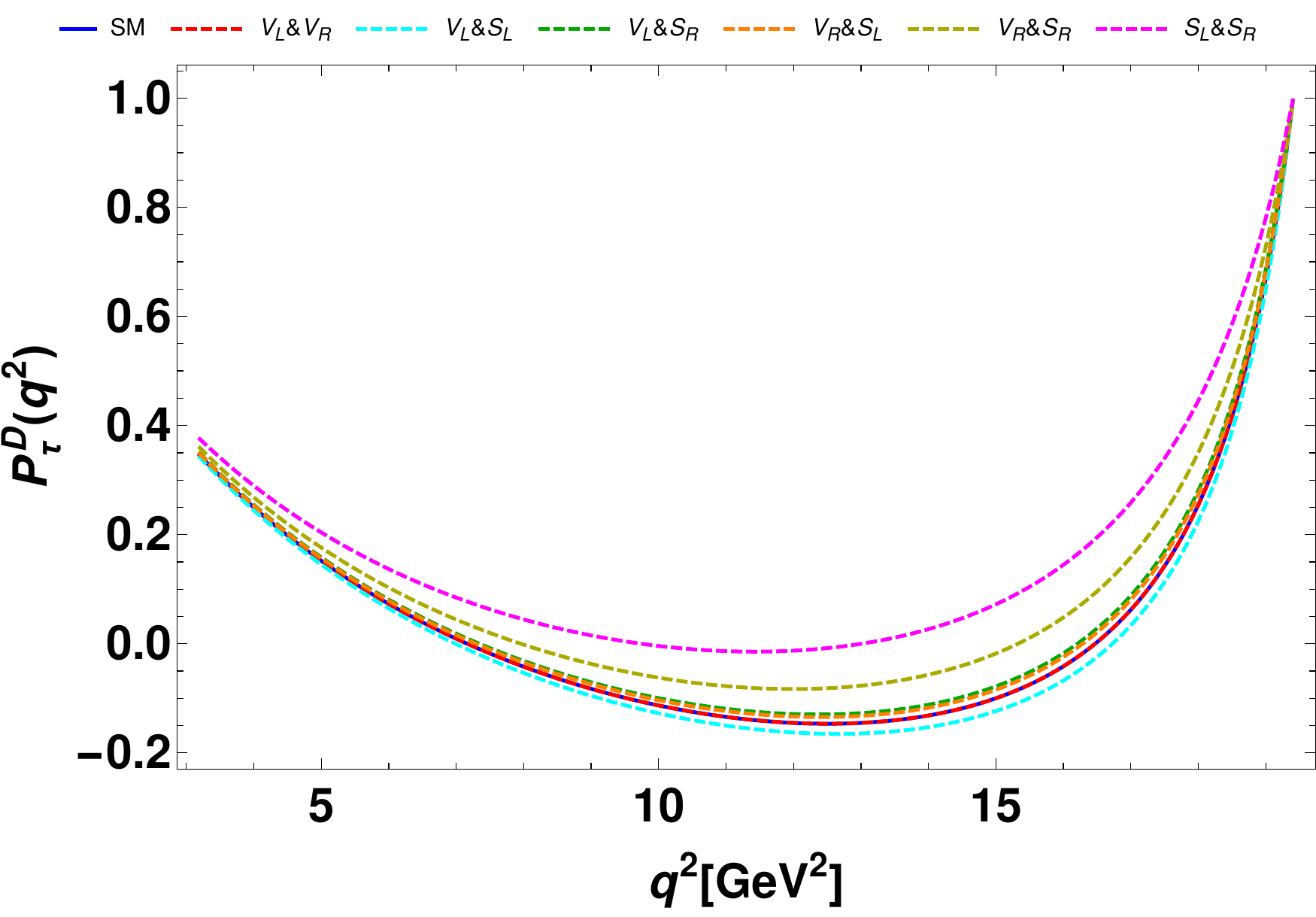}
\caption{The  $q^2$ variation of $\tau$ polarization  asymmetry of $B_c\to D\tau \nu_\tau$  process  for case A (left panel) and case B (right panel), respectively.} \label{Fig:Ptau}
\end{figure}
\begin{figure}[htb]
\includegraphics[scale=0.44]{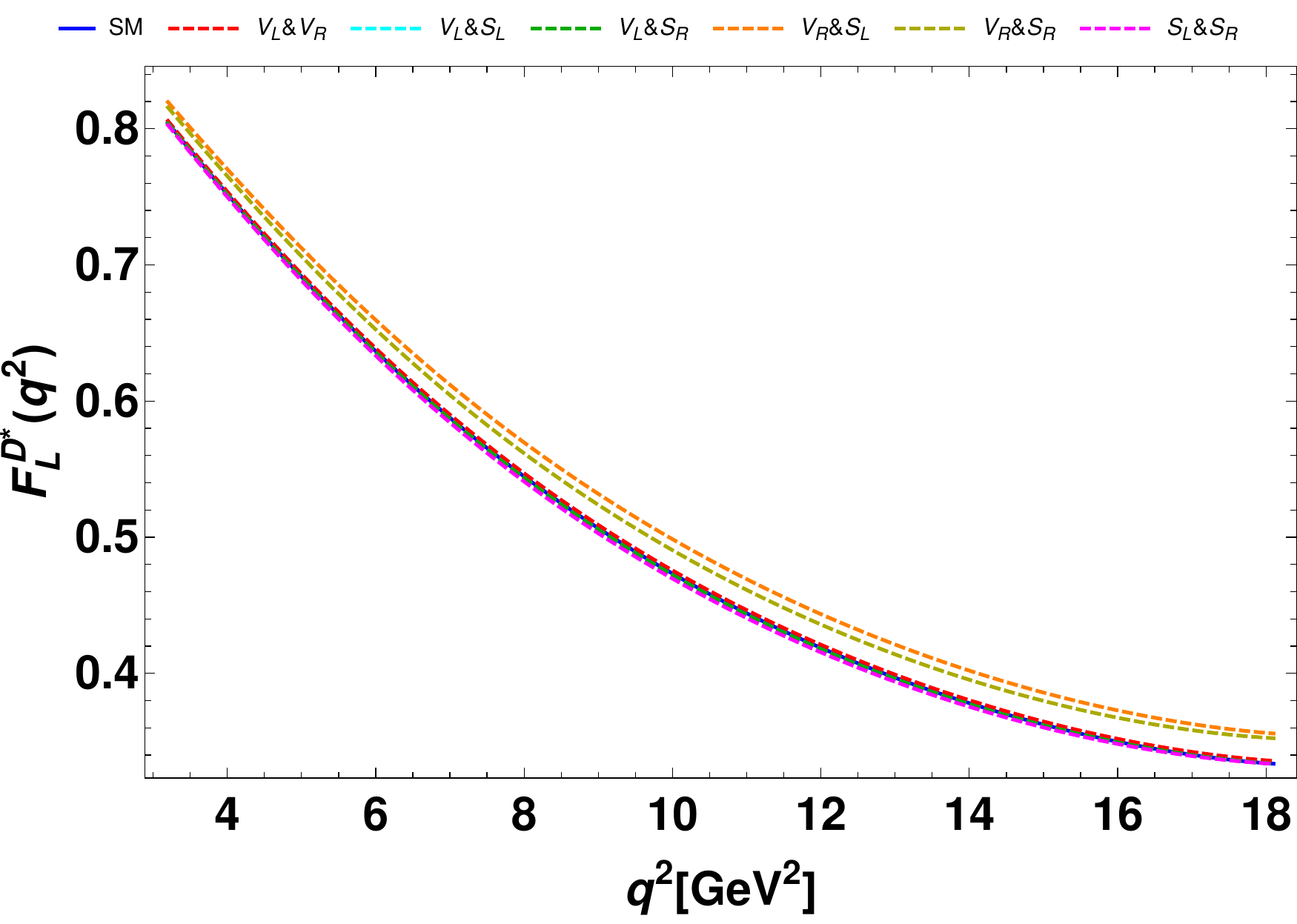}
\quad
\includegraphics[scale=0.44]{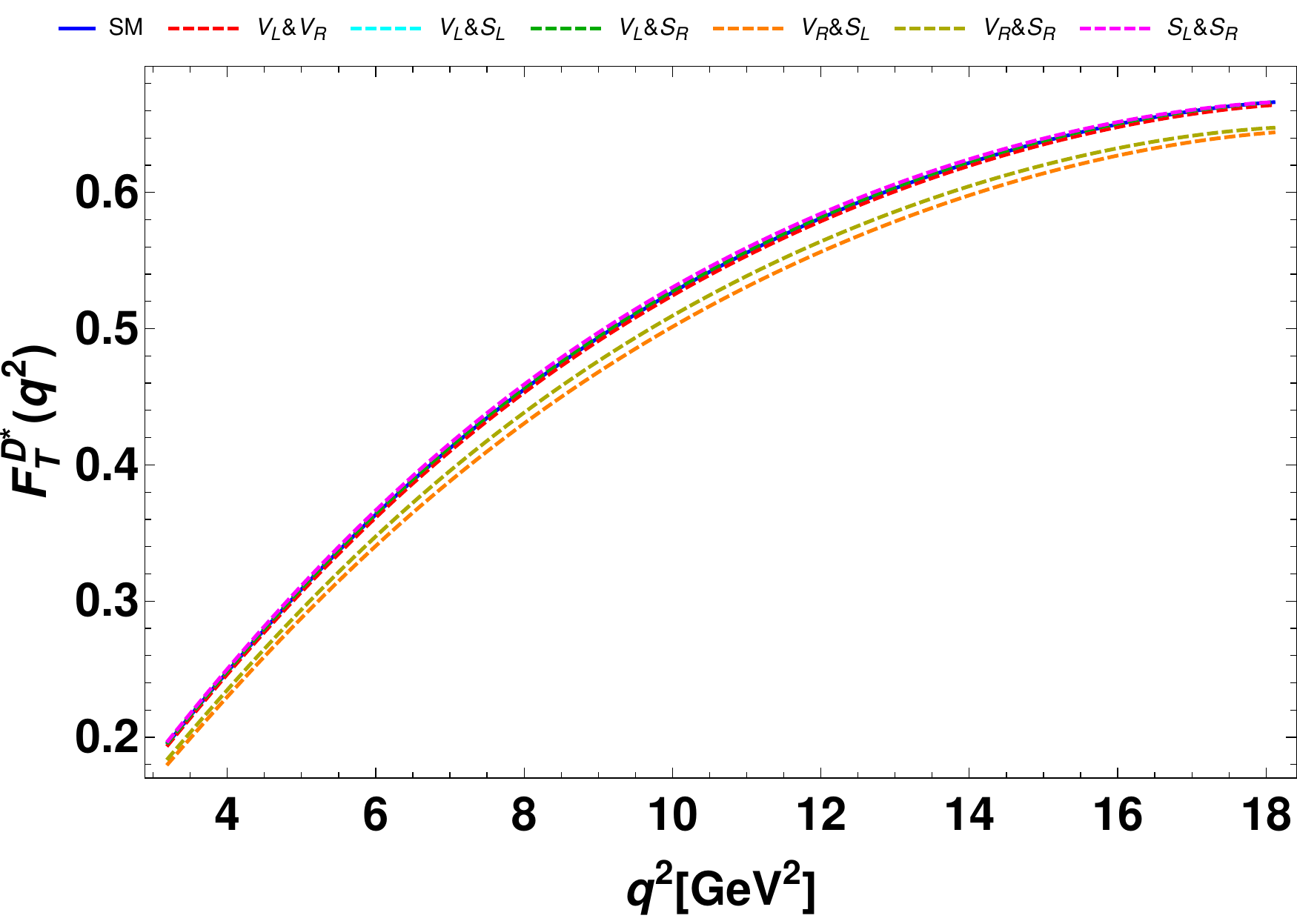}
\caption{The  $q^2$ variation of $D^*$ longitudinal (left panel) and transeverse (right panel) polarization  asymmetry of $B_c\to D^*\tau \nu_\tau$ process for case B.} \label{Fig:FL-FT}
\end{figure}

The $\tau$ longitudinal polarization asymmetry ($P_\tau^{D}$) of $B_c\to D\tau \nu_\tau$ decay mode with respect to $q^2$ in the presence of new complex (left panel) and real (right panel) coefficient are shown in Fig. \ref{Fig:Ptau}\,. We observe that the presence of only complex $S_L$ ($S_R$) coefficient in case A and the real 
$(S_L, S_R)$ ($(V_R, S_R)$) set of coefficients provide maximum deviation (small deviation) from the SM predictions in case B. Since there 
is no deviation in  $P_\tau^{D^*}$ observable of $B_c\to D^*\tau \nu_\tau$ process from SM predictions, for both case A and case B, we do not include it as Figure. Similarly, $D^*$ longitudinal and transverse polarization ($F_{L, T}^{D^*}$) of $B_c\to D^*\tau \nu_\tau$ decay mode remains unchanged by the presence of complex $V_L$, $V_R$, $S_L$ and $S_R$ coefficients in case A, which is also excluded in this work. However, in case B, the $q^2$ variation of $F_{L, T}^{D^*}$ with longitudinal polarization asymmetry in left panel and transverse polarization asymmetry in right panel are shown in Fig. \ref{Fig:FL-FT}\,.
 We see a small deviation in $F_{L,T}^{D^*}$ observable for $(V_R, S_L)$ and $(V_R, S_R)$ sets of real coefficients in case B.  

\begin{figure}[htb]
\includegraphics[scale=0.52]{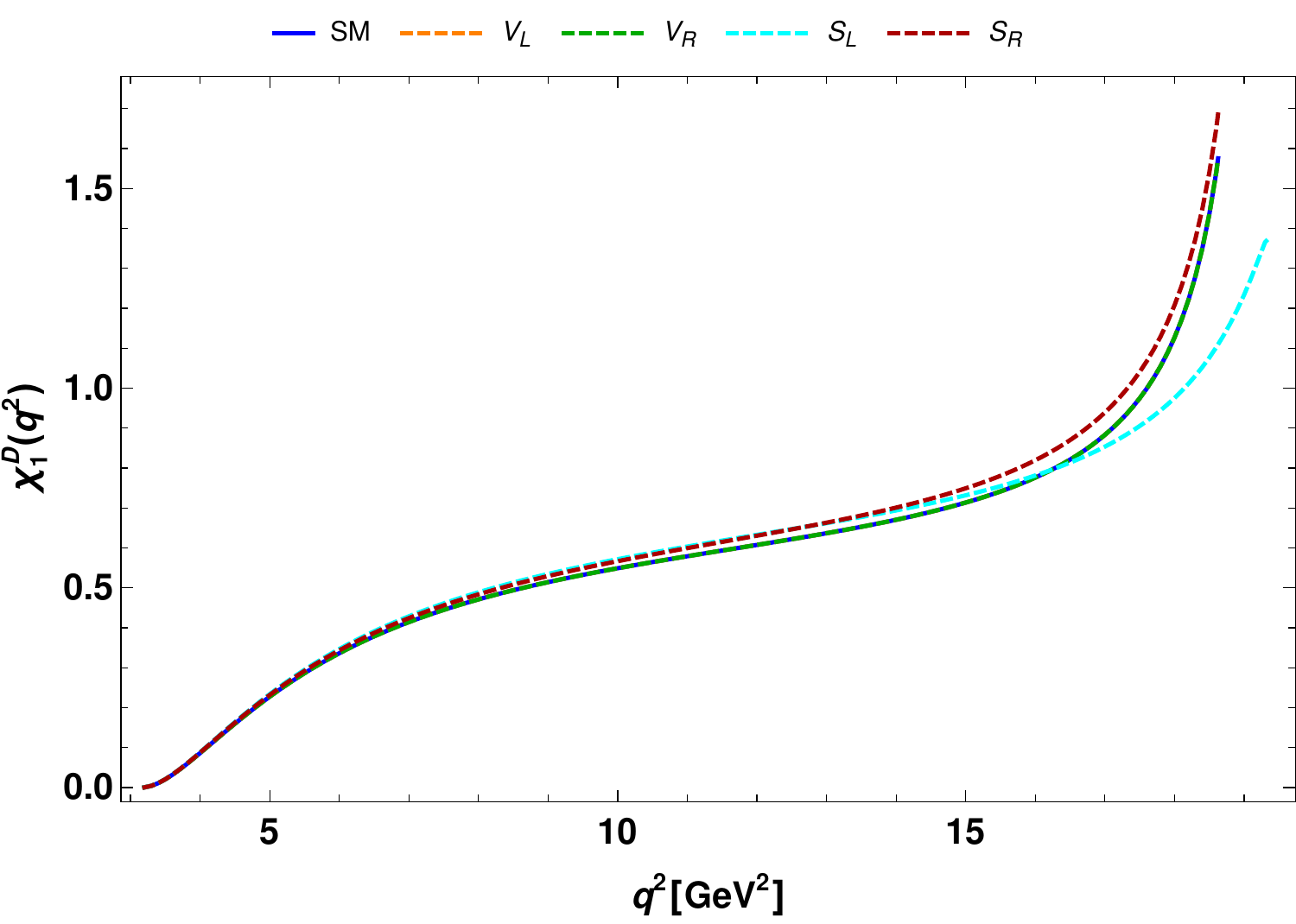}
\quad
\includegraphics[scale=0.45]{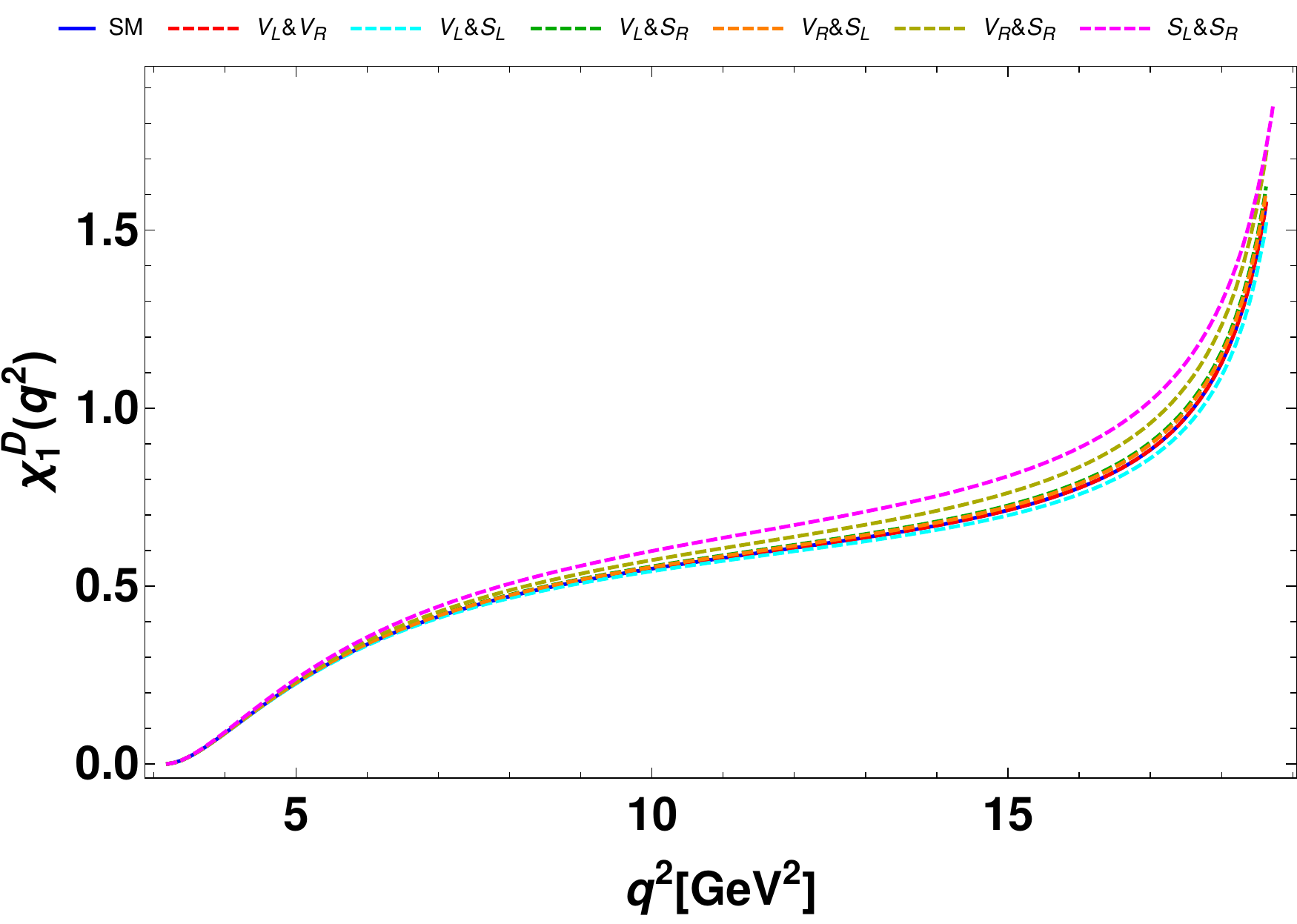}
\quad
\includegraphics[scale=0.52]{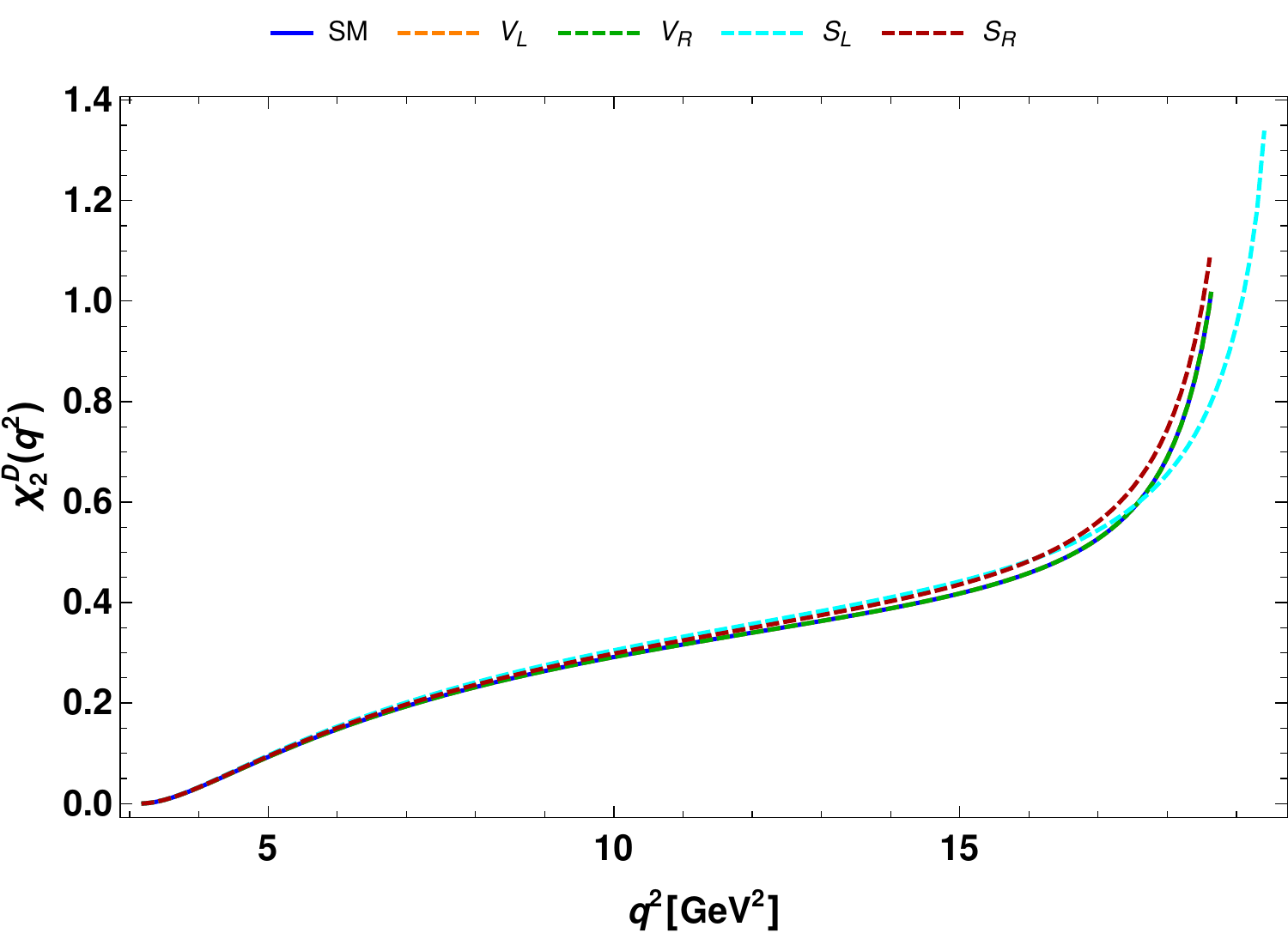}
\quad
\includegraphics[scale=0.45]{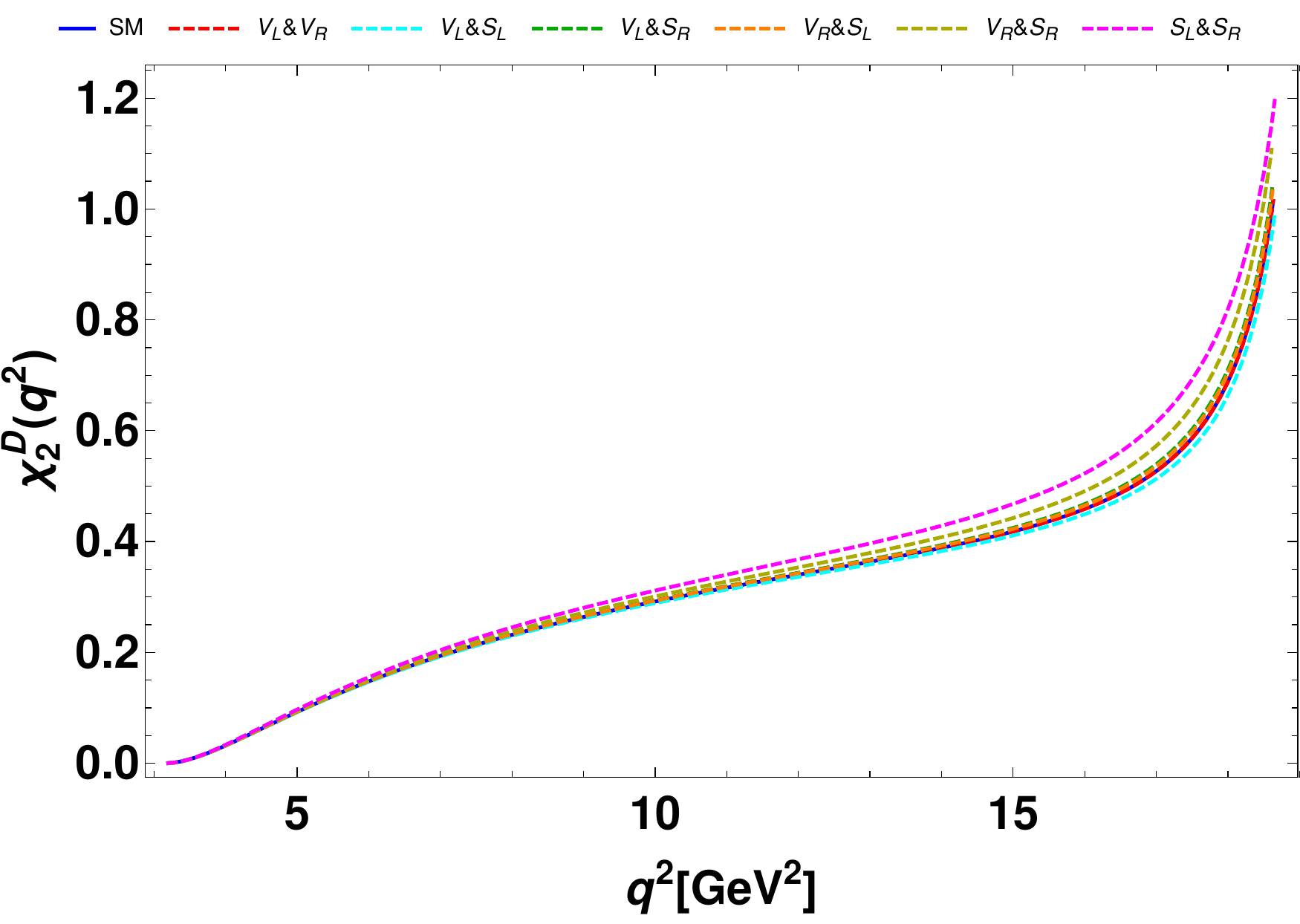}
\quad
\includegraphics[scale=0.52]{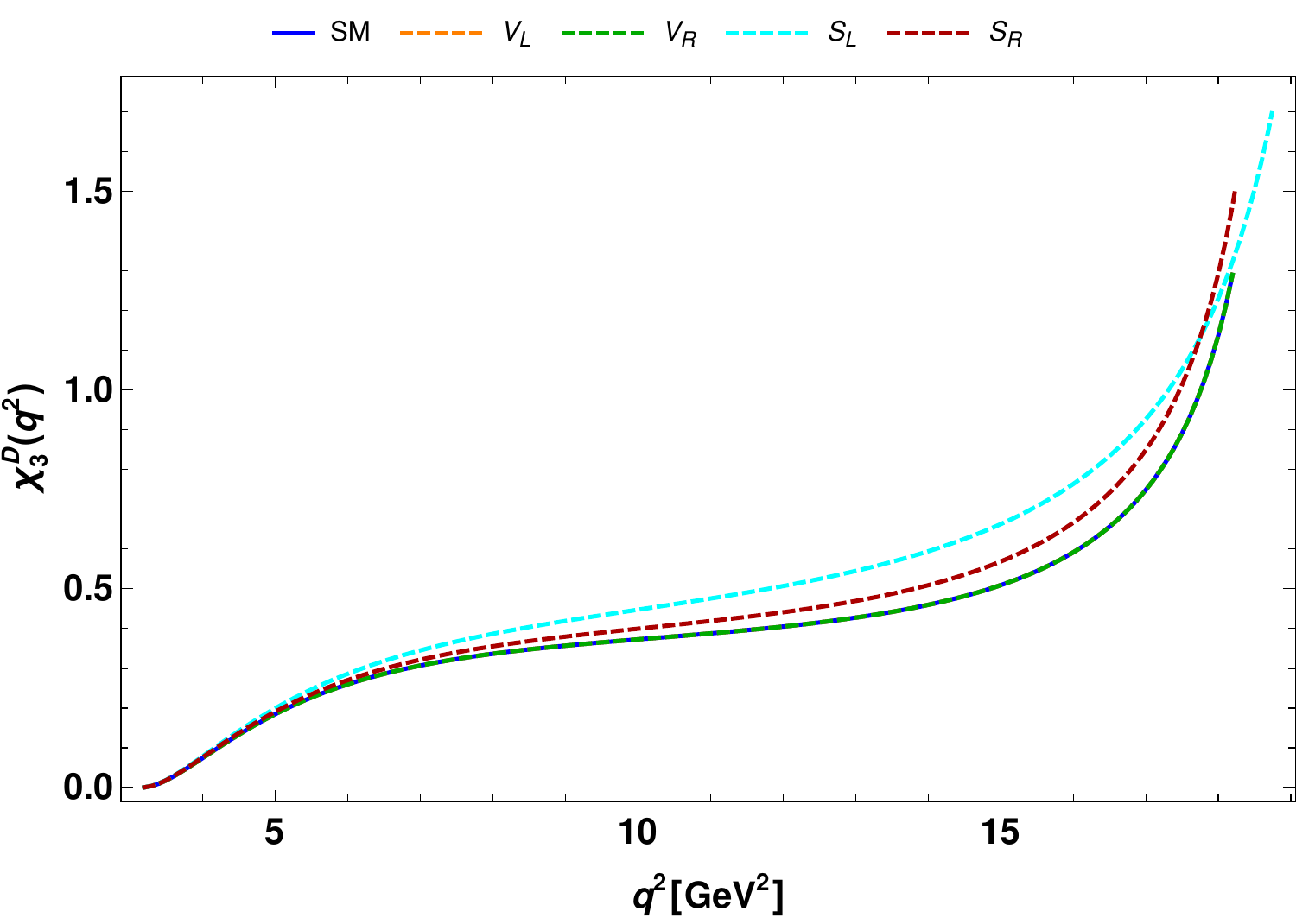}
\quad
\includegraphics[scale=0.45]{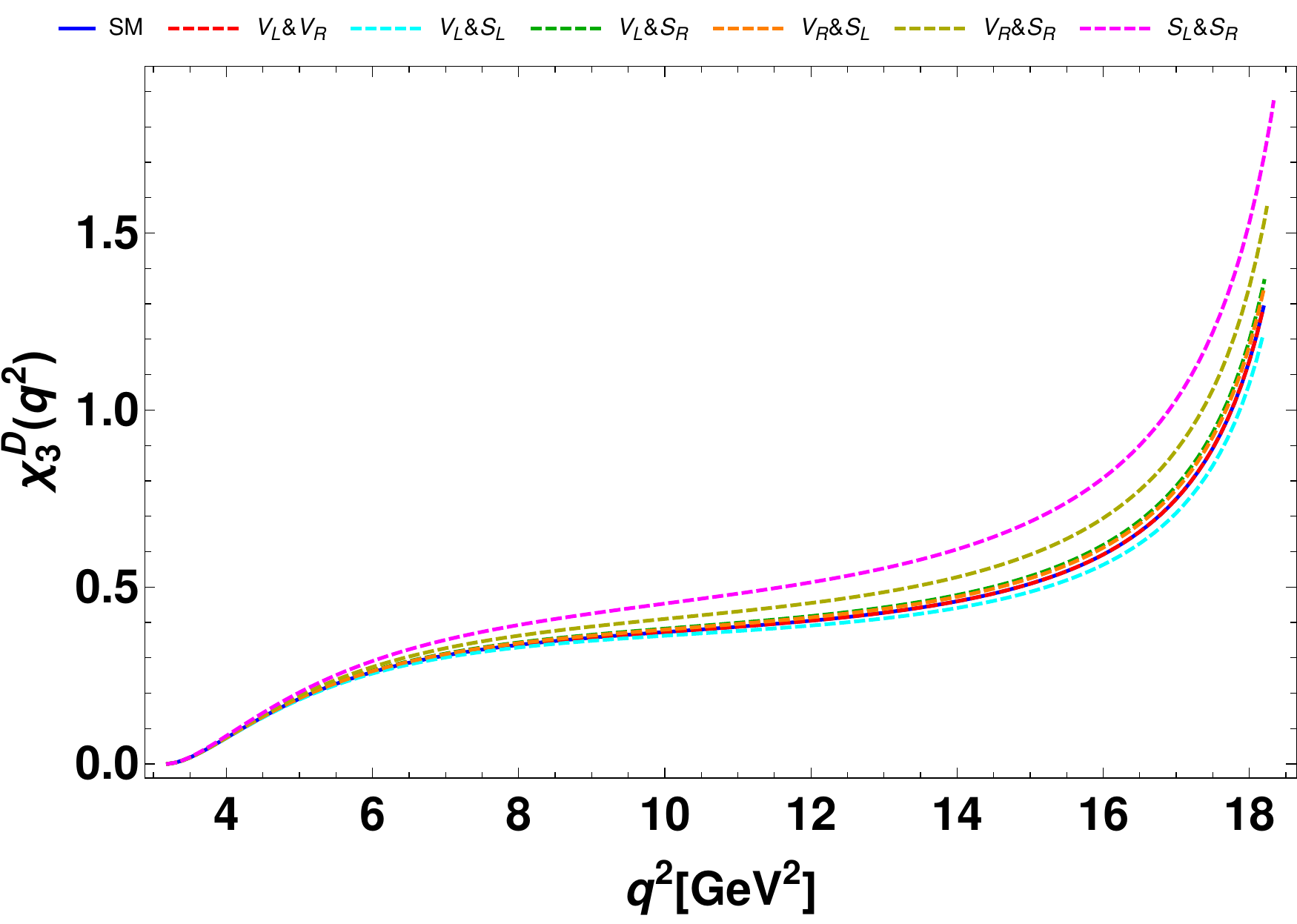}
\quad
\includegraphics[scale=0.52]{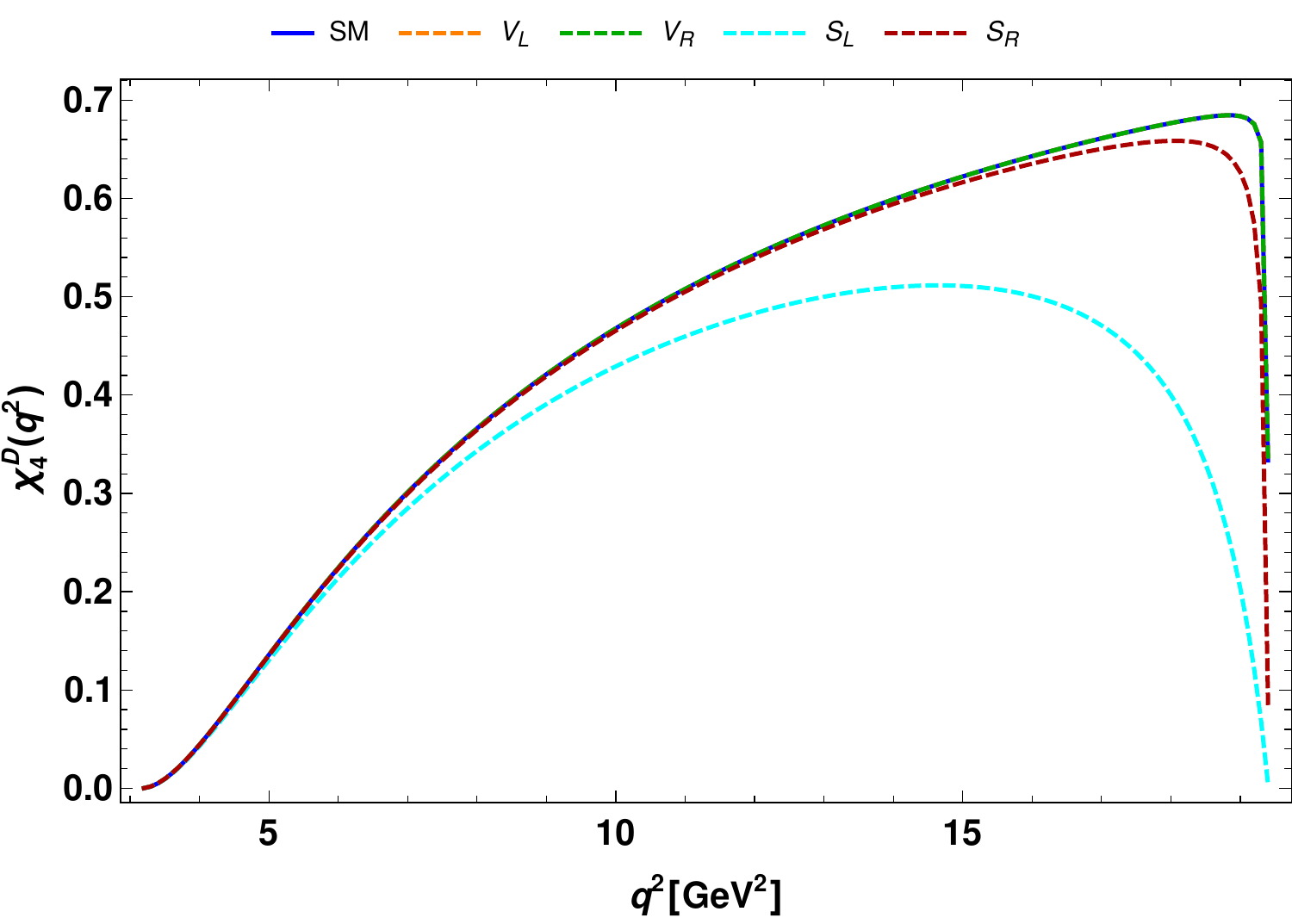}
\quad
\includegraphics[scale=0.45]{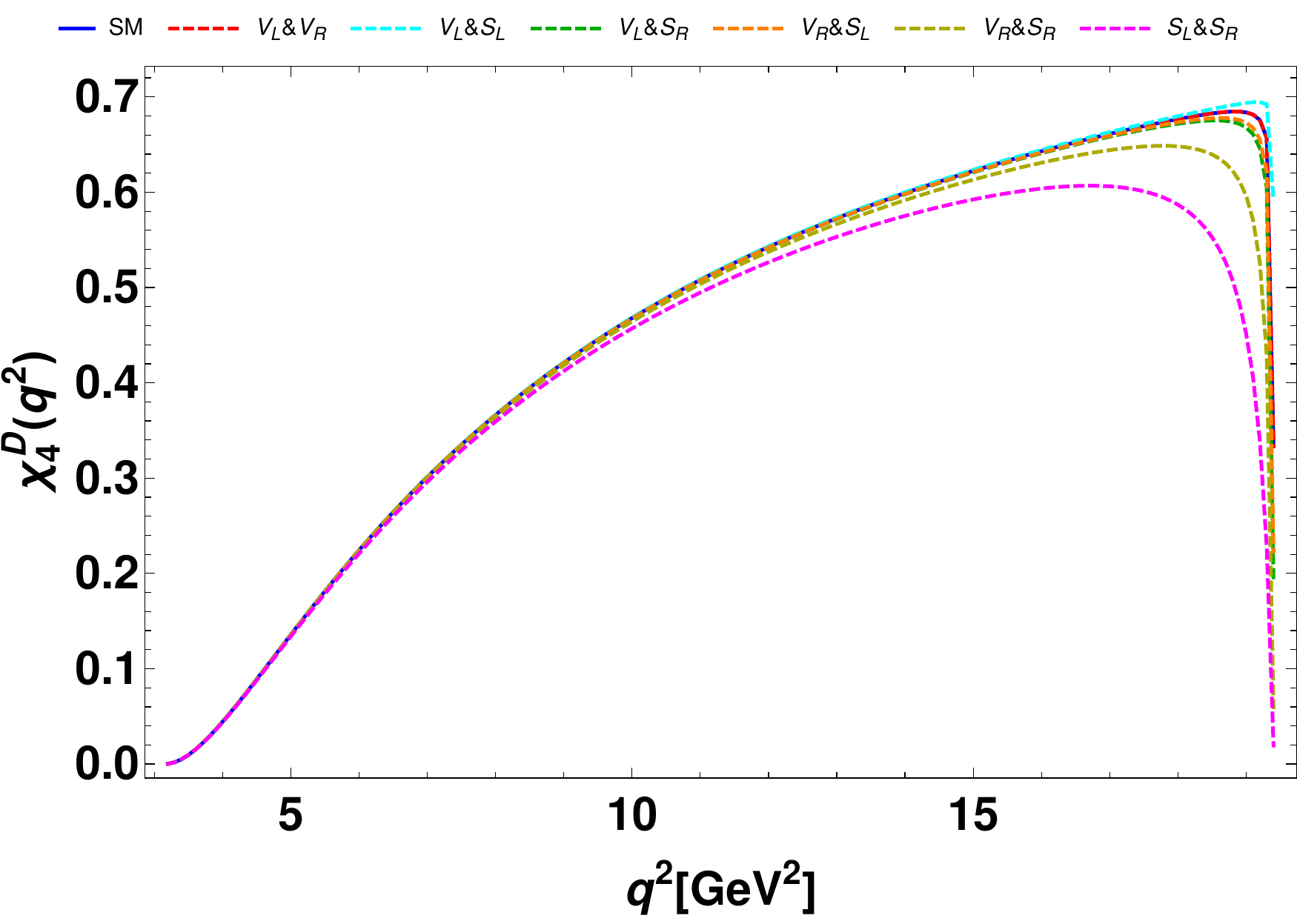}
\caption{The  $q^2$ variation of $\chi^D_{1,2,3,4}$  observables of $B_c\to D\tau \nu_\tau$   process for case A (left) and case B (right), respectively.} \label{Fig:chi-BD}
\end{figure}
 The variation $\tau$ forward and backward fractions ($\chi_{1,2}^{D}$) and $\tau$ spin 
 $1/2$ and $-1/2$ fractions ($\chi_{3,4}^{D}$) of $B_c\to D\tau \nu_\tau$ decay mode 
 with respect to $q^2$ in the presence of new complex and real 
 coefficient are given in Fig. \ref{Fig:chi-BD}\,. Here,
the plots for $q^2$ distribution of $\chi_1^D$ (top), $\chi_2^D$ (second from top), $\chi_1^D$ (third from top) and $\chi_4^D$ (bottom) observables for case A (left panel) and case B (right panel) are shown.
 We notice that $q^2$ the distribution looks quite similar in $\chi_{1}^D$, $\chi_2^D$,
 $\chi_3^D$, that differs from $q^2$ the distribution of $\chi_{4}^D$ in both case A and case B. 
 But the NP behaviour is quite similar 
 in $\chi_{1,2,3,4}^D$ once we include the new Wilson coefficients in both case A and case B. Thus, 
in case A, the Wilson coefficient $S_L$ ($S_R$) causes significant (small) deviation from SM prediction, 
while in case B $(S_L, S_R)$ ($(V_R, S_R)$) set provides significant (small) deviation from SM prediction 
for $\chi_{1,2,3,4}^D$. Again, we want to point out the fact that, $\chi_{4}^D$ deviate profoundly from SM predictions, where the peak is shifted to lower $q^2$ region when we add complex $S_L$ coefficient 
in case A as is evident from bottom left panel of Fig. \ref{Fig:chi-BD}\,.   

\begin{figure}[htb]
\includegraphics[scale=0.52]{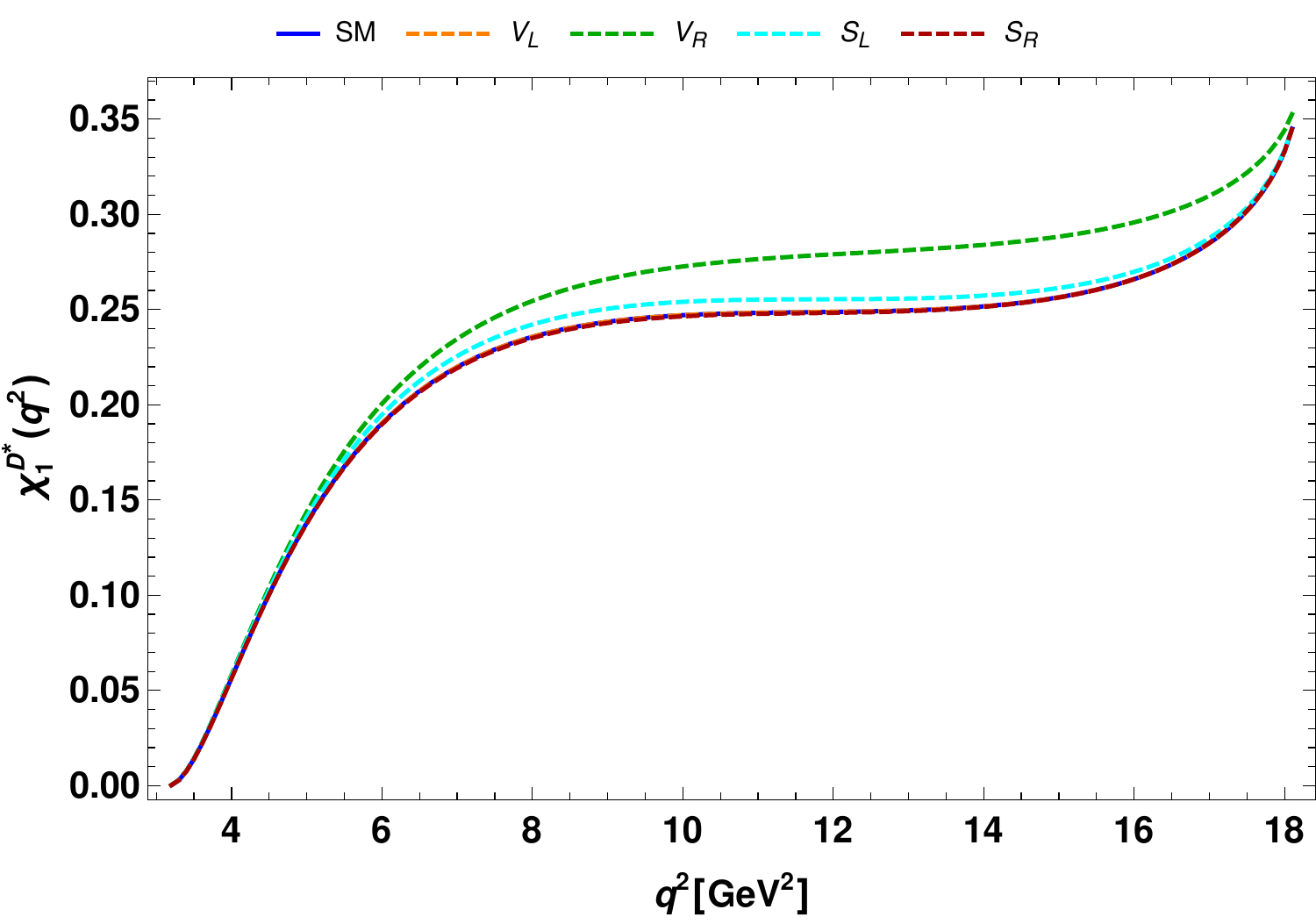}
\quad
\includegraphics[scale=0.52]{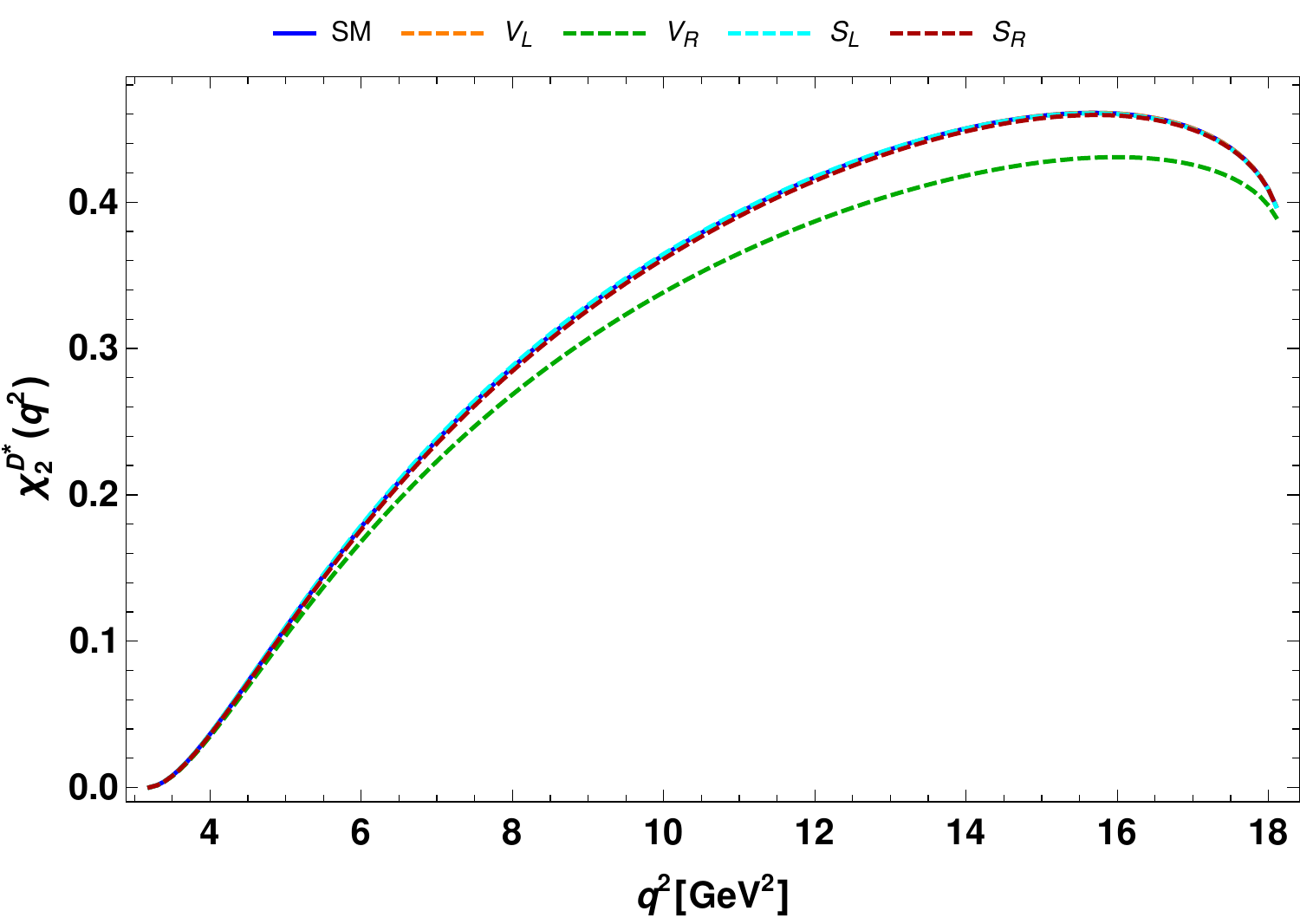}
\quad
\includegraphics[scale=0.52]{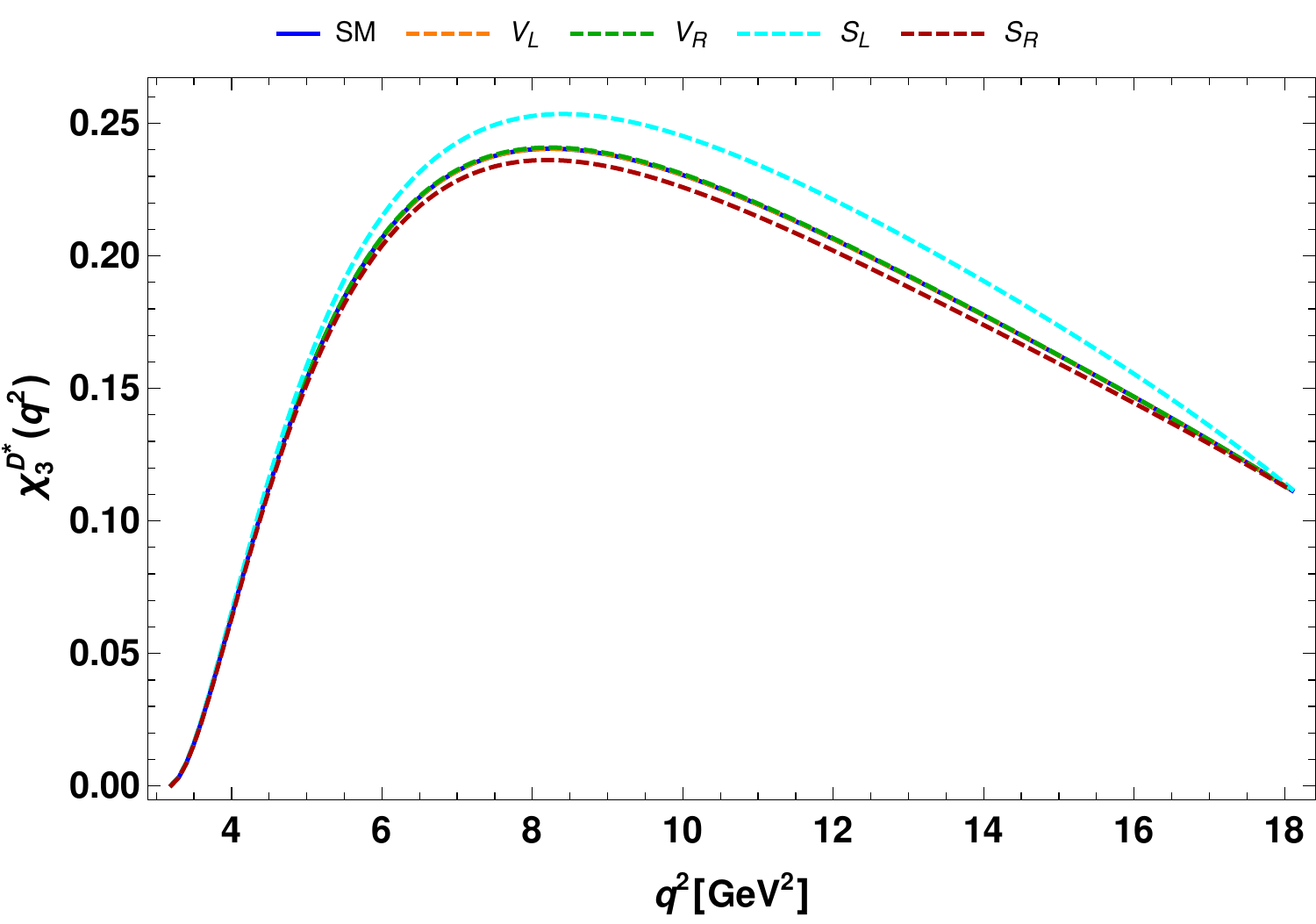}
\quad
\includegraphics[scale=0.52]{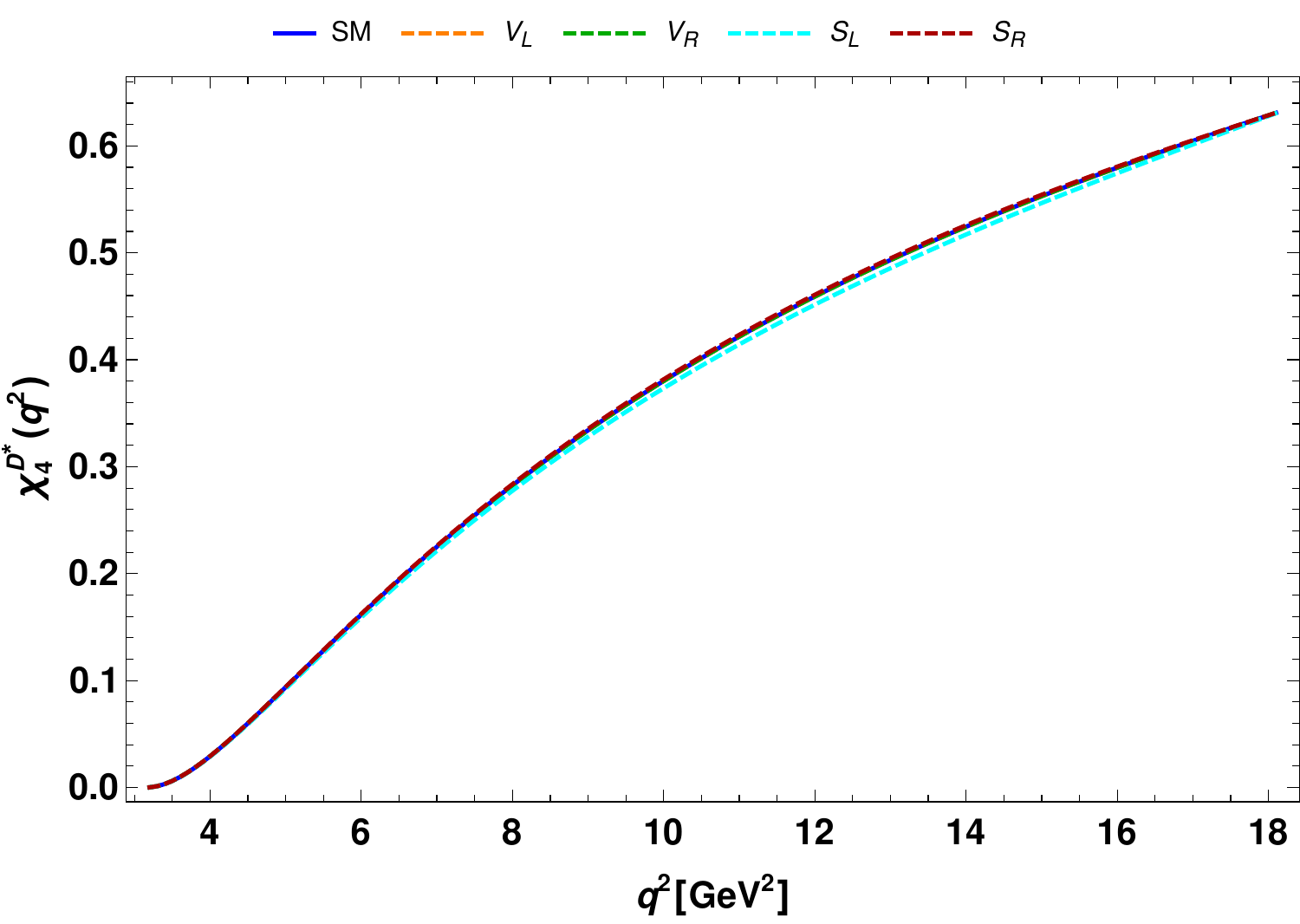}
\quad
\includegraphics[scale=0.52]{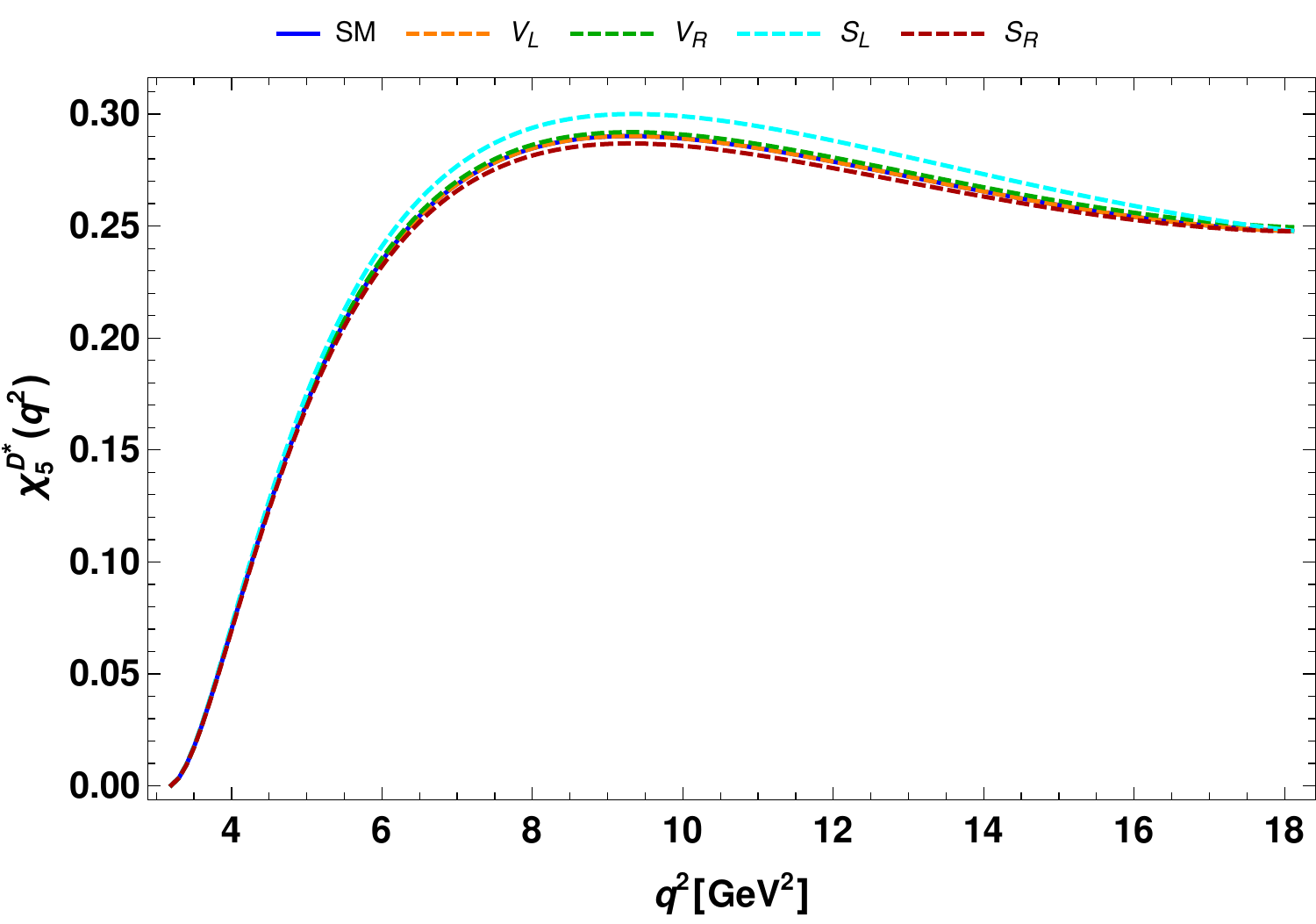}
\quad
\includegraphics[scale=0.52]{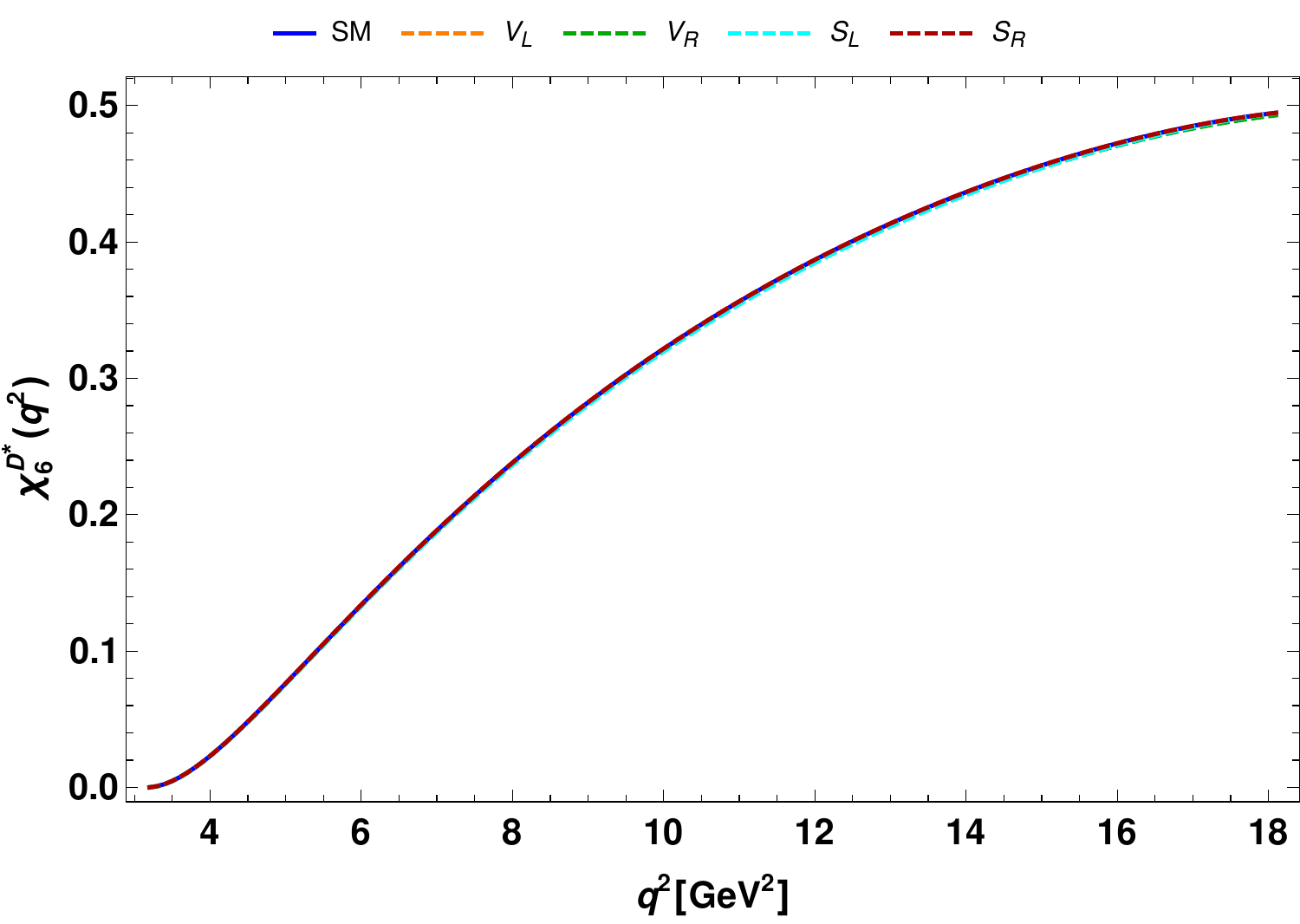}
\caption{The  $q^2$ variation of $\chi^{D^*_{1,2,3,4,5,6}}$  observables of $B_c\to D^* \tau \nu_\tau$   process for case A.} \label{Fig:CA-chi-BDstar}
\end{figure}

\begin{figure}[htb]
\includegraphics[scale=0.44]{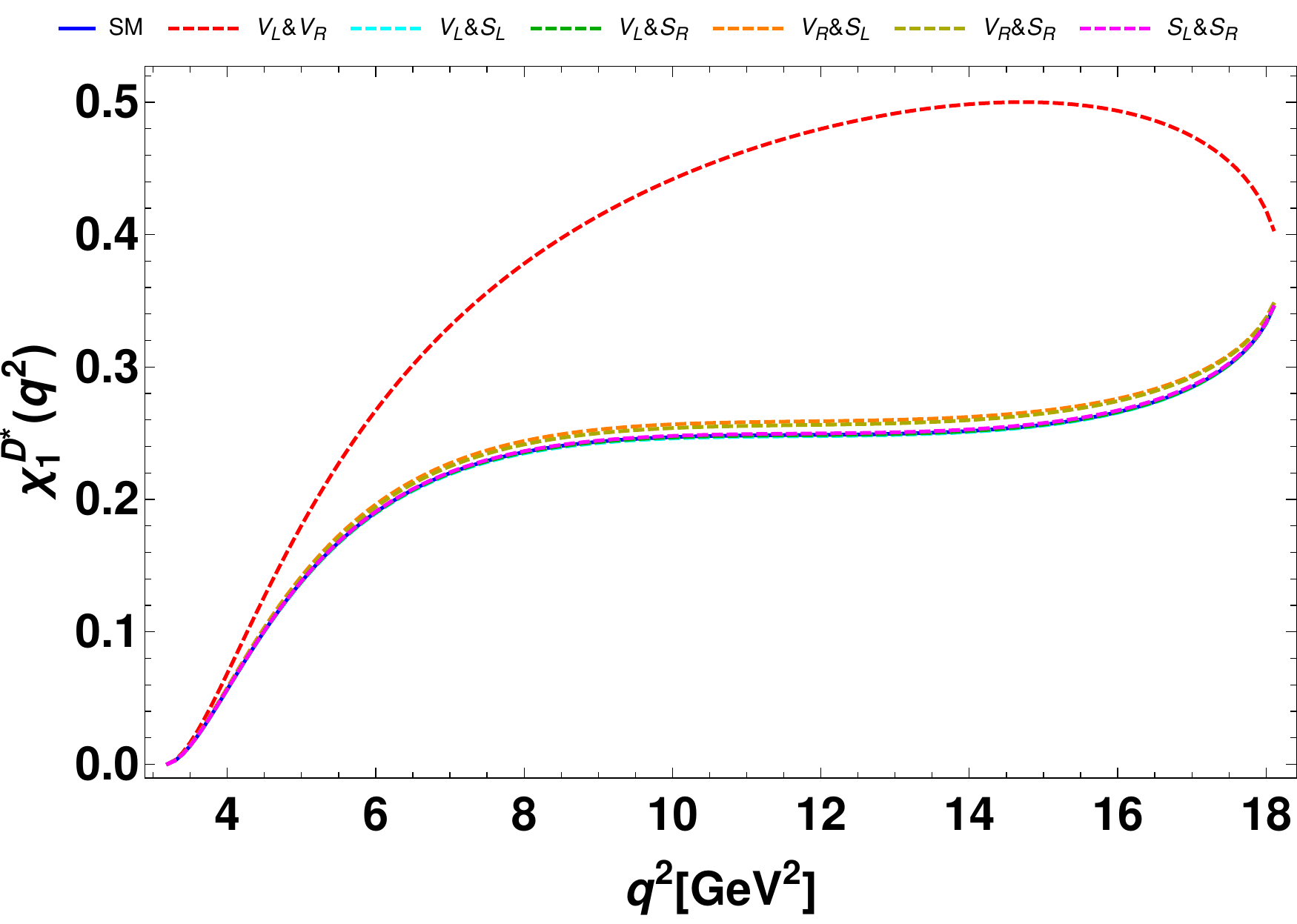}
\quad
\includegraphics[scale=0.44]{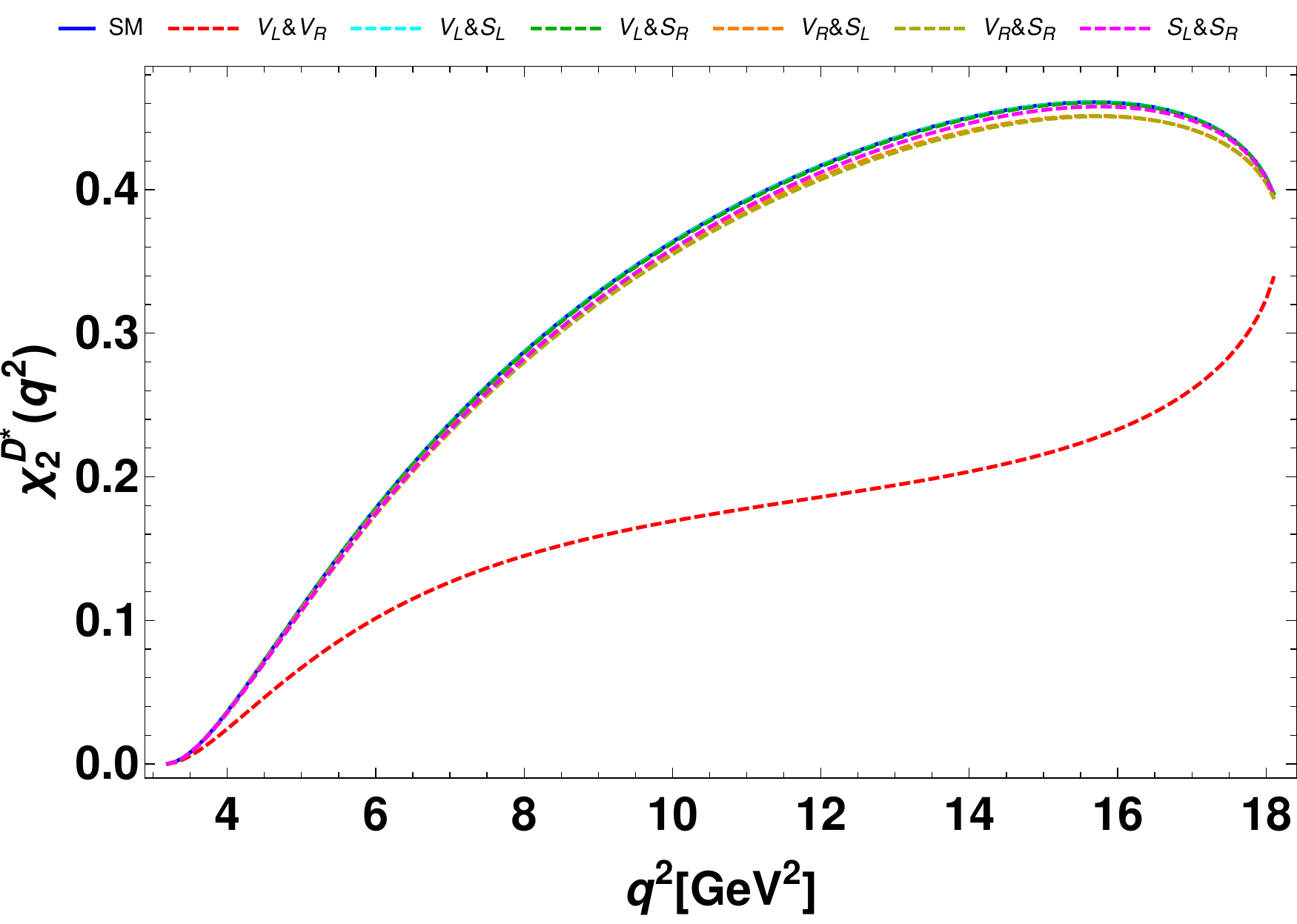}
\quad
\includegraphics[scale=0.44]{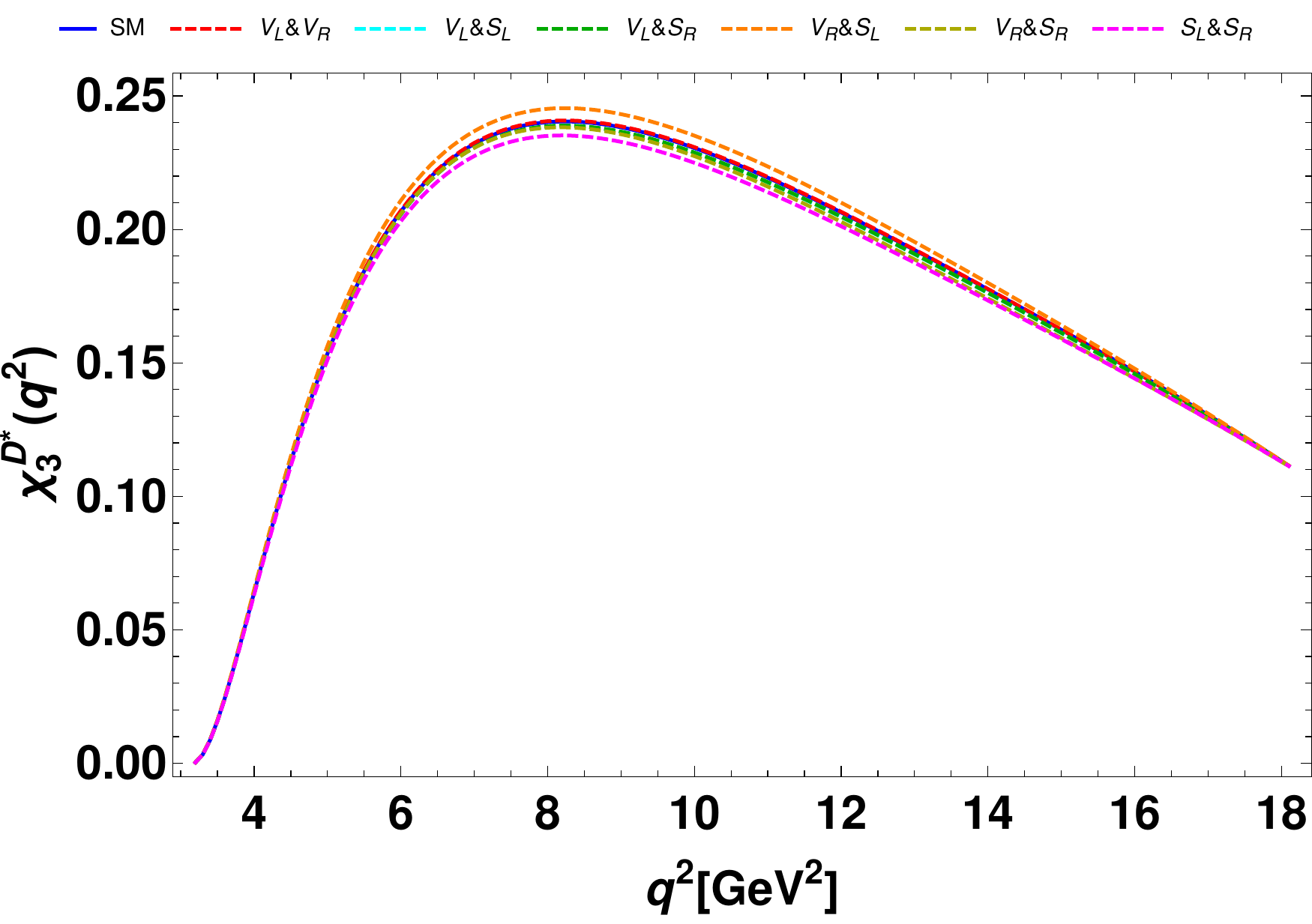}
\quad
\includegraphics[scale=0.44]{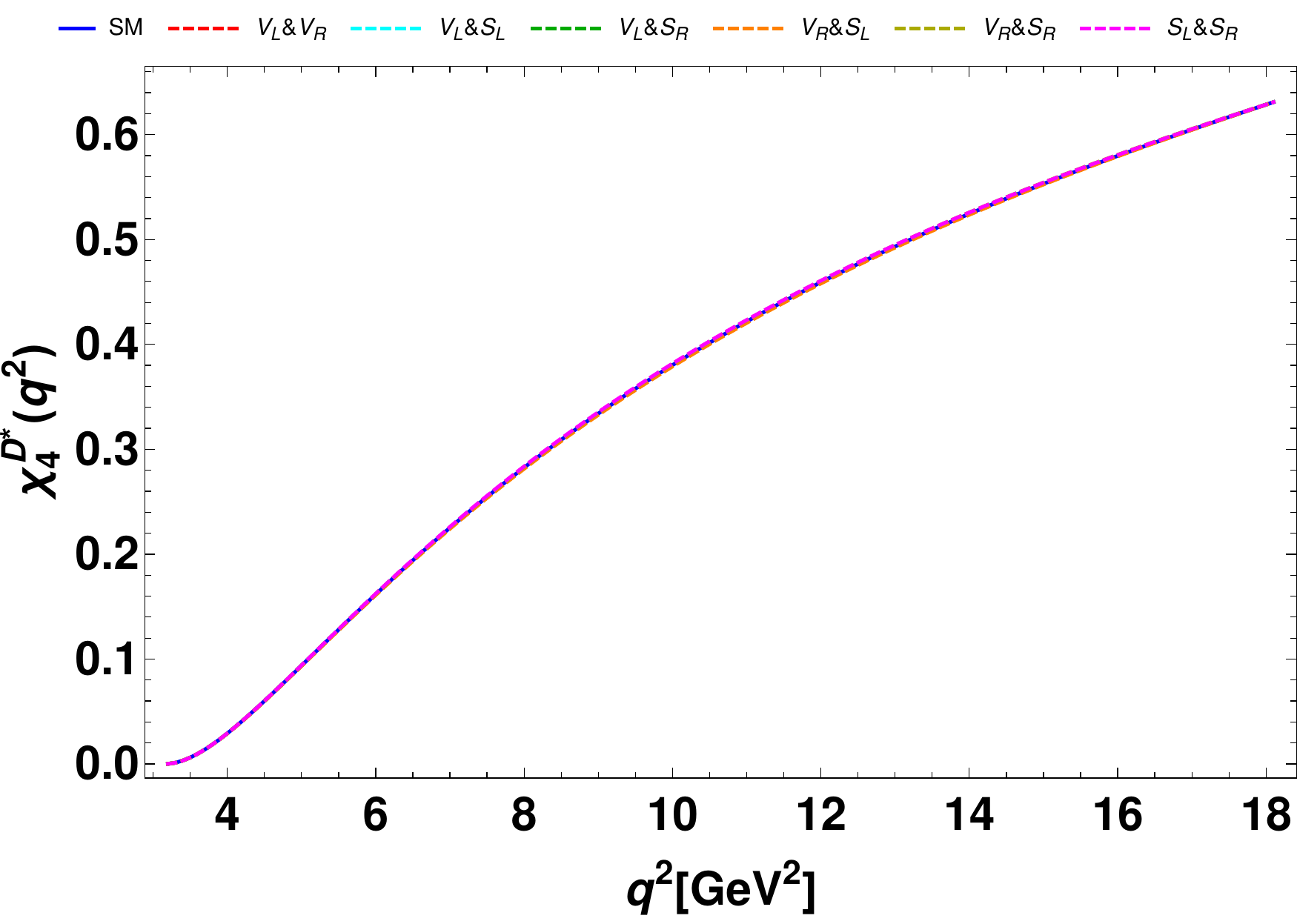}
\quad
\includegraphics[scale=0.44]{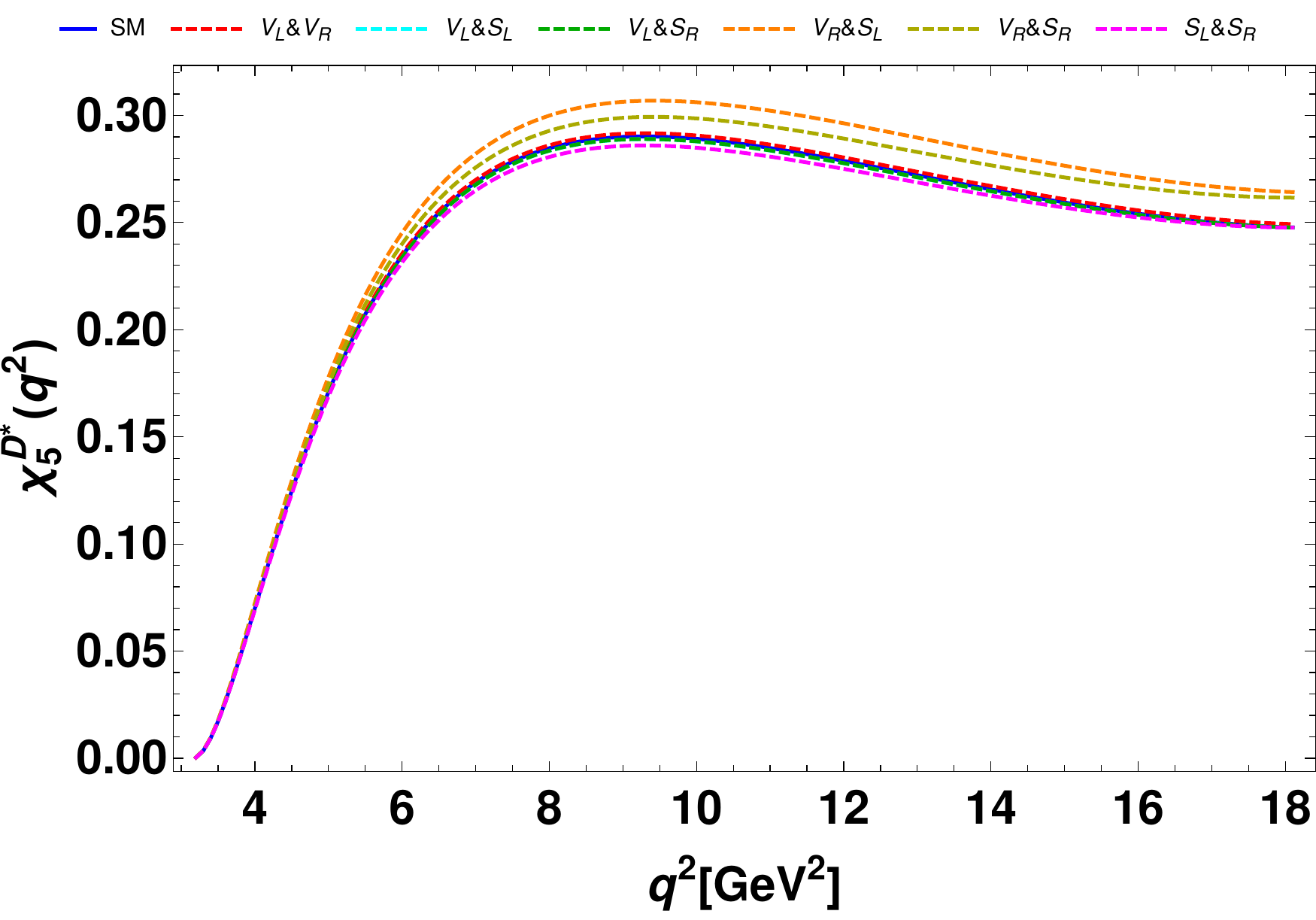}
\quad
\includegraphics[scale=0.44]{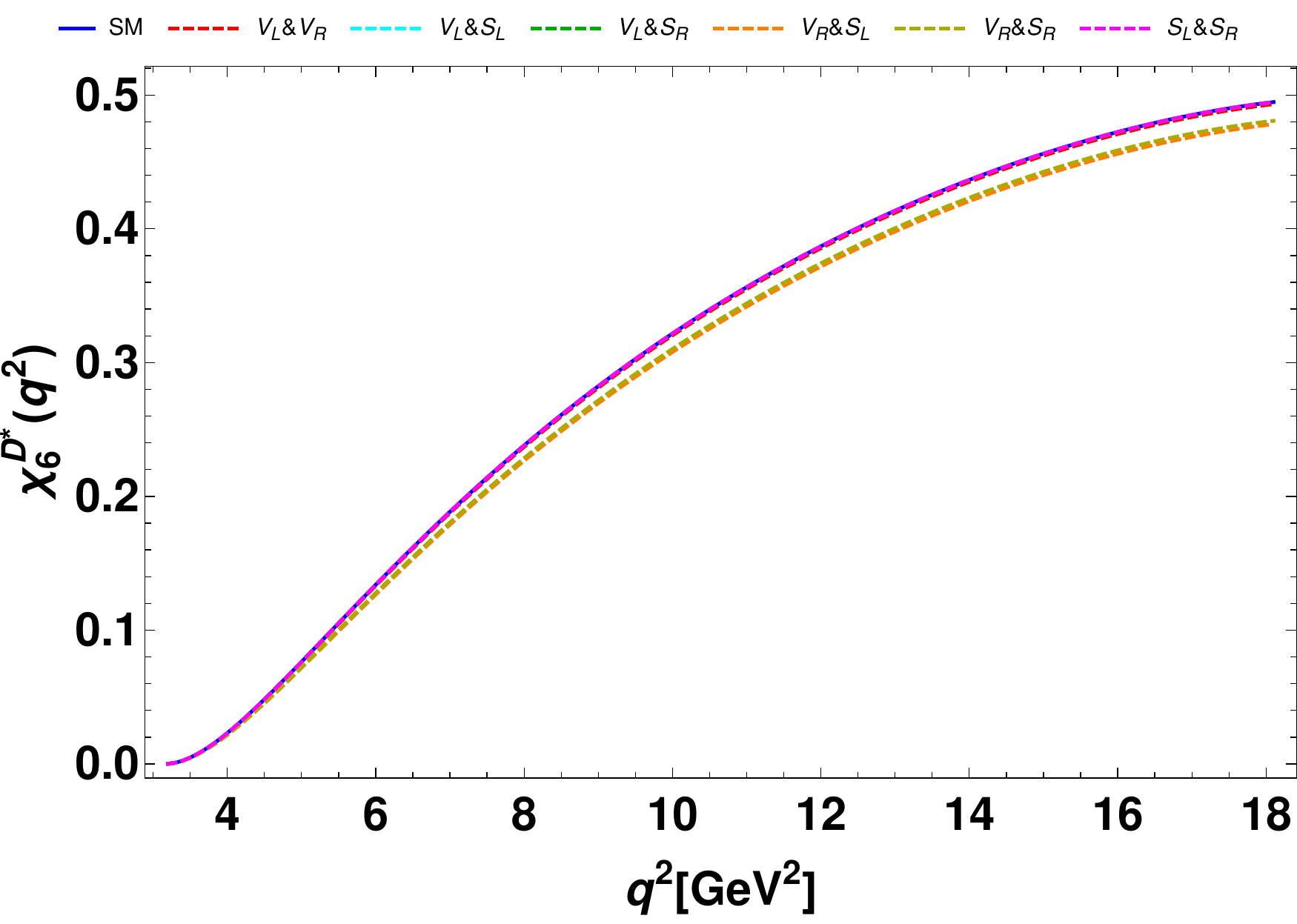}
\caption{The  $q^2$ variation of $\chi^{D^*_{1,2,3,4,5,6}}$  observables of $B_c\to D^* \tau \nu_\tau$   process for case B.} \label{Fig:CB-chi-BDstar}
\end{figure}

 The $\chi_{1,2,3,4,5,6}^{D^*}$ plots obtained by using the complex (case A) and real (case B) new Wilson coefficients are shown in Fig. \ref{Fig:CA-chi-BDstar} and \ref{Fig:CB-chi-BDstar}\,, respectively. 
 In Fig. \ref{Fig:CA-chi-BDstar} and \ref{Fig:CB-chi-BDstar}\, the $q^2$ the distributions are different 
 for 
 $\chi_{1}^{D^*}$, $\chi_2^{D^*}$, $\chi_3^{D^*}$, $\chi_{5}^{D^*}$, but similar for $\chi_4^{D^*}$ and 
 $\chi_6^{D^*}$. In Fig. \ref{Fig:CA-chi-BDstar}, we observe small deviation in $\chi_{1}^{D^*}$, 
 $\chi_2^{D^*}$ due to inclusion of the complex Wilson coefficient $V_R$ and in $\chi_3^{D^*}$, 
 $\chi_{5}^{D^*}$ (very little) due to $S_L$ coefficient but, no deviation in $\chi_4^{D^*}$ 
 and $\chi_6^{D^*}$ due to any of the complex coefficients for case A. 
Similarly, for case B, in Fig \ref{Fig:CB-chi-BDstar}\, we see $\chi_{1}^{D^*}$ and $\chi_2^{D^*}$ are seriously affected by $(V_L, V_R)$ set of real coefficients, $\chi_5^{D^*}$ and $\chi_6^{D^*}$ are very less 
affected by $(V_R, S_L)$, $(V_R, S_R)$ sets of coefficients, $\chi_{3}^{D^*}$ is negligibly 
affected only at the peak and $\chi_{4}^{D^*}$ is independent of all sets of real coefficients.  

Table \ref{Tab:CA-BD} and \ref{Tab:CA-BDstar} contain the numerical values of the branching ratios and all the discussed angular observables of $B_c \to D \tau \nu_\tau$ and  $B_c \to D^* \tau \nu_\tau$ decay processes in the presence of individual complex Wilson coefficient. The predicted values of branching ratio, forward-backward asymmetry, LNU parameters, $\tau$ and $D^*$ polarization asymmetry of  $B_c \to D$ ($B_c \to D^*$) mode for case B are represented in Table \ref{Tab:CB-BD} ( \ref{Tab:CB-BDstar} )\,.

\begin{table}[htb]
\centering
\caption{Predicted  numerical values of  branching ratio and all the discussed angular observables   of $B_c \to D \tau \nu_\tau$ process in the presence of individual complex Wilson coefficients. }\label{Tab:CA-BD}
\begin{tabular}{|c |c|c|c|c|c|}
\hline
~Observables~&~Values in SM~&~Values in $V_L$~&~Values in $V_R$~&~Values in $S_L$~&~Values in $S_R$~\\
\hline\hline
$\rm{Br}(B_c\to D \tau \bar \nu_\tau)\times 10^5$~&~$1.88\pm 0.131$~&~$2.165$~&~$2.132$~&~$2.282$~&~$1.972$~\\
$\langle A_{FB}^D \rangle$~&~$0.295$~&~$0.295$~&~$0.295$~&~$0.28$~&~$0.296$~\\
$\langle R_D \rangle$~&~$0.7$~&~$0.7$~&~$0.7$~&~$0.76$~&~$0.731$~\\
$\langle P_\tau^D\rangle$~&~$-0.046$~&~$-0.046$~&~$-0.046$~&~$0.138$~&~$0.0023$~\\
$\langle \chi_1^D \rangle$~&~$0.454$~&~$0.454$~&~$0.454$~&~$0.486$~&~$0.474$~\\
$\langle \chi_2^D \rangle$~&~$0.247$~&~$0.247$~&~$0.247$~&~$0.274$~&~$0.257$~\\
$\langle \chi_3^D \rangle$~&~$0.335$~&~$0.335$~&~$0.335$~&~$0.432$~&~$0.366$~\\
$\langle \chi_4^D \rangle$~&~$0.367$~&~$0.367$~&~$0.367$~&~$0.328$~&~$0.365$~\\
\hline
\end{tabular}
\end{table}

\begin{table}[htb]
\centering
\caption{Predicted  numerical values of  branching ratio and all the discussed angular observables   of $B_c \to D^* \tau \nu_\tau$ process in the presence of individual complex Wilson coefficients. }\label{Tab:CA-BDstar}
\begin{tabular}{|c |c|c|c|c|c|}
\hline
~Observables~&~Values in SM~&~Values in $V_L$~&~Values in $V_R$~&~Values in $S_L$~&~Values in $S_R$~\\
\hline\hline
$\rm{Br}(B_c\to D^* \tau \bar \nu_\tau)\times 10^5$~&~$2.3\pm 0.161$~&~$2.65$~&~$2.66$~&~$2.33$~&~$2.292$~\\
$\langle A_{FB}^{D^*} \rangle$~&~$-0.221$~&~$-0.221$~&~$-0.14$~&~$-0.211$~&~$-0.219$~\\
$\langle R_{D^*} \rangle$~&~$0.603$~&~$0.603$~&~$0.603$~&~$0.608$~&~$0.601$~\\
$\langle P_\tau^{D^*}\rangle$~&~$-0.41$~&~$-0.41$~&~$-0.41$~&~$-0.378$~&~$-0.418$~\\
$\langle F_L^{D^*}\rangle$~&~$0.412$~&~$0.412$~&~$0.415$~&~$0.42$~&~$0.41$~\\
$\langle F_T^{D^*}\rangle$~&~$0.588$~&~$0.588$~&~$0.585$~&~$0.58$~&~$0.59$~\\
$\langle \chi_1^{D^*} \rangle$~&~$0.235$~&~$0.235$~&~$0.26$~&~$0.24$~&~$0.235$~\\
$\langle \chi_2^{D^*} \rangle$~&~$0.368$~&~$0.368$~&~$0.344$~&~$0.368$~&~$0.366$~\\
$\langle \chi_3^{D^*} \rangle$~&~$0.1782$~&~$0.1782$~&~$0.1784$~&~$0.189$~&~$0.175$~\\
$\langle \chi_4^{D^*} \rangle$~&~$0.425$~&~$0.425$~&~$0.425$~&~$0.419$~&~$0.426$~\\
$\langle \chi_5^{D^*} \rangle$~&~$0.249$~&~$0.249$~&~$0.25$~&~$0.26$~&~$0.246$~\\
$\langle \chi_6^{D^*} \rangle$~&~$0.355$~&~$0.355$~&~$0.353$~&~$0.353$~&~$0.355$~\\
\hline
\end{tabular}
\end{table}

\begin{table}[htb]
\centering
\caption{Predicted  numerical values of  branching ratio and all the angular observables of $B_c \to D \tau \nu_\tau$ process for case B. }\label{Tab:CB-BD}
\begin{tabular}{|c |c|c|c|c|c|c|c|c|c|c|}
\hline
~Models~&~$\rm{Br}(B_c\to D\tau \bar \nu_l)\times 10^5$~&~$\langle A_{FB}^D \rangle$~&~$\langle R_D \rangle$~&~$\langle P_\tau^D\rangle$~&~$\langle \chi_1^D\rangle$~&~$\langle \chi_2^D\rangle$~&~$\langle \chi_3^D\rangle$~&~$\langle \chi_4^D\rangle$~\\
\hline\hline
~($V_L,V_R$)~&~$2.134$~&~$0.295$~&~$0.7$~&~$-0.046$~&~$0.454$~&~$0.247$~&~$0.335$~&~$0.367$~\\
~($V_L,S_L$)~&~$2.124$~&~$0.294$~&~$0.69$~&~$-0.065$~&~$0.446$~&~$0.244$~&~$0.323$~&~$0.367$~\\
~($V_L,S_R$)~&~$2.185$~&~$0.2952$~&~$0.712$~&~$-0.028$~&~$0.461$~&~$0.251$~&~$0.346$~&~$0.366$~\\
~($V_R,S_L$)~&~$1.712$~&~$0.295$~&~$0.71$~&~$-0.033$~&~$0.459$~&~$0.25$~&~$0.343$~&~$0.366$~\\
~($V_R,S_R$)~&~$1.84$~&~$0.297$~&~$0.742$~&~$0.02$~&~$0.481$~&~$0.261$~&~$0.378$~&~$0.363$~\\
~($S_L,S_R$)~&~$2.173$~&~$0.297$~&~$0.786$~&~$0.095$~&~$0.51$~&~$0.277$~&~$0.43$~&~$0.356$~\\
\hline
\end{tabular}
\end{table}

\begin{table}[htb]
\centering
\caption{Predicted  numerical values of  branching ratio and all the angular observables of $B_c \to D^* \tau \nu_\tau$ process for case B. }\label{Tab:CB-BDstar}
\begin{tabular}{|c |c|c|c|c|c|c|c|c|c|c|c|c|c|}
\hline
~Models~&~$\rm{Br}(B_c\to D^* \tau \bar \nu_\tau)$~&~$\langle A_{FB}^{D^*} \rangle$~&~$\langle R_{D^*} \rangle$~&~$\langle P_\tau^{D^*}\rangle$~&~$\langle F_L^{D^*}\rangle$~&~$\langle \chi_1^D\rangle$~&~$\langle \chi_2^D\rangle$~&~$\langle \chi_3^D\rangle$~&~$\langle \chi_4^D\rangle$~&~$\langle \chi_5^D\rangle$~&~$\langle \chi_6^D\rangle$~\\
\hline\hline

~($V_L,V_R$)~&~$2.652\times 10^{-5}$~&~$0.4$~&~$0.603$~&~$-0.41$~&~$0.414$~&~$0.421$~&~$0.182$~&~$0.178$~&~$0.425$~&~$0.25$~&~$0.353$~\\
~($V_L,S_L$)~&~$2.64 \times 10^{-5}$~&~$-0.222$~&~$0.602$~&~$-0.412$~&~$0.41$~&~$0.234$~&~$0.368$~&~$0.177$~&~$0.425$~&~$0.248$~&~$0.352$~\\
~($V_L,S_R$)~&~$2.62\times 10^{-5}$~&~$-0.22$~&~$0.602$~&~$-0.412$~&~$0.41$~&~$0.235$~&~$0.368$~&~$0.177$~&~$0.425$~&~$0.25$~&~$0.355$~\\
~($V_R,S_L$)~&~$2.49\times 10^{-5}$~&~$-0.195$~&~$0.61$~&~$-0.4$~&~$0.437$~&~$0.244$~&~$0.36$~&~$0.18$~&~$0.425$~&~$0.264$~&~$0.341$~\\
~($V_R,S_R$)~&~$2.44\times 10^{-5}$~&~$-0.197$~&~$0.602$~&~$-0.417$~&~$0.43$~&~$0.242$~&~$0.36$~&~$0.175$~&~$0.426$~&~$0.259$~&~$0.343$~\\
~($S_L,S_R$)~&~$2.3\times 10^{-5}$~&~$-0.215$~&~$0.6$~&~$-0.42$~&~$0.41$~&~$0.236$~&~$0.365$~&~$0.174$~&~$0.426$~&~$0.246$~&~$0.355$~\\
\hline
\end{tabular}
\end{table}

\section{Conclusion}
In this paper, we have investigated $B_c \to D^{(*)}\tau \bar \nu_\tau$ semileptonic decays mediated by 
$b \to u \tau \nu_\tau$  quark level transitions in an effective field theory approach. We consider the presence of an additional vector and scalar type interactions which can be either complex or real 
catagorized as two cases. In case A, existence of only  individual  new complex  coefficient and case B, existence of only two  new real coefficients and we perform a chi-square fitting to extract the best-fit values of these new Wilson coefficients from the experimental data on $R_{D^{(*)}}$, $R_{J/\psi}$, 
$R_\pi^l$, Br($B_{u,c} \to \tau \bar \nu_\tau$) and Br($B \to\pi \tau \bar \nu_\tau$) observables. 
We estimate the branching ratio, forward-backward 
 asymmetry ($\A_{\rm FB}$), lepton non-universality ($R_{D^{(*)}}^{B_c}$), 
 $\tau$ polarization asymmetry ($P_\tau^{D^{(*)}}$), $D^*$ longitudinal and transverse polarization 
 ($F_{L, T}^{D^*}$), $\tau$ forward and backward fractions ($\chi_{1,2}^{D^{(*)}}$), $\tau$ spin 
 $1/2$ and $-1/2$ fractions ($\chi_{3,4}^{D^{(*)}}$), $D^*$ longitudinal and transverse polarization 
 fractions ($\chi_{5,6}^{D^{*}}$) of $B_c \to D \tau \bar \nu_\tau$ and 
 $B_c \to D^* \tau \bar \nu_\tau$ decay modes for both case A and case B using
the best-fit values of real and complex Wilson coefficients. 

The presence of complex $S_L$ coefficient provide significant deviation from SM prediction in the 
differential branching ratio of $B_c \to D\tau \bar \nu_\tau$ process, where the peak of the 
distribution of differential branching ratio can shift to a higher $q^2$ region, while $V_L$, $V_R$ coefficients show a small deviation, and $S_R$ coupling results in no deviation for case A.
For case B, the differential branching ratio of $(B_c \to D \tau \bar \nu_\tau)$ has deviated 
from the SM prediction for all possible sets of new coefficients except $(V_R, S_R)$ set of 
real coefficients while the peak can be shifted to a higher 
value $q^2$ for ($S_L, S_R)$ set of coefficients.
Similarly the presence of complex $V_L$ and $V_R$ coefficients result in significant 
deviation in differential branching ratio for $B_c \to D^* \tau \bar \nu_\tau$ decay in case A 
while in case B, the sets of ($V_L,V_R$), $(V_L, S_L)$ and $(V_L,S_R)$ coefficients give 
significant deviation.
 
 We find almost no deviation from SM predictions in forward backward asymmetry for $B_c \to D \tau \bar \nu_\tau$ decay for both case A and case B except a very little deviation for complex $S_L$ coefficient higher $q^2$ region. This is expected, since it is a ratio, the NP 
dependency gets canceled in the ratio.
But the forward-backward asymmetry of 
$B_c \to D^* \tau \bar \nu_\tau$ process show profound deviation due to the presence of  
$V_R$ in case A and $(V_L, V_R)$ coefficients in case B.
  
The lepton non-universality parameter, $R_D^{B_c}$ of $B_c \to D \tau \bar \nu_\tau$ decay process 
is deviated slightly for complex $S_L$ coefficient in higher $q^2$ region in case A and for $(V_R, S_R)$, 
$(S_L, S_R)$ real coefficients in case B. But, $R_{D^*}^{B_c}$ of $B_c \to D^* \tau \bar \nu_\tau$ decay mode is not influenced by any of the NP coefficients in both case A and case B. It is worth mentioning that, the lepton non-universality parameter for both $B_c \to D \tau \bar \nu_\tau$ and 
$B_c \to D^* \tau \bar \nu_\tau$ decay modes, does not differ significantly with the NP coefficients as anticipated, because, the impact of NP coefficients gets canceled in the ratio of lepton non-universality parameter.
   
In the $\tau$ longitudinal polarization asymmetry ($P_\tau^{D}$) of $B_c\to D\tau \nu_\tau$  process, the presence of only complex $S_L$ in case A and the real 
$(S_L, S_R)$ set provide maximum deviation from the SM predictions in case B. There is no deviation in  $P_\tau^{D^*}$ observable of $B_c\to D^*\tau \nu_\tau$ process from SM predictions, for both case A and case B. Similarly, $D^*$ longitudinal and transverse polarization ($F_{L, T}^{D^*}$) of $B_c\to D^*\tau \nu_\tau$ decay mode is independent of all complex coefficients in case A. However, longitudinal and transverse polarization asymmetry 
($F_{L,T}^{D^*}$) slightly deviated from SM prediction for $(V_R, S_L)$ and $(V_R, S_R)$ sets of new coefficients in case B. 

Again, the
 spin $1/2$ and $-1/2$ fractions, $\chi_{4}^D$ of $B_c\to D\tau \nu_\tau$ decay mode deviate profoundly from SM predictions due to complex $S_L$ coefficient, where the peak is shifted to lower $q^2$ region in case A and due to real $(S_L, S_R)$ set of  coefficients in case B. 
In $B_c\to D^*\tau \nu_\tau$ decay, we observe small deviation in $\chi_{1}^{D^*}$, 
 $\chi_2^{D^*}$ due to inclusion of the complex Wilson coefficient $V_R$ and no deviation in $\chi_4^{D^*}$ 
 and $\chi_6^{D^*}$ due to any of the complex coefficients for case A. 
Similarly, for case B, we see $\chi_{1}^{D^*}$ and $\chi_2^{D^*}$ are seriously affected by $(V_L, V_R)$ set of real coefficients and $\chi_{4}^{D^*}$ is independent of all sets of real coefficients.  

The observables associated with $B_c\to D\tau \nu_\tau$ decay mode like differential branching ratio, 
forward backward asymmetry ($\A_{\rm FB}$), lepton non-universality parameter ($R_D^{B_c}$), $\tau$ longitudinal polarization asymmetry ($P_\tau^{D}$), spin $1/2$ and $-1/2$ fractions ($\chi_{4}^D$) are 
found to be mostly affected by complex $S_L$ coefficient in case A and real $(S_L, S_R)$ set of coefficients in case B. Similarly $B_c\to D^*\tau \nu_\tau$ decay, the observables differential branching ratio, forward backward asymmetry, $\chi_{1}^{D^*}$ and $\chi_2^{D^*}$ are influenced by complex $V_R$ coefficient while 
lepton non-universality parameter, $\chi_4^{D^*}$ and $\chi_6^{D^*}$ do not depend on any of the real cefficients in case A. In case B, forward backward asymmetry, $\chi_{1}^{D^*}$ and $\chi_2^{D^*}$ are seriously affected by $(V_L, V_R)$ set of real coefficients, but lepton non-universality parameter, 
$\chi_{4}^{D^*}$ are independent of all sets of real coefficients

At present, there exist a few
measurements of $B_c$ meson and its properties from Tevatron data. The LHCb experiments assure the first detailed study of $B_c$ meson. More precise
measurements of its mass and lifetime are now feasible, and several decay channels have
been witnessed for the first time.
LHCb experiment will provide additional
information with significant reduction of the uncertainties of the observables already measured
and measurement of new observables that can provide complementary information on  lepton flavor universality violation and the NP coefficients, thus we expect a better understanding of the different NP scenarios involved in
$b\to u\tau \nu_\tau$ transitions

\acknowledgments  

AB would like to thank Odisha State Higher Education Council (OSHEC), Government of Odisha for financial support through grant No. 16 Seed/2019/Physics-5.
 
\bibliography{btou}

\end{document}